\documentclass[%
 reprint,
 superscriptaddress,
 nofootinbib,
 amsmath,amssymb,
 aps,
]{revtex4-2}

\usepackage{graphicx}
\usepackage{dcolumn}
\usepackage{bm}
\usepackage{hyperref}
\usepackage[dvipsnames]{xcolor}

\newcommand{\hMpc}{h^{-1}\,\mathrm{Mpc}}
\newcommand{\Mpc}{\mathrm{Mpc}}
\newcommand{\hGpcvol}{h^{-3}\,\mathrm{Gpc}^{3}}
\newcommand{\hMpcvol}{h^{-3}\,\mathrm{Mpc}^{3}}
\newcommand{\hinvMpc}{h\,\mathrm{Mpc}^{-1}}
\newcommand{\hinvMpcvol}{h^3\,\mathrm{Mpc}^{-3}}

\newcommand{\bx}{\boldsymbol{x}}
\newcommand{\br}{\boldsymbol{r}}
\newcommand{\bk}{\boldsymbol{k}}
\newcommand{\bs}{\boldsymbol{s}}
\newcommand{\bu}{\boldsymbol{u}}
\newcommand{\bv}{\boldsymbol{v}}
\newcommand{\bq}{\boldsymbol{q}}
\newcommand{\ds}{\delta_{\mathrm{s}}}
\newcommand{\kf}{k_{\mathrm{f}}}
\newcommand{\VDG}{VDG$_{\infty}$~}

\newcommand{\beq}{\begin{equation}}
\newcommand{\eeq}{\end{equation}}
\newcommand{\beqa}{\begin{eqnarray}}
\newcommand{\eeqa}{\end{eqnarray}}

\newcommand{\kv}{\boldsymbol{k}}
\newcommand{\qv}{\boldsymbol{q}}

\newcommand{\PL}{P_{\rm lin}}
\newcommand{\PLIR}{\tilde{P}_{\rm lin,IR}}

\newcommand{\kmax}{k_{\rm{max}}}

 \def\la{\mathrel{\mathpalette\fun <}}
 
 \def\fun#1#2{\lower3.6pt\vbox{\baselineskip0pt\lineskip.9pt
        \ialign{$\mathsurround=0pt#1\hfill##\hfil$\crcr#2\crcr\sim\crcr}}}

\begin{document}

\preprint{APS/123-QED}

\title{Boosting galaxy clustering analyses with non-perturbative \\ modelling of redshift-space distortions}


\author{Alexander Eggemeier}
\email{aeggemeier@astro.uni-bonn.de}
\affiliation{Argelander Institut f\"ur Astronomie der Universit\"at Bonn, Auf dem H\"ugel 71, 53121 Bonn, Germany.}
\author{Nanoom Lee}
\affiliation{Center for Cosmology and Particle Physics, Department of Physics, New York University, New York, NY 10003, USA.}
\affiliation{William H. Miller III Department of Physics \& Astronomy, Johns Hopkins University, Baltimore, MD 21218, USA.}
\author{Rom\'an Scoccimarro}
\affiliation{Center for Cosmology and Particle Physics, Department of Physics, New York University, New York, NY 10003, USA.}
\author{Benjamin Camacho-Quevedo}
\affiliation{Institute of Space Sciences (ICE, CSIC), Campus UAB, Carrer de Can Magrans, s/n, 08193 Barcelona, Spain.}
\affiliation{Institut d’Estudis Espacials de Catalunya (IEEC), 08034 Barcelona, Spain.}
\author{Andrea Pezzotta}
\affiliation{INAF - Osservatorio Astronomico di Brera, via Emilio Bianchi 46, 23807 Merate, Italy.}
\affiliation{Max-Planck-Institut f\"ur extraterrestrische Physik, Postfach 1312, Giessenbachstr., 85748 Garching, Germany.}
\author{Martin Crocce}
\affiliation{Institute of Space Sciences (ICE, CSIC), Campus UAB, Carrer de Can Magrans, s/n, 08193 Barcelona, Spain.}
\affiliation{Institut d’Estudis Espacials de Catalunya (IEEC), 08034 Barcelona, Spain.}
\author{Ariel G. S\'anchez}
\affiliation{Max-Planck-Institut f\"ur extraterrestrische Physik, Postfach 1312, Giessenbachstr., 85748 Garching, Germany.}
\affiliation{Universit\"ats-Sternwarte M\"unchen,  Fakult\"at f\"ur Physik, Ludwig- Maximilians-Universit\"at M\"unchen, Scheinerstrasse 1, 81679 M\"unchen, Germany.}

\date{\today}

\begin{abstract}
  Redshift-space distortions (RSD), caused by the peculiar velocities of galaxies, are a key modelling challenge in galaxy clustering analyses, limiting the scales from which cosmological information can be reliably extracted.  Unlike dynamical or galaxy bias effects, RSD imprint features that are sensitive to non-linearities across all scales.  Yet, no distinction between these effects is made by the state-of-the-art analytical approach---the effective field theory (EFT)---which applies the same perturbative expansion to each of them.  This paper explores an alternative approach, where the non-perturbative nature of RSD is partially preserved, and compares its effectiveness against the EFT in analysing power spectrum and bispectrum multipoles from synthetic samples of luminous red galaxies, using the projected sensitivity of a Stage-IV galaxy survey.  Our results demonstrate that this distinct treatment of RSD improves the robustness of model predictions for both statistics, extending the validity range of the EFT from approximately $0.2\,\hinvMpc$ to $0.35\,\hinvMpc$ for the one-loop power spectrum and from $0.1\,\hinvMpc$ to $0.14\,\hinvMpc$ for the tree-level bispectrum.  This leads to a significant enhancement in the precision of cosmological parameter constraints, with uncertainties on the Hubble rate, matter density, and scalar amplitude of fluctuations reduced by $20$-$40\,\%$ for the power spectrum multipoles alone compared to the EFT, and by $25$-$50\,\%$ for joint analyses with the bispectrum.  The RSD treatment proposed here may thus play a crucial role in maximising the scientific return of current and future galaxy surveys.  To support this advancement, all models for the power spectrum and bispectrum used in this work are made available through an extended version of the Python package \texttt{COMET}.
\end{abstract}

\maketitle


\section{Introduction}
\label{sec:introduction}

The spatial distribution of galaxies in the Universe holds a wealth of information about its initial conditions, expansion history, and growth of structures.  This information is typically extracted by summarising the galaxy distribution into statistical measures, such as the two- and three-point correlation functions (or, equivalently, the power spectrum and bispectrum in Fourier space), and comparing these measurements with predictions from suitable models.  The challenge lies in crafting models that maximise the range of scales over which the data can be analysed, while meeting the stringent accuracy requirements demanded by the precision of the current generation of Stage-IV galaxy surveys, including the Dark Energy Spectroscopic Instrument (DESI) \cite{ColAghAgu1612} and Euclid \cite{EucMelAbd2409}. 

The effective field theory (EFT) of large-scale structure is a powerful perturbative framework that delivers accurate predictions for galaxy clustering on linear and quasi-linear scales \cite[for a review, see][]{Iva2212}.  In recent years, it has therefore become the primary method for deriving cosmological constraints from the full shape of clustering statistics, as underlined by numerous analyses utilising data from the (extended) Baryon Oscillation Spectroscopic Survey (BOSS/eBOSS) \cite[e.g.,][]{dAGleKok2005,IvaSimZal2005}, as well as the recent results published by the DESI collaboration \cite{AdaAguAhl2411}.  The key to the EFT's success over previous standard perturbation theory approaches lies in the systematic addition of all contributions to galaxy clustering that  encapsulate the impact of small-scale physics on large scales, guided by the constraint that they obey fundamental symmetries such as the equivalence principle and Galilean invariance as well as rotational and shift symmetries at each order of perturbation theory.  This guarantees that any dependence on complicated small-scale physics can then be effectively absorbed into a set of free parameters, which are determined from the data alongside the parameters of the cosmological model.

In the modelling of galaxy clustering statistics one has to account for various sources of non-linearity: gravitational evolution of the density fluctuations, the relationship between galaxies and the underlying dark matter field (galaxy bias), and the distortion of galaxy positions along the line of sight due to their peculiar velocities (redshift-space distortions, or RSD).  Each of these sources has an associated scale that determines when the non-linear effects become significant.  For instance, non-linear effects due to gravitational dynamics typically occur on scales where density fluctuations approach unity, while non-linear RSD, which give rise to the ``fingers of God'' (FoG) feature~\cite{Jac7202,DeGelHuc8603,VogParGel9205}, depend on the magnitude of the velocity dispersion.  The perturbative expansion in the EFT formally treats (small scale) velocity dispersion as a quantity of the same order in perturbation theory as large-scale density fluctuations. The associated non-linear scales are taken to be small enough that scales of interest are within the regime of validity of Taylor expansion. 

However, these assumptions are unlikely to be strictly correct. In linear theory, it is well-known that the Kaiser formula for the power spectrum multipoles \cite{Kai8707} fails on larger scales compared to the validity of the linear power spectrum in real space (i.e., in the absence of RSD) \cite{ColFisWei9404,FisNus9603,TayHam9610,HatCol9805,Sco0410}.  Empirical evidence also indicates that this is the case for the EFT, whose power spectrum predictions on small scales are more accurate in real space  \cite{EggScoCro2011,PezCroEgg2108,EucPezMor2407} than in redshift space \cite{dAGleKok2005,NisDAIva2012,ChuDolIva2101,MauLaiNor2406}.  A similar discrepancy appears to affect higher-order statistics, too, comparing for example bispectrum analyses in real space \cite{EggScoSmi2106} and redshift space \cite{IvaPhiNis2203}.  All this suggests that non-linearities due to RSD develop on larger scales than those from the other sources.  In practice, this means that more conservative scale cuts are required in the analysis of real data, hence limiting the constraining power.  This is particularly apparent for the bispectrum, which in combination with the power spectrum cannot significantly reduce the uncertainties on cosmological parameters compared to the power spectrum alone \cite{PhiIva2202,IvaPhiCab2304}.  Even the inclusion of corrections from the next higher perturbative order \cite{DADonLew2405,DADonLew2407} does not lead to substantial improvements as this requires a large number of additional nuisance parameters in the EFT.

This motivates the main question for the present paper: can we enhance galaxy clustering analyses by modelling RSD non-perturbatively and, in turn, obtain more precise constraints on cosmological parameters?  Various non-perturbative approaches to RSD have been proposed in the past, based on different simplifications of the streaming model, which describes the exact relationship between galaxy clustering statistics in real and redshift space \cite{Pee80,Fis9508,Sco0410,KurPor2007}.  One such approach is the Gaussian streaming model \cite{Fis9508,ReiWhi1111,WanReiWhi1401}, which approximates the central quantity for the redshift-space mapping---the probability distribution of velocities---as a Gaussian, while another approach showed how the non-linear effects of the mapping can be summarised in an effective non-Gaussian  damping function on large scales \cite{Sco0410}.  In this study, we follow the latter approach, which has previously also served as the foundation of the ``TNS'' model \cite{TarNisSai1009}.  However, unlike typical TNS implementations \cite[e.g.,][]{BeuSaiSeo1409,BeuSeoSai1704,GilPerVer1702,GilBauPav2009}, we incorporate additional developments of the EFT for non-RSD effects, and a theoretically motivated damping function, as detailed in \cite{EggCamPez2212}.

Our aim is to compare the cosmological constraints of this model with those of the standard EFT approach when both are applied to the same data and using a consistent set of analysis settings, such as the number of fitted parameters and their priors.  We conduct this comparison separately for the power spectrum and its combination with the bispectrum, which we measure from synthetic galaxy catalogues that mimic the selections of the BOSS survey, but without the complication of systematic effects.  To make this analysis instructive for current Stage-IV galaxy surveys, we scale measurement uncertainties to approximate what is expected at the end of their missions.  As a representative implementation of the EFT model we closely follow the \texttt{Class-PT} code \cite{ChuIvaPhi2009}, which makes identical assumptions about RSD as other implementations like \texttt{PyBird} \cite{DASenZha2101}, \texttt{velocileptors} \cite{CheVlaCas2103}, \texttt{Class-OneLoop} \cite{LinDizRad2407}, and \texttt{Folps} \cite{NorAviGil2411}, all of which have been shown to yield consistent results \cite{MauLaiNor2406}.

The paper is organised as follows.  In Sec.~\ref{sec:theory}, we provide a summary and comparison of the theoretical models used in this work.  Section~\ref{sec:data} details the measurements obtained from our synthetic galaxy samples and their covariances, followed by a description of the Bayesian inference pipeline and the metrics used to assess model performance in Sec.~\ref{sec:pipeline}.  Next, in Sec.~\ref{sec:validation} we discuss various approximations to improve the computational efficiency of the bispectrum model and their validation for the adopted measurement uncertainties.  Our results are presented in Sec.~\ref{sec:results}, focusing first on the performance of the models for the power spectrum alone, before considering joint analyses with the bispectrum.  Finally, we conclude in Sec.~\ref{sec:conclusions}.


\section{Modelling clustering statistics in redshift space}
\label{sec:theory}

In this work, we focus on the modelling of the galaxy power spectrum $P$ and bispectrum $B$, which represent the correlation functions of density fluctuations $\delta$ in Fourier space.  They are defined as
\begin{align}
  \langle \delta(\bk_1)\,\delta(\bk_2) \rangle &\equiv (2\pi)^3 \, P(\bk_1) \, \delta_{\rm D}(\bk_{12})\,, \\
  \langle \delta(\bk_1)\,\delta(\bk_2)\,\delta(\bk_3) \rangle &\equiv (2\pi)^3 \, B(\bk_1,\bk_2,\bk_3) \, \delta_{\rm D}(\bk_{123})\,,
\end{align}
where $\delta_{\rm D}$ denotes the Dirac delta function and $\bk_{1 \ldots n} \equiv \bk_1 + \ldots + \bk_n$.  In one of the models considered in our study, the EFT model, the galaxy power spectrum can be broadly decomposed into four different terms:
\begin{equation}
  \label{eq:theory.PEFT_overview}
  P_{\rm EFT}(\bk) = P_{\rm tree}(\bk) + P_{\rm 1loop}(\bk) + P_{\rm stoch}(\bk) + P_{\rm ctr,EFT}(\bk)\,.
\end{equation}
The first two include dynamical and galaxy bias contributions, as well as the perturbative expansion of RSD at either leading (``tree'') or next-to-leading (``one-loop'') order, which will be detailed in Sec.~\ref{sec:theory.RSD} and
\ref{sec:theory.bias} below.  The latter two terms account for the effects of stochasticity in the density and velocity fields, and further small-scale corrections through the so-called \emph{counterterms}, discussed in Sec.~\ref{sec:theory.stochasticity} and \ref{sec:theory.counterterms}.  The bispectrum in the EFT model follows a similar structure, although in this case, we will omit the one-loop contribution, so that
\begin{equation}
  \label{eq:theory.BEFT_overview}
  \begin{split}
    B_{\rm EFT}(\bk_1,\bk_2,\bk_3) = B_{\rm tree}(\bk_1,\bk_2,\bk_3) \\ + B_{\rm stoch,EFT}(\bk_1,\bk_2,\bk_3) + B_{\rm ctr,EFT}(\bk_1,\bk_2,\bk_3)\,.
  \end{split}
\end{equation}

The second model we consider builds upon the EFT framework but does not fully expand the mapping from real to redshift space.  Instead, it exploits that the contributions most sensitive to non-linearities can be grouped into a single quantity that is identified as the generating function of velocity differences between pairs of galaxies (\emph{velocity difference generator}, or VDG, for short).  It causes a damping of the power spectrum or bispectrum---as expected from the FoG effect---and it is approximated in this work by its behaviour in the infinite pair separation limit, which we denote as $W^P_{\infty}$ or $W^B_{\infty}$.  This leads to the following general structure of the model:
\begin{equation}
  \label{eq:theory.PVDG_overview}
  \begin{split}
  P_{\rm VDG}(\bk) = W_{\infty}^P(\bk) \, \Big[P_{\rm tree}(\bk) + P_{\rm 1loop}(\bk) \Big. \\ \Big. + P_{\rm ctr,VDG}(\bk) - \Delta P(\bk)\Big] + P_{\rm stoch}(\bk)\,,
  \end{split}
\end{equation}
with an additional correction term $\Delta P(\bk)$ derived in Sec.~\ref{sec:theory.VDG}, which does not appear in the EFT model.  This correction is relevant only at the one-loop order, such that the bispectrum is written as\footnote{At tree-level we only consider bispectrum counterterms for the EFT model, but not for the \VDG model, see discussion in Sec.~\ref{sec:theory.counterterms}.}
\begin{equation}
  \label{eq:theory.BVDG_overview}
  \begin{split}
    B_{\rm VDG}(\bk_1,\bk_2,\bk_3) = W^B_{\infty}(\bk_1,\bk_2,\bk_3) \, B_{\rm tree}(\bk_1,\bk_2,\bk_3) \\ + B_{\rm stoch,VDG}(\bk_1,\bk_2,\bk_3)\,.
  \end{split}
\end{equation}
Importantly, the stochastic contributions are not damped by $W^P_{\infty}$ and $W^B_{\infty}$, but unlike for the power spectrum, they differ between the two models for the bispectrum, as explained in Sec.~\ref{sec:theory.stochasticity}.  To distinguish this model from the EFT and other variants, such as TNS, we will refer to it as \VDG in the remainder of this work.  All expressions for the power spectrum and bispectrum presented in this section are implemented in an extended version of our publicly available Python package \texttt{COMET}\footnote{\url{https://comet-emu.readthedocs.io/en/latest/index.html}} \cite{EggCamPez2212}, which makes use of Gaussian processes in combination with the evolution mapping methodology \cite{San2012,SanRuiJar2108} for efficient and accurate predictions over a wide parameter space.

Throughout this work, we use the Fourier transform convention
\begin{equation}
  \label{eq:theory.Fourier_convention}
  \delta(\bx) = \int_{\bk} \mathrm{e}^{-\mathrm{i}\bk \cdot \bx} \, \delta(\bk)\,,
\end{equation}
where integrals over $\bk$-space quantities are written with the short-hand notation $\int_{\bk_1,\ldots,\bk_n} \equiv \int \mathrm{d}^3k_1/(2\pi)^3 \cdots \int \mathrm{d}^3k_n/(2\pi)^3$.  In configuration space, we analogously use $\int_{\bx_1,\ldots,\bx_n} \equiv \int \mathrm{d}x_1 \cdots \int \mathrm{d}x_n$.

\subsection{The redshift-space mapping and the VDG}
\label{sec:theory.RSD}

\subsubsection{Basics}
\label{sec:basics}

In the plane-parallel approximation, the mapping of comoving positions $\bx$ to redshift-space comoving positions is given by 
\beq
\bs = \bx -f \, u_z(\bx)\, \hat{z}\,,
\label{RSDmap}
\eeq
where the line of sight (LOS) component of the velocity field $\bv(\bx)$ has been written as $v_z(\bx) \equiv - f\,{\cal H}\,u_z(\bx)$, with $f$ the velocity field growth rate and ${\cal H}^{-1}$ the comoving Hubble scale, so that $\nabla\cdot \bu = \delta$ in linearised perturbation theory ($\delta$ denotes the matter density fluctuations).  The redshift-space density field for galaxies in Fourier space can be written as~\cite{ScoCouFri9906},
\beq
\delta_s(\bk) = \int_{\bx} \mathrm{e}^{-\mathrm{i} \bk\cdot\bx}\, \mathrm{e}^{\lambda u_z(\bx)} \, [\delta_g(\bx)+f \nabla_z u_z(\bx)]\,,
\label{deltaRSDk}
\eeq
with $\lambda\equiv \mathrm{i}f k_z$. The contribution in square brackets enhances the redshift-space density perturbations compared to the true galaxy density perturbations $\delta_g$ along the LOS and gives rise to the Kaiser formula~\cite{Kai8707} at large scales in linear theory, when the exponential is set to unity. The (oscillating) exponential factor, instead, {\em suppresses} the redshift-space density perturbation (compared to the square brackets) along the LOS as small scales (large $|\lambda|$) are probed. 

Equation~\ref{deltaRSDk} can be rewritten expanding the exponential and writing the non-linear density and velocity fields in terms of their perturbation theory (PT) expansions. This results in the standard expansion of $\delta_s$ in terms of the redshift-space PT kernels $Z_n$ and the linearised {\em matter} density perturbations $\delta_{\rm lin}$,
\beq
\begin{split}
\delta_s(\bk) = (2\pi)^3\sum_{n=1}^\infty \int_{\bq_1,\ldots,\bq_n} \delta_{\rm D}\left(\bk-\bq_{1 \ldots n}\right) \, Z_n(\bq_1,\ldots,\bq_n)\ \\ \times \, \delta_{\rm lin}(\bq_1) \ldots  \delta_{\rm lin}(\bq_n)\,,
\end{split}
\label{deltasZns}
\eeq
where the $Z_n$ kernels contain both galaxy bias (see Sec.~\ref{sec:theory.bias}) and RSD effects, and we quote the ones relevant for this work in Appendix~\ref{sec:app.PTkernels}.

The expansion of the redshift-space density in Eq.~(\ref{deltasZns}) leads to the standard formulas for the galaxy power spectrum and bispectrum, which we recall in the next subsection. However, there is motivation to consider an alternative approach for deriving the power spectrum and bispectrum. This is because the redshift-space PT kernels $Z_n$ conflate two different non-linearities and treat them equivalently: the non-linear dynamics (encoded by the fields $\delta_g$ and $u_z$) and the non-linear redshift-space mapping (encoded by the exponential in Eq.~\ref{deltaRSDk}). As highlighted in the introduction, it is well known that standard redshift-space PT calculations have a more restricted range of validity than in real space. Therefore, it may be beneficial to treat the redshift-space mapping distinctly from the non-linear dynamics, to find an improved model performance in redshift-space. This can be accomplished by directly evaluating the power spectrum and bispectrum from Eq.~(\ref{deltaRSDk}) before applying PT. For the power spectrum, this gives~\cite{Sco0410}
\beq
P(\bk) = \int_{\br}  \mathrm{e}^{-\mathrm{i} \bk\cdot\br}\, \langle  \mathrm{e}^{\lambda \Delta u_z} \, {\cal D}(\bx) {\cal D}(\bx') \rangle\,,
\label{PkRSD}
\eeq
where $\Delta u_z\equiv u_z(\bx)-u_z(\bx')$, $\br \equiv \bx-\bx'$, and the field ${\cal D}\equiv \delta_g+f \nabla_z u_z$ is the redshift-space density perturbation in the Kaiser limit (but with fully non-linear $\delta_g$ and $u_z$ fields). For the bispectrum, one analogously obtains
\beqa
B(\bk_1,\bk_2) &=&  \int_{\br_{13},\br_{23}}  \mathrm{e}^{-\mathrm{i} (\bk_1\cdot\br_{13}+\bk_2\cdot\br_{23})}\nonumber \\ & &
 \times \, \langle  \mathrm{e}^{\lambda_1 \Delta u_{13z}+\lambda_2 \Delta u_{23z}} \, {\cal D}(\bx_1) {\cal D}(\bx_2){\cal D}(\bx_3) \rangle\,, \nonumber \\ & &
\label{BkRSD}
\eeqa
with $\bk_{123}=0$, $\br_{ij}\equiv \bx_i-\bx_j$, $\lambda_j \equiv \mathrm{i} f \bk_j\cdot \hat{z}$, and $\Delta u_{ijz} \equiv u_z(\bx_i)-u_z(\bx_j)$. Expressions like Eqs.~(\ref{PkRSD}-\ref{BkRSD}) are useful because they offer the possibility to go beyond the standard perturbative approximation of the redshift-space mapping, and to see this, consider the power spectrum: the relevant correlator satisfies the \emph{exact} relation~\cite{Sco0410}
\beqa
\langle  \mathrm{e}^{\lambda \Delta u_z} \, {\cal D}(\bx) {\cal D}(\bx') \rangle &=&  W(\lambda,\br) \times \Big[
\langle  \mathrm{e}^{\lambda \Delta u_z} \, {\cal D}(\bx) {\cal D}(\bx') \rangle_c  \nonumber \\ &&+ \langle  \mathrm{e}^{\lambda \Delta u_z} \, {\cal D}(\bx)  \rangle_c \ \langle  \mathrm{e}^{\lambda \Delta u_z} \,  {\cal D}(\bx') \rangle_c  \Big]\,,
 \nonumber \\ & & 
\label{CumEx}
\eeqa
where $\langle\ldots\rangle_c$ denotes a connected expectation value (or cumulant), and the prefactor represents the  VDG, defined as
\beq
W(\lambda,\br) \equiv \langle  \mathrm{e}^{\lambda \Delta u_z} \rangle= \exp{\langle  \mathrm{e}^{\lambda \Delta u_z} \rangle_c}\,. 
\label{VDGdef}
\eeq
By taking successive derivatives of $W$ ($\ln W$) with respect to $\lambda$, one can generate all the moments (cumulants) of velocity differences. However, instead of expanding $W$ naively as a power series in $\lambda$ (which leads to the same expressions as SPT using the $Z_n$ PT kernels), in this paper we are particularly interested in modelling $W$ as a prescribed function of $\lambda$.  This approach may improve the prediction of redshift-space statistics when compared against simulations~\cite{Sco0410,TarNisSai1009,SanScoCro1701,AdaAguAhl2411,ScoNaN}. 

The reasoning behind this, as pointed out in~\cite{Sco0410}, is that the cumulant expansion in Eq.~(\ref{CumEx}) separates the contribution that is rather sensitive to small scales (the VDG) from the remaining contributions that are large-scale dominated. This is because the velocity difference at scale $r$ receives no contributions from modes with $kr\la 1$, unlike usual two-point correlation functions, which are dominated by such modes~\cite{Sco0410}. Hence, in practice, it is difficult to obtain genuinely non-perturbative results for the VDG; what we mean here by ``non-perturbative" is that the VDG function is not a power series expansion in $\lambda$ but rather a function that is motivated by a resummation of quadratic non-linearities (see Sec.~\ref{sec:theory.VDG}).

\subsubsection{Standard Perturbation Theory (SPT)}
\label{sec:theory.SPT}

For reference, let us write down the usual SPT expressions for the power spectrum and bispectrum. These follow from  keeping calculations of correlators of $\delta_s$ to a fixed order in PT and thus can be expressed directly in terms of the $Z_n$ PT kernels and the linear power spectrum $P_{\rm lin}$, assuming Gaussian primordial fluctuations~\cite{BerColGaz0209}. For the tree-level and one-loop power spectrum we have:
\begin{align}
  P_{\rm tree}^{\rm SPT}(\bk) &= [Z_1(\bk)]^2 P_{\rm lin}(k) \,, \label{eq:theory.SPT.Pspt_tree}\\
    P_{\rm 1loop}^{\rm SPT}(\bk) &= \int_{\bq} 2 [Z_2(\bk-\bq,\bq)]^2 \, P_{\rm lin}(|\bk-\bq|) \, P_{\rm lin}(q) \nonumber \\ &+ \int_{\bq} 6Z_3(\bk,\bq,-\bq) \, Z_1(\bk)  \, P_{\rm lin}(q) \, P_{\rm lin}(k) \,, \label{eq:theory.SPT.Pspt_1loop}
\end{align}
whereas the tree-level bispectrum is given by~\cite{ScoCouFri9906},
\begin{equation}
  \begin{split}
B_{\rm tree}^{\rm SPT}(\bk_1,\bk_2,\bk_3)= 2 Z_1(\bk_1) \, Z_1(\bk_2) \, Z_2(\bk_1,\bk_2) \\ \times \, P_{\rm lin}(k_1) \, P_{\rm lin}(k_2) + {\rm cyc.}\,,
  \end{split}
\label{Bspt}
\end{equation}
and ``cyc." denotes cyclic permutations of the three wave vectors. As mentioned above, these expressions can be alternatively derived from Eqs.~(\ref{PkRSD}) and (\ref{BkRSD}) for the power spectrum and bispectrum, by expanding the exponentials in power series to the desired order in $\lambda$. In particular, this corresponds to expanding the VDG (see Eq.\ref{VDGdef}).

An effect that is only poorly captured by SPT up to one-loop order, is the broadening and damping of the baryon acoustic oscillation (BAO) feature due to large-scale flows \cite[see, e.g.,][]{MeiWhiPea9904,EisSeoWhi0708,CroSco0801}.  This is because the displacements generated by long-wavelength perturbations affect the amplitude of the BAO feature by an amount larger than any dynamical corrections.  Treating the effect of the large-scale displacements perturbatively therefore leads to significant inaccuracies, but \cite{SenZal1502} showed that their contributions can be resummed to all orders in PT.  This procedure is known as infrared (IR) resummation and is typically used to enhance the conventional SPT predictions.

A common approach for implementing IR resummation exploits that the large-scale displacements leave the broadband shape of the power spectrum $P_{\rm nw}$ invariant, such that the overall effect is a damping of only its wiggly component\footnote{There is no unique way of isolating the wiggly component from the smooth part of linear power spectrum.  In this work, we follow the procedure outlined in Appendix A of \cite{VlaSelChu1603}.} $P_{\rm w}$ \cite{BalMirSim1508,BlaGarIva1607,IvaSib1807}.  The linear, IR resummed, power spectrum is thus given by
\begin{equation}
  \label{eq:theory.SPT.Plin_IR}
  P_{\rm lin,IR}(\bk) = P_{\rm nw}(k) + \mathrm{e}^{-k^2\,\Sigma_{\rm tot}^2(\mu)}\,P_{\rm w}(k)\,,
\end{equation}
where $\Sigma_{\rm tot}^2$ is the two-point function of the large-scale displacement field in redshift space, evaluated at the BAO scale $l_{\rm BAO} = 110\,\hinvMpc$.  It can be decomposed into two terms,
\begin{equation}
  \label{eq:theory.SPT.Sigma_tot}
  \Sigma_{\rm tot}^2(\mu) = \left[1 + f \mu^2 (2+f)\right] \Sigma^2 + f^2\mu^2 \left(\mu^2-1\right) \delta\Sigma^2\,,
\end{equation}
with
\begin{align}
  \Sigma^2 &= \frac{1}{6\pi^2} \int_0^{k_{\rm s}} \mathrm{d}q \, \left[1 - j_0(q\,l_{\rm BAO}) + 2j_2(q\,l_{\rm BAO})\right] P_{\rm nw}(q)\,, \\
  \delta\Sigma^2 &= \frac{1}{2\pi^2} \int_{0}^{k_{\rm s}} \mathrm{d}q \, j_2(q\,l_{\rm BAO}) \, P_{\rm nw}(q)\,,
\end{align}
and the integral cutoff $k_{\rm s}$ distinguishes between the long- and short-wavelength regimes.  Ideally, this scale should depend on $k$, as all modes $q < k$ need to be resummed, but it is well approximated by a fixed intermediate scale.  Following \cite{IvaSib1807}, we adopt the value\footnote{Note that in order to exploit the evolution mapping methodology \cite{San2012,SanRuiJar2108}, \texttt{COMET} internally employs $\mathrm{Mpc}$ units opposed to $\hMpc$. Hence, the scale $k_{\rm s}$ is also defined in $\mathrm{Mpc}^{-1}$.} $k_{\rm s} = 0.14\,\mathrm{Mpc}^{-1}$.  Note that the $\mu$-dependent terms in Eq.~(\ref{eq:theory.SPT.Sigma_tot}) arise due to RSD, which amplify the large-scale displacements along the LOS.

The full IR-resummed power spectrum and bispectrum are then obtained by replacing all occurrences of the linear power spectrum in Eqs.~(\ref{eq:theory.SPT.Pspt_tree})-(\ref{Bspt}) with its IR-resummed equivalent.  The tree-level power spectrum is thus
\begin{equation}
  \label{eq:theory.SPT.Ptree_IR}
  P_{\rm tree}(\bk) = \left[Z_1(\bk)\right]^2 \, P_{\rm lin,IR}(\bk)\,,
\end{equation}
and similarly for the tree-level bispectrum.  As shown in \cite{BalMirSim1508,IvaSib1807}, at one-loop order there is an additional correction term, so that the IR-resummed one-loop contribution reads:
\begin{equation}
  \label{eq:theory.SPT.P1loop_IR}
  \begin{split}
    P_{\rm 1loop}(\bk) = P_{\rm 1loop}^{\rm SPT}\left[P_{\rm lin,IR}\right](\bk) \\ + k^2\,\Sigma_{\rm tot}^2(\mu)\,\mathrm{e}^{-k^2\,\Sigma_{\rm tot}^2(\mu)} \, \left[Z_1(\bk)\right]^2 P_{\rm w}(k)\,,
  \end{split}
\end{equation}
where $P_{\rm 1loop}^{\rm SPT}\left[P_{\rm lin,IR}\right]$ denotes the two one-loop integrals from Eq.~(\ref{eq:theory.SPT.Pspt_1loop}), evaluated using $P_{\rm lin,IR}$.  This term is more conveniently computed in the following approximation:
\begin{equation}
  \label{eq:theory.SPT.P1loop_for_eval}
  \begin{split}
    P_{\rm 1loop}^{\rm SPT}\left[P_{\rm lin,IR}\right](\bk) \approx P_{\rm 1loop}^{\rm SPT}\left[P_{\rm nw}\right](k) + \mathrm{e}^{-k^2\,\Sigma_{\rm tot}^2(\mu)} \\ \times \left\{P_{\rm 1loop}^{\rm SPT}\left[P_{\rm nw} + P_{\rm w}\right] - P_{\rm 1loop}^{\rm SPT}\left[P_{\rm nw}\right]\right\}(k)\,,
  \end{split}
\end{equation}
introducing an error that is negligible at one-loop order.

\subsubsection{Velocity Difference Generating Functions (VDGs)}
\label{sec:theory.VDG}

The large-scale power spectrum and bispectrum multipoles are dominated by the large-scale behavior of the VDGs, which in turn are sensitive to small-scale Fourier modes as discussed above. This sensitivity is already present in the infinite separation limit, where points are effectively uncorrelated but receive contributions from non-linear structures, e.g. virialised halos. In this paper, we assume that $W$ can be approximated by its behavior at infinity, where it becomes a function of $\lambda$'s alone. In particular, for the power spectrum we have,
\beq
W(\lambda,\br) \simeq W^P_\infty(\lambda) = {1\over \sqrt{1-\lambda^2 a_{\rm vir}^2}} \exp {\left(\frac{\lambda^2\sigma_v^2}{1-\lambda^2 a_{\rm vir}^2}\right)}\,,
\label{Winfty}
\eeq
where $a_{\rm vir}$ is a free parameter that contributes to the velocity dispersion and controls the kurtosis of the VDG at infinity, and $\sigma_v^2 = \langle u_z^2 \rangle$ is the usual zero-lag velocity correlation in linear theory.  It describes the velocity dispersion due to bulk motion and is computed from the velocity divergence power spectrum $P_{\theta\theta}$ as follows:
\begin{equation}
  \label{eq:sigmav}
  \sigma_v^2 = \frac{1}{3} \int_{\bk} \frac{P_{\theta\theta}(k)}{k^2} = \frac{1}{3} \int_{\bk} \frac{P_{\rm lin}(k)}{k^2} \,.
\end{equation}
Note that the VDG at infinity is a non-Gaussian function of $\lambda$, closely related to the non-Gaussianity of the pairwise velocity distribution function at infinity~\cite{Sco0410,ScoNaN}. The parametrisation in Eq.~(\ref{Winfty}) can be obtained by resumming quadratic non-linearities as advocated in~\cite{Sco0410} and has already been used in various previous analyses of BOSS data \cite{SanScoCro1701,GriSanSal1701,HouSanSco1810,NevBurde2010,HouSanRos2011,SemSanPez2204,SemSanPez2210}.  A detailed derivation will be given in~\cite{ScoNaN}, along with a demonstration showing that this parametrisation accurately reproduces the infinite separation limit of the pairwise VDG function, as measured directly from simulations.

For the bispectrum, the corresponding expression to Eq.~(\ref{CumEx}) involves the three-point VDG,
\beq
W(\lambda_1,\lambda_2,\br_{13},\br_{23}) \equiv \langle  \mathrm{e}^{\lambda_1 \Delta u_{13z}+\lambda_2 \Delta u_{23z}} \rangle\,, 
\label{VDGBdef}
\eeq
and following the power spectrum case, we assume that it can be approximated by its behavior at infinity, which we take to be
\beq
W^B_{\infty}(\lambda_{123}) = {1\over (1 -\lambda_{123}^2 a_{\rm vir}^2)^{3/2}}\ \exp{\left(\lambda_{123}^2\ \sigma_v^2\over {1 - \lambda_{123}^2 a_{\rm vir}^2}\right)}, 
\label{WBinfty}
\eeq
where $\lambda_{123}^2 \equiv ( \lambda_1^2 + \lambda_2^2 + \lambda_3^2 )/2$, and note that closed triangles impose $-\lambda_3=\lambda_1+\lambda_2$. Unlike the power spectrum case, Eq.~(\ref{WBinfty}) is not a rigorous result, but it is a simple expression that matches the rigorous calculation of the bispectrum when quadratic non-linearities in velocity differences are resummed at all scales of interest~\cite{Noo2405}. 

\begin{figure}[t!]
  \centering
  \includegraphics{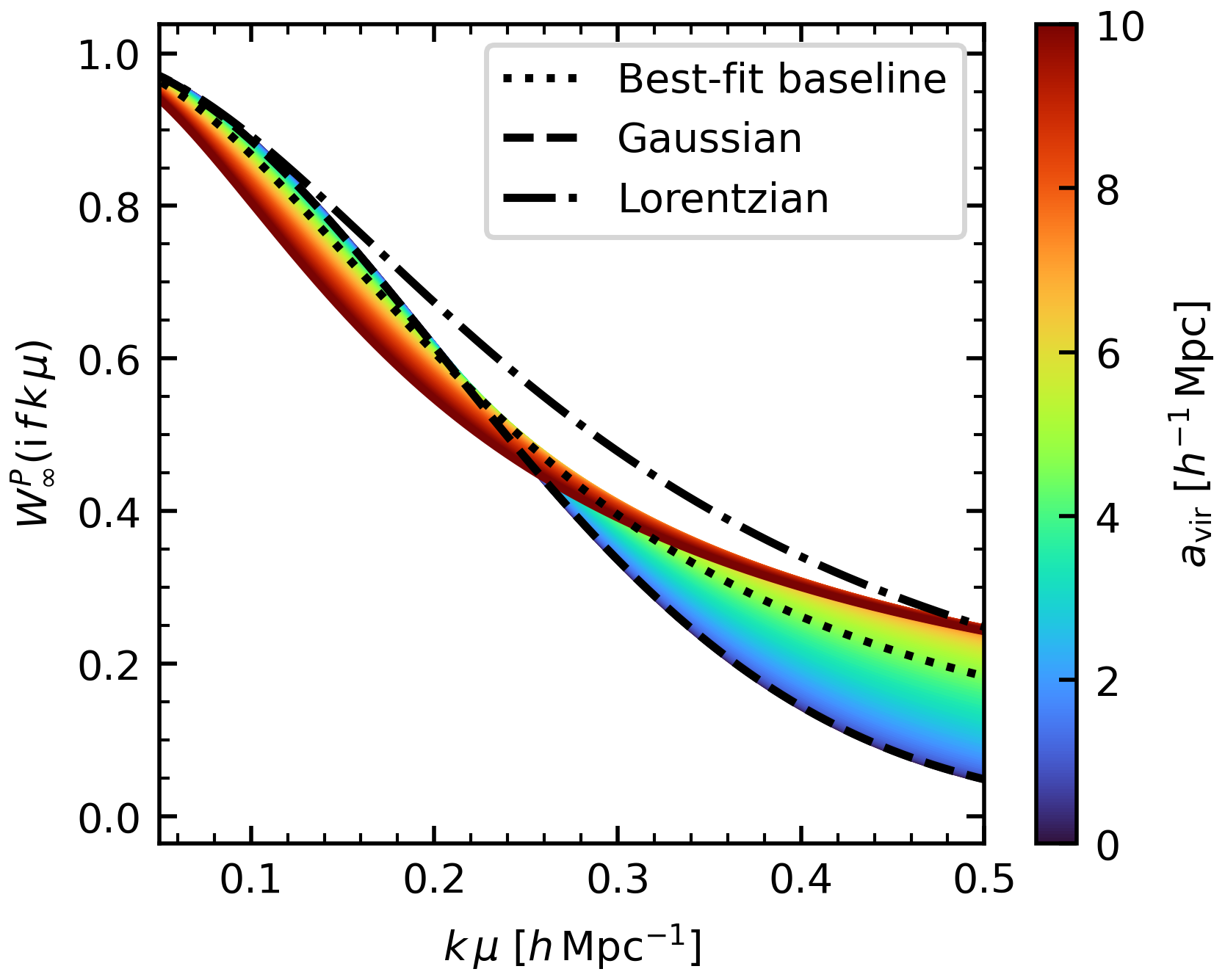}
  \caption{The VDG in the infinite separation limit, which acts as a damping function on the redshift-space power spectrum. The dotted line shows the physically motivated VDG, Eq.~(\ref{Winfty}), as a function of scale $k\,\mu$ corresponding to a best-fit parameter $a_{\rm vir}$ for CMASS-type galaxies, while the coloured curves illustrate different values of $a_{\rm vir}$ (for fixed $f$ and $\sigma_v$) as labeled. In addition, we show the Gaussian (dashed) and Lorentzian (dot-dashed) damping functions for the same $f$ and $\sigma_v$ values.}
  \label{fig:Wdamp_vs_kmu}
\end{figure}

The TNS model, frequently employed in the analysis of past galaxy surveys \cite[e.g.][]{BeuSaiSeo1409,BeuSeoSai1704,GilPerVer1702,GilBauPav2009}, also builds on the cumulant expansion (Eq.~\ref{CumEx}) and approximates the VDG by a damping function.  However, while the parametrisation discussed above (hereafter referred to as our baseline) is derived in the infinite separation limit, other, more \emph{ad-hoc}, approaches have traditionally been used for TNS.  The two most common choices have been either a Gaussian or Lorentzian function, which in terms of $\lambda$ can be written as
\begin{align}
  W^P_{\rm G}(\lambda) &= \exp{\left(\lambda^2\,\sigma_v^2\right)}\,, \label{eq:theory.VDG.Winfty_G} \\
  W^P_{\rm Lor}(\lambda) &= \frac{1}{1 - \lambda^2\,\sigma_v^2}\,, \label{eq:theory.VDG.Winfty_Lor}
\end{align}
where $\sigma_v$ is usually treated as a free parameter, rather than the prediction for the bulk velocity dispersion in linear theory.  To illustrate how these two functions compare to the baseline parametrisation, we plot them against $k \mu$ in Fig.~\ref{fig:Wdamp_vs_kmu} for the fiducial values of $\sigma_v$ and $f$.  The colours indicate the dependence on $a_{\rm vir}$ for Eq.~(\ref{Winfty}), while the dotted line corresponds to a representative best-fit $a_{\rm vir}$ value that we obtain in Sec.~\ref{sec:results.pk} when analysing the synthetic BOSS CMASS sample.  We see that the Gaussian (equivalent to $a_{\rm vir} = 0$) initially aligns closely with the best-fit baseline, but significantly underestimates the tail of the VDG and hence leads to excessive damping for $k \mu \gtrsim 0.25\,\hinvMpc$.  In contrast, the Lorentzian damping effect is too little on all scales $k\mu \gtrsim 0.15\,\hinvMpc$.  Due to the different scale-dependencies, adjusting the value of $\sigma_v$ can only improve agreement with the baseline in a narrow range of scales, leading to a lack or excess of damping elsewhere.  We will see in Sec.~\ref{sec:results.pk.VDGdamping} and \ref{sec:results.pk.counterterms} that this can partially be accounted for by the counterterms, but at the expense of introducing a running with scale of the associated parameters.

In the \VDG model, all remaining terms from the cumulant expansion, i.e., the terms in square brackets in Eq.~(\ref{CumEx}), are treated perturbatively.  The exponentials are expanded consistently up to one-loop order\footnote{For a treatment that resums quadratic non-linearities for these terms as well, see~\cite{ScoNaN}.}, so that evaluation of the correlators yields the same contributions as in the EFT model (including stochasticity and counterterms discussed below), except for the terms $\Delta P(\bk)$ that arise from the expansion of the VDG.  To compute these terms up to one-loop order, it is sufficient to expand the exponential as $\langle\mathrm{e}^{\lambda\,\Delta u_z}\rangle \approx 1 + \lambda^2/2 \langle\Delta u_z^2\rangle$, which leads to
\begin{align}
  \label{eq:DeltaP}
  \Delta P(\bk) &= \int_{\boldsymbol{r}} \mathrm{e}^{\mathrm{i} \bk \cdot \boldsymbol{r}}\,\left[\frac{\lambda^2}{2} \left<\Delta u_z^2\right> \left<{\cal D}(\bx)\,{\cal D}(\bx')\right>\right] 
  \nonumber \\
  &= \lambda^2\,\sigma_v^2\,P_{{\cal D}{\cal D}}(\bk) - \lambda^2 \int_{\qv} \frac{q_z^2}{q^4} P_{\theta \theta}(q)\,P_{{\cal D}{\cal D}}(\bk-\qv)\,,
\end{align}
where $P_{{\cal D}{\cal D}}$ is the power spectrum of the ${\cal D}(\bx)$ field.  In the above expression, it is sufficient to evaluate it at leading order, which implies $P_{{\cal D}{\cal D}}(\bk) = P_{\rm tree}(\bk)$.  Finally, since $W_{\infty}^P$ does not depend on scale $\boldsymbol{r}$, it can be factored out of the integral in Eq.~(\ref{PkRSD}), from which follows Eq.~(\ref{eq:theory.PVDG_overview}).  The same reasoning also applies to the bispectrum, with the difference that at tree-level there are no corrections from the expansion of the VDG.

\subsection{Galaxy bias expansion}
\label{sec:theory.bias}

The relationship between the distribution of galaxies (or any other luminous tracer) and the underlying matter field is described by the galaxy bias expansion in perturbation theory \cite[for a review, see][]{DesJeoSch1802}.  This expansion is expressed as a series of terms (often called \emph{operators}), each representing different contributions from the matter density, tidal field, and their derivatives, which capture the effect of the large-scale environment on the formation and evolution of galaxies.  At each order of perturbation theory, these terms can be rigorously derived based on the symmetries of the equations of motion, in particular the equivalence principle and Galilean invariance \cite{McDRoy0908,ChaScoShe1204,AssBauGre1408,MirSchZal1507,Sen1511,DesJeoSch1802,EggScoSmi1906}.

There is no unique definition of the operators and in this work, we adopt the approach from \cite{EggScoSmi1906} and express the galaxy over-density as follows:
\begin{equation}
  \label{eq:theory.bias_expansion}
  \begin{split}
    \delta_g = \; &b_1\,\delta + \frac{b_2}{2}\delta^2 + \gamma_2\,{\cal G}_2(\Phi_v) + \gamma_{21}\,{\cal G}_{21}(\varphi_2,\varphi_1) \\ &+ b_{\nabla^2}\,\nabla^2\delta + \ldots\,.
  \end{split}
\end{equation}
Each term is evaluated at position $\bx$ and redshift $z$, and the galaxy bias parameters $b_1$, $b_2$, $\gamma_2$, $\gamma_{21}$, and $b_{\nabla^2}$ absorb any dependencies that characterise the observed population of galaxies.  The two Galilean operators, ${\cal G}_2$ and ${\cal G}_{21}$, in Eq.~(\ref{eq:theory.bias_expansion}) encapsulate the effect of large-scale tides at second and third order, and they are defined as (with repeated indices being summed over):
\begin{align}
  {\cal G}_2(\Phi_v) &\equiv \nabla_i\nabla_j\Phi_v \, \nabla_i\nabla_j\Phi_v - \left(\nabla^2\Phi_v\right)^2\,, \\
  {\cal G}_{21}(\varphi_2,\varphi_1) &\equiv \nabla_i\nabla_j\varphi_2 \, \nabla_i\nabla_j\varphi_1 - \nabla^2 \varphi_2 \, \nabla^2 \varphi_1\,.
\end{align}
Here, $\Phi_v = \nabla^{-2}\,\theta$ is the velocity potential, while $\varphi_2$ and $\varphi_1$ are the Lagrangian perturbation theory potentials, which satisfy
\begin{equation}
  \label{eq:theory.phiLPT}
  \nabla^2 \varphi_1 = -\delta\,, \quad \nabla^2 \varphi_2 = -{\cal G}_2(\varphi_1)\,.
\end{equation}
The last term in Eq.~(\ref{eq:theory.bias_expansion}) is a higher-derivative contribution, which accounts for the fact
that gravitational collapse is not a local process, but occurs over finite patches of typical size $R_*$.  This spatial non-locality implies that the galaxy over-density at position $\bx$ is not only related to the matter properties at the same location, but should rather depend on these properties across the full extent of the patch.  However, perturbatively, this non-locality can be reduced to terms scaling as $R_*^2\,\nabla^2 {\cal O}$ with ${\cal O} \in \{\delta\,,\,\delta^2\,,\,{\cal G}_2\,,\ldots\}$ \cite{Des0811,McDRoy0908,DesCroSco1011,DesJeoSch1802}.  Assuming that $R_*$ is of similar order as the non-linearity scale, all terms except $R^2\,\nabla^2\delta$ are irrelevant (in the sense of being subdominant) for the computation of the one-loop power spectrum and tree-level bispectrum and have therefore been omitted from Eq.~(\ref{eq:theory.bias_expansion}).  The magnitude of $R_*$ will vary for different galaxy populations and its dependence is absorbed into the bias parameter $b_{\nabla^2}$.

\subsection{Stochasticity}
\label{sec:theory.stochasticity}

\subsubsection{Real space}
\label{sec:theory.stoch.real}

Due to the highly non-linear nature of galaxy formation, the galaxy density is not only influenced by large-scale environments, but also small-scale perturbations.  For Gaussian initial conditions, small-scale perturbations are not correlated over long distances, such that on large scales they manifest as an additional stochastic contribution, $\epsilon_g$, to the galaxy bias relation \cite{DekLah9907,TarSod9909,Mat9911}.  As shown in \cite{DesJeoSch1802}, this stochastic field can be expanded to the relevant order for this work as follows:
\begin{equation}
  \label{eq:theory.epsilon_g}
  \epsilon_g(\bx) = \epsilon(\bx) + \left(\epsilon_\delta\delta\right)(\bx) + \ldots\,,
\end{equation}
where the fields $\epsilon$ and $\epsilon_\delta$ have zero ensemble average and do not correlate with any operators in the large-scale bias expansion.  In the large-scale limit, these fields are fully described by their joint one-point probability density function, allowing us to determine their two- and three-point statistics in Fourier space in terms of the stochastic amplitudes $N_0^P$, $N_{2,0}^P$, $M_0^B$, and $N_0^B$, which we define as:
\begin{align}
  \langle \epsilon(\bk)\,\epsilon(\bk')\rangle' &= \frac{1}{\bar{n}} \left(N_0^P + N_{2,0}^P\,k^2 + \ldots\right)\,, \label{eq:theory.stoch_epseps}\\
  \langle \epsilon(\bk)\,\epsilon_\delta(\bk)\rangle' &= \frac{b_1\,M_0^B}{2\bar{n}} + \ldots\,, \label{eq:theory.stoch_epsepsdelta}\\
   \langle \epsilon(\bk)\,\epsilon(\bk')\,\epsilon(\bk'')\rangle' &= \frac{N_0^B}{\bar{n}^2} + \ldots\,, \label{eq:theory.stoch_epsepseps}
\end{align}
where $\bar{n}$ denotes the mean number density of the galaxy sample and the prime on the angle brackets indicates that we have dropped a factor of $(2\pi)^3$ and the momentum conserving Dirac delta function from the right-hand side.  The spatial non-locality discussed in the previous section also affects the stochastic terms, giving rise to terms such as $R_*^2\,\nabla^2\epsilon$, which we have included via the scale-dependent correction in Eq.~(\ref{eq:theory.stoch_epseps}).

\subsubsection{Redshift space}
\label{sec:theory.stoch.RSD}

In redshift space and in the context of the EFT, it has been argued that the small-scale, virialised motions of galaxies act like a stochastic component to the velocity field \cite{PerSenJen1610,DesJeoSch1812}.  Through the redshift-space mapping, the LOS gradient of this stochastic component, $\nabla_z\epsilon_{u_z}$, can then couple with the stochasticity in the galaxy density.  At leading order, this results in an additional term in the power spectrum:
\begin{align}
  \langle \epsilon(\bk)\,\nabla_z\epsilon_{u_z}(\bk')\rangle' = \frac{N_{2,2}^P}{\bar{n}}\,k^2\,\mu^2 + \ldots\,,
\end{align}
with an additional free parameter $N_{2,2}^P$.  Due to the equivalence principle, the stochasticity in the LOS velocity field must be suppressed by a factor of $k_z$ \cite{DesJeoSch1802}, such that the contribution to the power spectrum scales as $k_z^2$. 

Using the \VDG model, we can provide an alternative derivation of how stochasticity is induced by small-scale motions of galaxies.  The corresponding velocity components in this case are captured through the non-perturbative VDG function, which is why we do not account for any additional stochastic contribution to the velocity field itself.  Stochasticity, therefore, only affects the galaxy density within the configuration-space correlators that appear in Eq.~(\ref{CumEx}) (terms in square brackets).  Since the stochastic terms in configuration space are localised at separations $\br = 0$, we can no longer approximate the VDG by its infinite separation limit, as done for the other terms in Sec.~\ref{sec:theory.VDG}.  Instead, we solve the expression using that the Fourier transform of Eq.~(\ref{eq:theory.stoch_epseps}) gives
\begin{equation}
  \label{eq:theory.stoch_epseps_config}
  \langle \epsilon(\bx)\,\epsilon(\bx+\br) \rangle = \frac{1}{\bar{n}} \left(N_0^P - N_{2,0}^P\,\nabla^2\right)\,\delta_{\rm D}(\br)\,,
\end{equation} 
so that after plugging into Eq.~(\ref{PkRSD}) and integrating by parts we obtain:
\begin{align}
  P_{\rm stoch,VDG}(\bk) &= \int \mathrm{d}^3r \, \mathrm{e}^{\mathrm{i}\bk \cdot \br} \, W(\lambda\,|\,\br) \, \langle \epsilon(\bx)\,\epsilon(\bx + \br) \rangle \nonumber \\
                                &= \frac{1}{\bar{n}} \left(N_0^P + N_{2,0}^P\,k^2\right) \lim_{r \to 0} \, W(\lambda\,|\,\br) \nonumber \\
  & \hspace{1.2em} - \frac{N_{2,0}^P}{\bar{n}} \, \lim_{r \to 0} \left[\nabla^2\,W(\lambda\,|\,\br)\right]\,.
\end{align}
The zero-lag limits of the VDG function and its second derivative are given by:
\begin{align}
  \lim_{r \to 0} W(\lambda \, | \, \br) &={}1 \\
  \lim_{r \to 0} \left[\nabla^2\,W(\lambda \, | \, \br)\right] &={}\frac{5}{3} \, \lambda^2 \, \langle (\nabla_zu_{z})^2 \rangle \,,
\end{align}
where we have used that velocity differences vanish in the zero-lag limit; this demonstrates that the redshift-space mapping leads to an additional LOS-dependent stochastic term.  Its amplitude is determined by the variance of $\nabla_z\,u_z$, which is sensitive to small-scale perturbations and requires us to introduce an additional free stochastic parameter, consistent with the stochastic interpretation given at the beginning of this section.  Importantly, in contrast with what has typically been assumed in past analyses using the \VDG or the related TNS model \cite[e.g.][]{BeuSaiSeo1409,SanScoCro1701,DeRuhRai2012}, and what we reported in our previous paper \cite{EggCamPez2212}, there is \emph{no FoG damping} of the stochastic contributions\footnote{It is possible to show that a more rigorous treatment of the VDG including shell-crossing does not change these conclusions.}.  For that reason, we adopt the same expression for both, the EFT and \VDG model\footnote{We have made the additional (arbitrary) redefinition $N^P_{2,0} \to N^P_{2,0} + N^P_{2,2}/2$, so that the $N^P_{2,2}$ term predominantly only affects the power spectrum quadrupole.}
\begin{equation}
  \label{eq:theory.stoch.Pggstoch}
  P_{\rm stoch}(\bk) = \frac{1}{\bar{n}}\left(N^P_0 + N^P_{2,0}\,k^2 + N^P_{2,2}\,{\cal L}_2(\mu)\,k^2\right)\,.
\end{equation}

The stochastic contributions in the redshift-space bispectrum can be derived in a fully analogous manner.  First, we isolate all terms involving the stochastic fields $\epsilon$ and $\epsilon_\delta$ that result from the cumulant expansion of Eq.~(\ref{BkRSD}), and using Eqs.~(\ref{eq:theory.stoch_epseps})-(\ref{eq:theory.stoch_epsepseps}), gives:
\begin{align}
  \label{eq:theory.stoch.Bstoch_VDG}
  B_{\rm stoch,VDG}(\bk_1,\bk_2,\bk_3) ={}&\frac{N^B_0}{\bar{n}^2} \lim_{\substack{r_{13} \to 0, \\ r_{23} \to 0}} W(\lambda_1,\lambda_2 \, | \, \br_{13},\br_{23})  \nonumber \\
                                              &\hspace{-6.5em}+ \bigg\{\frac{1}{\bar{n}}\left[b_1\,M^B_0 + f\,\mu_1^2\,N^P_0\right]\,Z_1(\bk_1)\,P_{\rm lin,IR}(\bk_1) \nonumber \\
                                              &\hspace{-6.5em}\times \, \lim_{\substack{r_{13} \to \infty, \\ r_{23} \to 0}} W(\lambda_1,\lambda_2 \, | \, \br_{13},\br_{23}) + \mathrm{cyc.} \bigg\}\,.
\end{align}
The first term arises from self-triplets, i.e. configurations whose three positions coincide, while the second term captures self-pairs.  In the latter case, we apply the same approximation as before, taking the infinite separation limit of the VDG for the non-zero pair separation vectors.  The two different limits of the three-point VDG can thus be expressed as 
\begin{align}
  \lim_{\substack{r_{13} \to 0, \\ r_{23} \to 0}} W(\lambda_1,\lambda_2 \, | \, \br_{13},\br_{23}) &= 1 \\
  \lim_{\substack{r_{13} \to \infty, \\ r_{23} \to 0}} W(\lambda_1,\lambda_2 \, | \, \br_{13},\br_{23}) &= \frac{1}{\left(1 - \lambda_1^2 a_{\rm vir}^2\right)^{3/2}}  \nonumber \\ &\times \exp{\left(\frac{\lambda_1^2\,\sigma_v^2}{1 - \lambda_1^2 a_{\rm vir}^2}\right)} \,.
\end{align}
For the EFT model, we additionally neglect the large-scale damping effect, which leads to the following stochasticity bispectrum:
\begin{equation}
  \label{eq:theory.stoch.Bstoch_EFT}
  \begin{split}
  B_{\rm stoch,EFT}(\bk_1,\bk_2,\bk_3) = \frac{N_0^B}{\bar{n}^2} + \frac{1}{\bar{n}}\Big\{ \left[b_1\,M_0^B + f\,\mu_1^2\,N^P_0\right] \, \\ \times \, Z_1(\bk_1)\,P_{\rm lin,IR}(\bk_1) + \mathrm{cyc.} \Big\}\,.
  \end{split}
\end{equation}
It is worth noting that our findings indicate that the term scaling as $f\, \mu_i^2 \, Z(\bk_i) \, \PL(k_i)$ in Eqs.~(\ref{eq:theory.stoch.Bstoch_VDG}) and (\ref{eq:theory.stoch.Bstoch_EFT}) is naturally generated by the redshift-space mapping, albeit with a fixed amplitude determined by the large-scale power spectrum stochasticity $N^P_0/\bar{n}$ (as found also by \cite{IvaPhiNis2203,PhiIvaCab2206,IvaPhiCab2304}).  This is in contrast with \cite{DesJeoSch1812}, which argues that only anisotropic selection effects can produce this term.

\subsection{Counterterms}
\label{sec:theory.counterterms}

The final ingredient for describing the statistics of tracers in redshift space is the so-called \emph{counterterms}.  In any effective theory that aims to describe physics only up to a certain wave number cutoff scale, counterterms are essential for regularising the effects of scales beyond this cutoff.  It was demonstrated in \cite{SenZal1409,DesJeoSch1812} that the counterterms required to regularise the redshift-space power spectrum at the one-loop level scale as $\sim \mu^{2n}\,k^2\,\PL(k)$ with $n=0,1,2$.  For that reason, we introduce three more free parameters $c_0$, $c_2$, and $c_4$, and define the leading order (LO) counterterms for the power spectrum as
\begin{equation}
  \label{eq:theory.ctr.Pctr_LO}
  P_{\rm ctr}^{\rm LO}(\bk) \equiv -2 \sum_{n=0}^2 c_{2n}\,{\cal L}_{2n}(\mu) \, k^2 \,P_{\rm lin,IR}(\bk)\,,
\end{equation}
such that each counterterm predominantly only contributes to one of the multipoles.

Apart from absorbing potential cutoff dependencies, these counterterms also model various physical effects.  This includes the higher-derivative bias contribution $\nabla^2\delta$ discussed in Sec.~\ref{sec:theory.bias}, which scales identically to the $n=0$ counterterm and implies that its bias coefficient, $b_{\nabla^2\delta}$, can be absorbed in the $c_0$ parameter.  The same term also describes the leading effect from the breakdown of the perfect fluid approximation due to shell crossing \cite{PueSco0908,BauNicSen1207,CarHerSen1209}.  Finally, \cite{DesJeoSch1812} showed that the lowest order contributions from velocity bias---differences in the velocity field of galaxies and the underlying matter field---can be accounted for by the $n=1$ and $n=2$ counterterms.

In the EFT framework, where the VDG function is expanded perturbatively, the counterterms moreover capture partially the FoG damping caused by small-scale velocity dispersion.  This is because the perturbative expansion includes terms that scale as $\sim (f\,k\,\mu\,\sigma_v)^{2n}\,P_{\rm tree}(\bk)$ for $n \geq 1$, which depend on the (non-linear) velocity dispersion $\sigma_v$.  Since one-loop perturbation theory does not account for the non-linear corrections to $\sigma_v$, these differences are effectively absorbed by the counterterms.  The significance of these contributions relative to the other effects described by the counterterms may lead to parameter values that differ between the EFT and \VDG models.  This will be explored empirically in Sec.~\ref{sec:results.pk.counterterms}.

A difference in the non-linearity scales of the redshift-space mapping and the dynamics would imply that higher-order corrections ($n \geq 2$) in the expansion of the VDG may not be subdominant compared to the remaining terms in the one-loop power spectrum.  This was already recognised in \cite{IvaSimZal2005}, which introduced an additional ``next-to-leading order'' (NLO) counterterm in an attempt to address this issue.  This counterterm is defined here as
\begin{equation}
  \label{eq:theory.ctr.Pctr_NLO}
  P_{\rm ctr}^{\rm NLO}(\bk) = c_{\rm nlo}\,(f\,k\,\mu)^4\,P_{\rm tree}(\bk)\,,
\end{equation}
so that the total counterterm power spectrum in the EFT model is given by:
\begin{equation}
  \label{eq:theory.ctr.Pctr}
  P_{\rm ctr,EFT}(\bk) = P_{\rm ctr}^{\rm LO}(\bk) + P_{\rm ctr}^{\rm NLO}(\bk)\,.
\end{equation}
However, since this counterterm is solely motivated by the perturbative expansion of the VDG, we do not include it in the \VDG model, such that in that case we simply have $P_{\rm ctr,VDG}(\bk) = P^{\rm LO}_{\rm ctr}(\bk)$.

\begin{table}[b]
  \caption{Cosmological parameters of the \textsc{Minerva} simulation suite.}
  \begin{ruledtabular}
    \begin{tabular}{lccccc}
      Parameter & $\omega_c$ & $\omega_b$ & $h$ & $n_s$ & $\sigma_8$ \\[0.2em]
      Value & 0.11544 & 0.02222 & 0.695 & 0.9632 & 0.828 \\
    \end{tabular}
  \end{ruledtabular}  
  \label{tab:minerva_params}
\end{table}

The tree-level bispectrum does not require counterterms to regularise cutoff dependencies or model any of the other effects previously discussed.  However, within the EFT framework one can again consider NLO terms arising from the expansion of the VDG function, which may be able to capture potentially significant FoG effects.  In \cite{IvaPhiNis2203,IvaPhiCab2304} it is assumed that these terms effectively lead to a modification of the $Z_1$ kernel as follows:
\begin{equation}
  \label{eq:theory.ctr.Z1ctr}
  Z_1(\bk) \to Z_1(\bk) - c_1^B\,k^2\,\mu^2 - c_2^B\,k^2\,\mu^4\,,
\end{equation}
where $c_1^B$ and $c_2^B$ are additional free parameters.  This modification alters the expressions of the tree-level bispectrum (Eq.~\ref{Bspt}) and its stochasticity (Eq.~\ref{eq:theory.stoch.Bstoch_EFT}), but it is not possible to link the resulting corrections to an expansion of the VDG function.  This is why, in addition, we consider an alternative counterterm that is consistent with the approach taken for the power spectrum and obtained by a direct expansion of $W(\lambda_1,\lambda_2\,|\,\br_{13},\br_{23})$:
\begin{align}
  \label{eq:theory.ctr.Bctr_EFT}
    B_{\rm ctr,EFT}(\bk_1,\bk_2,\bk_3) ={} &c_{\rm nlo}^B\,f^2\,\bigg\{\left[k\mu\right]^2_{123} \, B_{\rm tree}(\bk_1,\bk_2,\bk_3)\bigg. \nonumber \\ &+ \left[\frac{k_1^2\,\mu_1^2}{\bar{n}} \left(b_1\,M_0^B + f\,\mu_1^2\,N_0^P\right) \right. \nonumber \\  &\bigg.\left. \times\, Z_1(\bk_1)\,P_{\rm lin,IR}(\bk_1) + \mathrm{cyc.}\right]\bigg\}\,,
\end{align}
using the short-hand $\left[k\mu\right]^2_{123} \equiv k_1^2\,\mu_1^2 + k_2^2\,\mu_2^2 + k_3^2\,\mu_3^2$.  Similar to the power spectrum, we only use Eq.~(\ref{eq:theory.ctr.Z1ctr}) or Eq.~(\ref{eq:theory.ctr.Bctr_EFT}) in case of the EFT model.


\section{Synthetic data}
\label{sec:data}

\subsection{Galaxy catalogues}
\label{sec:catalogues}

In this work, we analyse the clustering of two simulated populations of galaxies that closely mimic the properties of the CMASS and LOWZ samples of BOSS.  The catalogues are constructed from the \textsc{Minerva} dark matter only simulations \cite{GriSanSal1604,LipSanCol1901}, which were run with the \textsc{Gadget-2} code \cite{Spr0512} and evolve $1000^3$ particles in a periodic box of side length $1500\,\hMpc$.  Their initial conditions were generated at redshift $z_{\mathrm{ini}} = 63$ with second-order Lagrangian perturbation theory and adopt a flat $\Lambda$ cold dark matter cosmology with parameter values given in Table~\ref{tab:minerva_params}.

Halo catalogues have been produced with a friends-of-friends halo finder in conjunction with \textsc{Subfind} \cite{SprWhiTor0112} and were subsequently populated with galaxies at redshifts $z = 0.57$ and $0.3$ using a five-parameter halo occupation distribution (HOD) model tuned to match the two-point clustering of the observed samples \cite[for more details, see][]{GriSanSal1604,LipSanCol1901}.  The resulting catalogues have mean number densities of $\bar{n} \sim 4.0 \times 10^{-4}\,\hinvMpcvol$ and $\sim 3.1 \times 10^{-4}\,\hinvMpcvol$, and despite lacking any survey-specific selection and systematic effects, we will refer to them simply as CMASS and LOWZ, respectively, for the remainder of this paper.  In total, the \textsc{Minerva} suite consists of 300 independent realisations, providing a combined volume of $1012.5\,\hGpcvol$, which is sufficiently large to suppress sample variance on scales of interest to this analysis.

\subsection{Power spectrum and bispectrum measurements}
\label{sec:measurements}

We adopt the distant-observer approximation and distort the positions of the galaxies in each of the catalogues in $z$-direction according to the real- to redshift-space mapping (see Eq.~\ref{RSDmap}).  We then condense the information in the two-dimensional power spectrum into Legendre multipoles using the following estimator \cite[e.g.,][]{Sco1510}:
\begin{equation}
  \label{eq:data.Pl}
  \hat{P}_{\ell}(k) \equiv \frac{2\ell + 1}{N_P(k)} \sum_{\qv \in k} \, {\cal L}_{\ell}(\hat{\qv} \cdot \hat{\boldsymbol{z}}) \, \left| \ds(\qv) \right|^2\,,
\end{equation}
where the summation is taken over all modes $\qv$ that fall into a spherical shell centred on $k$ with bin width $\Delta k_P$, and $N_P(k)$ denotes the number of these modes per shell.  Choosing the fundamental frequency of the simulation box, $\kf \approx 0.00419\,\hinvMpc$, as the bin width, we measure the first three non-zero moments, i.e., the monopole $(\ell=0)$, quadrupole $(\ell=2)$, and hexadecapole $(\ell=4)$, between $k_{\mathrm{min}} = \kf$ and $k_{\mathrm{max}} = 0.4\,\hinvMpc$, which results in 95 bins each. 

For the bispectrum, we follow \cite{ScoCouFri9906,Sco1510} and characterise the orientation of a triangle configuration $\{\kv_1, \kv_2, \kv_3\}$ with respect to the LOS by means of the polar angle $\theta = \arccos{(\hat{\kv}_1 \cdot \hat{\boldsymbol{z}})}$ and the azimuthal angle $\phi$, which describes the rotation of $\kv_2$ around $\kv_1$.  We average over $\phi$ and extract the monopole and quadrupole moments of the bispectrum as follows:
\begin{equation}
  \label{eq:data.Bl}
  \begin{split}
    \hat{B}_{\ell}(k_1,k_2,k_3) \equiv \frac{2\ell + 1}{N_B(k_1,k_2,k_3)} \sum_{\qv_1 \in k_1}\,\sum_{\qv_2 \in k_2}\,\sum_{\qv_3 \in k_3} \delta^{\mathrm{K}}_{\qv_{123},0} \\ \times\, {\cal L}_{\ell}(\hat{\qv}_1 \cdot \hat{\boldsymbol{z}}) \, \ds(\qv_1) \, \ds(\qv_2) \, \ds(\qv_3)\,,
  \end{split}
\end{equation}
where the Kronecker-delta, $\delta^{\mathrm{K}}$, represents the statistical homogeneity constraint that the three wave vectors form a closed triangle, i.e., $\qv_{123} \equiv \qv_1 + \qv_2 + \qv_3 = 0$, and $N_B(k_1,k_2,k_3)$ is the total number of such fundamental triangles per bin $\{k_1,k_2,k_3\}$.  In practice, we make use of the efficient estimator that expands the Kronecker-delta into plane waves and identifies closed fundamental triangles via fast Fourier transforms \cite{SefCroSco1608,Sco1510}.  For our bispectrum measurements we choose a bin width of $\Delta k_B = 2\,\kf$ and consider all triangle bins that satisfy $k_1 \geq k_2 \geq k_3$ and $k_1 \leq k_2 + k_3$ with $k_{\mathrm{max}} = 0.143\,\hinvMpc$, leading to 597 configurations per multipole.  We associate a triangle index to each configuration via a nested loop over the three wave modes, in which $k_3$ varies most quickly.

In order to minimise the effect of sample variance on the comparison of the different redshift-space distortion models, we average each measurement over all 300 realisations and take these mean quantities as our mock data vectors in the analysis below.

\subsection{Covariances}
\label{sec:covariances}

For both, the power spectrum and bispectrum multipoles, we adopt analytical Gaussian covariance matrices in order to quantify the measurement uncertainties and correlation structure of our data vectors.  In this approximation the covariance between two power spectrum measurements with multipole numbers $\ell_1$ and $\ell_2$, and bin numbers $i$ and $j$, is given by:
\begin{equation}
  \label{eq:data.Pcov}
  \begin{split}
    &\mathrm{Cov}\left[\hat{P}_{\ell_1}^i,\,\hat{P}_{\ell_2}^j\right] \approx \frac{2 (2\ell_1 + 1) (2\ell_2+1)}{N_P(k_i)} \, \delta^{\mathrm{K}}_{ij} \\ &\quad\times \, \sum_{\ell_3 = 0}^{\infty} \sum_{\ell_4=0}^{\ell_3} R^P_{\ell_1 \ell_2 \ell_4 (\ell_3-\ell_4)} \, P^{\mathrm{tot}}_{\ell_4}\left(k_{\mathrm{eff}}^{(i)}\right) \, P^{\mathrm{tot}}_{\ell_3 - \ell_4}\left(k_{\mathrm{eff}}^{(j)}\right) \,,
  \end{split}
\end{equation}
where we have defined $P^{\mathrm{tot}}_{\ell}(k) \equiv P_{\ell}(k) + \delta^{\rm K}_{\ell 0}/\bar{n}$ and $R^P_{\ell_1 \ell_2 \ell_3 \ell_4}$ is a factor that results from replacing the discrete sum over wave vectors in the estimator, Eq.~(\ref{eq:data.Pl}), with its continuum limit and integrating over the LOS dependence.  It can be expressed as a product of Wigner-$3j$ symbols
\begin{equation}
  \label{eq:data.Cl1l2l3l4}
  R^P_{\ell_1 \ell_2 \ell_3 \ell_4} = \sum_{\substack{\ell = \max{\{|\ell_1 - \ell_2|,} \\ |\ell_3 - \ell_4|\} }}^{\min{\{\ell_1 + \ell_2,\\ \ell_3 + \ell_4}\}}
  \left(
    \begin{matrix}
      \ell_1 & \ell_2 & \ell \\
      0 & 0 & 0 
    \end{matrix}
  \right)^2
  \left(
    \begin{matrix}
      \ell_3 & \ell_4 & \ell \\
      0 & 0 & 0 
    \end{matrix}
  \right)^2\,,
\end{equation}
but for practical purposes we truncate the series expansion in Eq.~(\ref{eq:data.Pcov}) at $\ell=4$, which translates into $R^P_{\ell_1 \ell_2 \ell_3 \ell_4} = 0$ for $\ell_3 + \ell_4 > 4$.  We have, furthermore, approximated the remaining radial integral over the size of the bin with an evaluation at the bin-averaged, \emph{effective}, modes $k_{\mathrm{eff}} \equiv N_P^{-1} \sum_{\qv \in k} q$.  It is possible to partly account for non-linear (non-Gaussian) contributions to the covariance by using a non-linear model for the power spectrum multipoles in Eq.~(\ref{eq:data.Pcov}).  In practice, we do this by fitting the three power spectrum multipoles using the \VDG model with an initial guess of the covariance matrix and use the best-fit nuisance parameters to produce an updated model for the covariance.

\begin{figure}[t]
  \centering
  \includegraphics{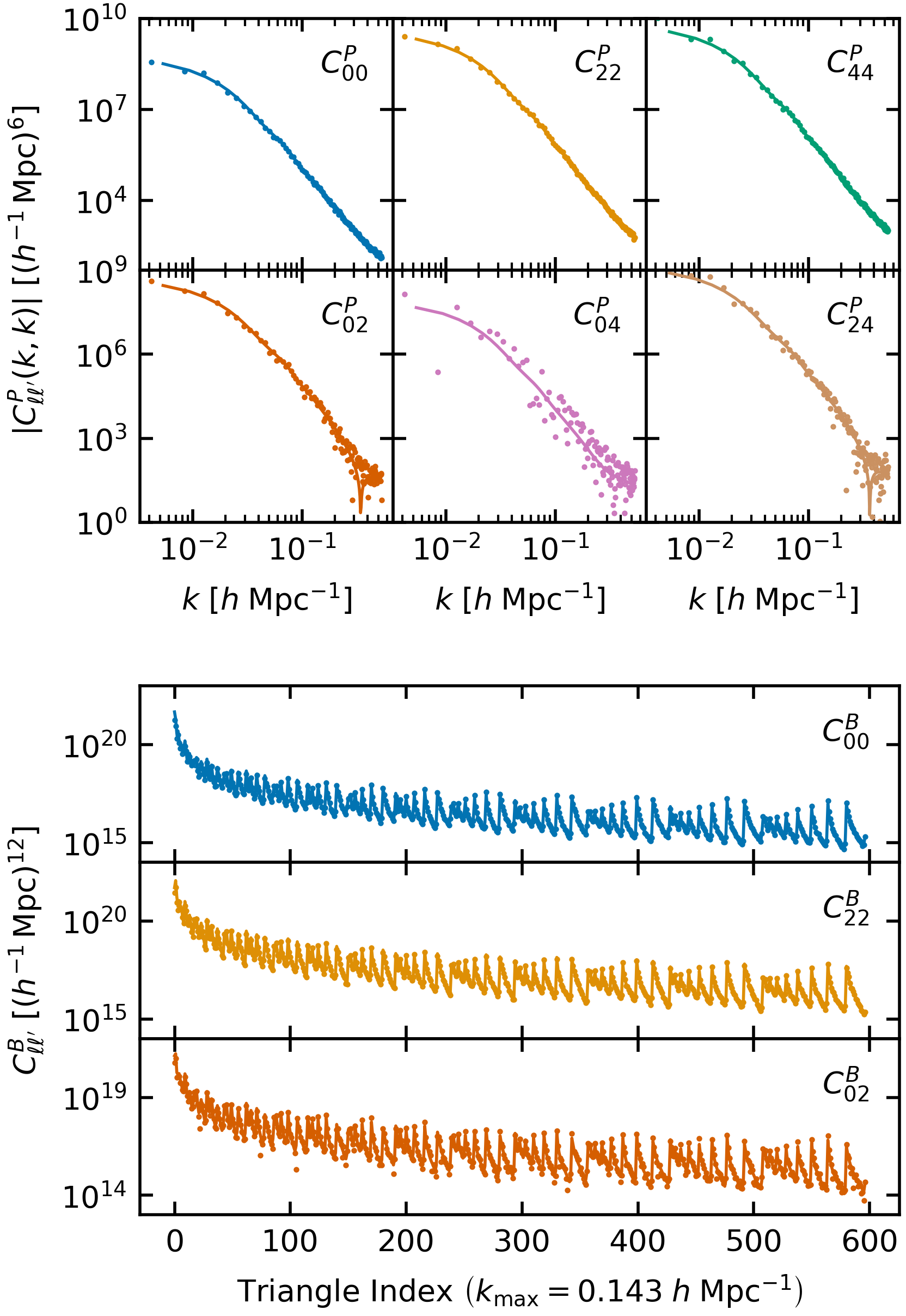}
  \caption{Comparison of measured auto- and cross-covariances for different multipole numbers with our analytical Gaussian predictions. Top panels show the results for the CMASS power spectrum, bottom panels for the bispectrum.}
  \label{fig:PBell_covariance_cmass}
\end{figure}

A comparison between this model and the estimated auto- and cross-covariances from the pool of 300 realisations is shown in the top panels of Fig.~\ref{fig:PBell_covariance_cmass} for the CMASS sample.  Up to scatter due to the limited number of realisations and discreteness effects, in particular for $C^P_{04}$ and $C^P_{24}$, the analytic model provides an excellent description of the measured covariances over the entire range of scales considered in this work (the agreement for the LOWZ sample is qualitatively identical).  Cross-correlation coefficients between different bins, which vanish in our covariance model, stay below 10\,\% in the numerical covariances, implying that their impact on our results would only be of subordinate importance\footnote{Note that this is largely a consequence of the periodic boundary conditions.  The presence of a survey window function induces stronger correlations between neighbouring bins, see e.g., \cite{WadSco2012}.}.

The Gaussian bispectrum covariance between multipoles $\ell_1$ and $\ell_2$, and triangle bins $\{k_{1}^{(i)},k^{(i)}_{2},k^{(i)}_{3}\}$ and $\{k^{(j)}_{1},k^{(j)}_{2},k^{(j)}_{3}\}$ can be expressed as
\begin{align}
  \label{eq:data.Bcov}
    &\mathrm{Cov}\left[\hat{B}_{\ell_1}^{(i)},\,\hat{B}_{\ell_2}^{(j)}\right] \approx \; \frac{(2\ell_1+1)(2\ell_2+1)}{N_B(k_{1}^{(i)},k_{2}^{(i)},k_{3}^{(i)})\,V} \, \delta^{\mathrm{K}}_{ij} \, \sum_{\ell_3=0}^{\infty} \sum_{\ell_4=0}^{\infty} \sum_{\ell_5=0}^{\infty} \nonumber \\
    &\hspace{1em}\times \, P^{\mathrm{tot}}_{\ell_3}\left(k_{\mathrm{eff},1}^{(i)}\right) \, P^{\mathrm{tot}}_{\ell_4}\left(k_{\mathrm{eff},2}^{(i)}\right) \, P^{\mathrm{tot}}_{\ell_5}\left(k_{\mathrm{eff},3}^{(i)}\right) \nonumber \\
    &\hspace{1em}\times \, R^B_{\ell_1 \ell_2 \ell_3 \ell_4 \ell_5}\left(k_{\mathrm{eff},1}^{(i)},k_{\mathrm{eff},2}^{(i)},k_{\mathrm{eff},3}^{(i)}\right)\,,
\end{align}
where $V$ denotes the survey volume.  As for the power spectrum, we have approximated all discrete sums over wave vectors in the estimator, Eq.~(\ref{eq:data.Bl}), with their continuum limit and replaced the average over the triangular configurations with an evaluation at the effective modes, which in case of the bispectrum are defined as
\begin{equation}
  \label{eq:data.keff}
  k_{\mathrm{eff},i} \equiv \frac{1}{N_B(k_1,k_2,k_3)} \sum_{\qv_1 \in k_1}\,\sum_{\qv_2 \in k_2}\,\sum_{\qv_3 \in k_3} \delta^{\mathrm{K}}_{\qv_{123},0}\,q_i\,.
\end{equation}
The expression for the mixing coefficient $R^B_{\ell_1 \ell_2 \ell_3 \ell_4 \ell_5}$ is given in Appendix~\ref{sec:app.Bcov}.  Using the exact number of fundamental triangles $N_B$ in Eq.~(\ref{eq:data.Bcov}), the same non-linear model for the power spectrum multipoles as we did in case of the power spectrum covariance, and truncating the sums over multipoles beyond the hexadecapole, we obtain the predictions for the monopole and quadrupole auto-covariance and their cross-covariance shown in the lower panels of Fig.~\ref{fig:PBell_covariance_cmass}. \\

Compared to the numerical estimates, the model provides again an accurate description, with relative differences staying below 20\,\% for the vast majority of entries in $C^B_{00}$ and $C^B_{22}$, while for $C^B_{02}$ that fraction is greater due to the increased scatter in the measured covariances.  These findings also extend to the LOWZ sample and are consistent with those of \cite{RizMorPar2301}.  We find that the largest differences occur for squeezed triangle configurations, which, as shown by \cite{BiaCasNor2209,SalCasNor2408}, are most heavily affected by non-Gaussian contributions to the covariance.  However, since we only consider triangles with a maximum side length of $0.143\,\hinvMpc$ we do not reach highly squeezed configurations and hence do not observe the strong deviations from the Gaussian predictions that are reported in their work.  The same extends to the cross-correlation between different triangle configurations: in accordance with \cite{BiaCasNor2209,SalCasNor2408} we find a prevalence of stronger correlations among configurations that share a long wave mode, but only a negligible number of configurations have cross-correlation coefficients that surpass 30\,\%.  Altogether, we therefore expect the inaccuracies in our covariance modelling to have limited impact. \\

Finally, we note that the Gaussian covariances, both for the power spectrum and bispectrum, scale inversely with the sample volume.  We use this fact to rescale our uncertainties, such that they are more representative of those expected from current galaxy survey data.  Specifically, we assume a Euclid redshift shell of $\Delta z$ = 0.2, centred on $z = 0.9$ with a survey area of $15,000\,\mathrm{deg}^2$ and a sample number density of $\bar{n} = 2.04 \times 10^{-3}\,\hinvMpcvol$.  We compute the effective volume of this shell using \cite{Teg9512}
\begin{equation}
  \label{eq:data.Veff}
  V_{\mathrm{eff}} = \left(\frac{\bar{n}\,P_0}{1 + \bar{n}\,P_0}\right)^2\,V\,,
\end{equation}
where we set $P_0 = 10^{4}\,\hMpcvol$.  We analogously compute the effective volume of our CMASS and LOWZ samples and rescale our covariances by  $V_{\mathrm{eff}}^{\mathrm{CMASS}}/V_{\mathrm{eff}}^{\mathrm{Euclid}}$ and $V_{\mathrm{eff}}^{\mathrm{LOWZ}}/V_{\mathrm{eff}}^{\mathrm{Euclid}}$, which results in the factors $1/2.42$ and $1/2.51$, respectively.  The sample volume that these covariances correspond to is thus much smaller than the total sample volume that our mean data vectors originate from.  This, however, will greatly facilitate our intended comparison of the different RSD models, as sample variance effects on the parameter posteriors will be suppressed.

\section{Analysis setup}
\label{sec:pipeline}

In this section, we describe the setup of our Bayesian analysis procedure, and discuss how our modelling and prior choices compare with previous studies.  We also introduce the metrics that we extract from the posteriors and which we use in Sec.~\ref{sec:results} to compare the different models.

\subsection{Likelihood function and sampler}
\label{sec:likelihood}

We use a standard Bayesian inference technique to analyse the synthetic CMASS and LOWZ data and extract posterior constraints on a set of cosmological, galaxy bias, and RSD parameters.  This requires the assumption of a likelihood function for the data, which we take to be a multivariate Gaussian.  Since we ignore any cross-correlation between the power spectrum and bispectrum measurements, the total likelihood is given by the product of the individual ones,
\begin{equation}
  \label{eq:pipe.like_tot}
  {\cal L}_{\mathrm{tot}} = {\cal L}_{P} \, \times \, {\cal L}_{B}\,,
\end{equation}
with
\begin{equation}
  \label{eq:pipe.like_ind}
  -2 \log{{\cal L}_{X}} = \sum_{i,j=1}^{N^X_{\mathrm{bin}}} \left(\hat{X}_i - X_i\right) \, C_{X,ij}^{-1} \left(\hat{X}_j - X_j\right)\,,
\end{equation}
where $X$ can either stand for the power spectrum or bispectrum and quantities with a hat denote the measurements, otherwise the model predictions.  The vector $\boldsymbol{X}$ is the concatenation of the different multipoles and the total number of bins is determined by the wave mode cutoff $k_{\mathrm{max}}$.  We will explore multiple different choices of cutoffs, which may differ between the power spectrum and bispectrum, and also between different multipoles.  We use the notation $k^X_{\mathrm{max}} = \left(k^X_{\mathrm{max},0},k^X_{\mathrm{max},2},k^X_{\mathrm{max},4}\right)$ to specify cutoffs for the individual multipoles, while a single number for $k^X_{\mathrm{max}}$ implies the same cutoff for all multipoles.

Our model fits are conducted by minimising the negative log-likelihood in a Markov chain Monte Carlo (MCMC) approach.  We use the nested sampling algorithm provided by \texttt{MultiNest} \cite{FerHob0801,FerHobBri0910,FerHobCam1911} and its \texttt{Python} front end \texttt{PyMultiNest} \cite{BucGeoNan1404} with a total of 500 live points, a sampling efficiency of 0.5, and an evidence tolerance of 0.4.  The Markov chains are subsequently processed with \texttt{getdist} \cite{Lew1910} to extract the marginalised parameter constraints and posterior density estimates.

\subsection{Modelling and prior choices}
\label{sec:priors}

For our analysis, we consider the two different models discussed in Sec.~\ref{sec:theory}: the EFT model, where the VDG function (responsible for the FoG effect) is expanded perturbatively, and the \VDG model, where this function is kept non-perturbative.  In Table~\ref{tab:priors} we list the various parameters that are being varied in our fits, using circles to indicate parameters that only appear in the respective power spectrum models, while triangle symbols indicate parameters that are additionally required for the joint analysis of the power spectrum and bispectrum.

\begin{table}
  \centering
  \caption{Fitting parameters included in the different models and the lower and upper limits of their priors (all priors are uniform).  Circles indicate parameters that are included in the fits involving the power spectrum, triangle symbols indicate those that are only relevant for the bispectrum model.  The open and filled symbols in case of the EFT stand for the different counterterm modelling options (see text for details).}
  \begin{ruledtabular}
    \begin{tabular}{cccc}
      Parameter & Prior boundaries & EFT & VDG$_{\infty}$ \\ \colrule\\[-0.6em]
      $h$ & [0.5, 1.0] & $\circ$ $\bullet$ & $\circ$ \\[0.3em]
      $\omega_c$ & [0.085, 0.155]& $\circ$ $\bullet$ & $\circ$ \\[0.3em]
      $10^{9} \times A_s$ & [1.055, 3.17] & $\circ$ $\bullet$ & $\circ$ \\[0.3em] \colrule\\[-0.6em]
      $b_1$ & [0.25, 4.00] & $\circ$ $\bullet$ & $\circ$ \\[0.3em]
      $b_2$ & [$-5$, 5] & $\circ$ $\bullet$ & $\circ$ \\[0.3em]
      $\gamma_2$ & [$-4.0$, 0.5] & $\circ$ $\bullet$ & $\circ$ \\[0.3em]
      $\gamma_{21}$ & Coevolution &  &  \\[0.3em]
      $c_0$ [$\mathrm{Mpc}^2$] & [$-5$, 5] $\times\,10^3$ & $\circ$ $\bullet$ & $\circ$ \\[0.3em]
      $c_2$ [$\mathrm{Mpc}^2$] & [$-5$, 5] $\times\,10^3$ & $\circ$ $\bullet$ & $\circ$ \\[0.3em]
      $c_4$ [$\mathrm{Mpc}^2$] & [$-5$, 5] $\times\,10^3$ & $\circ$ $\bullet$ & $\circ$ \\[0.3em]
      $c_{\mathrm{nlo}}$ [$\mathrm{Mpc}^4$] & [$-1$, 1] $\times\,10^5$ & $\circ$ $\bullet$ & \\[0.3em]
      $c^B_1$ [$\mathrm{Mpc}^2$] & [$-1$, 1] $\times\,10^5$ & $\vartriangle$ \textcolor{white}{$\blacktriangle$} & \\[0.3em]
      $c^B_2$ [$\mathrm{Mpc}^2$] & [$-1$, 1] $\times\,10^5$ & $\vartriangle$ \textcolor{white}{$\blacktriangle$} & \\[0.3em]
      $c^B_{\mathrm{nlo}}$ [$\mathrm{Mpc}^2$] & [$-1$, 1] $\times\,10^5$ & \textcolor{white}{$\vartriangle$} $\blacktriangle$ &  \\[0.3em]
      $a_{\mathrm{vir}}$ [$\mathrm{Mpc}$] & [0, 10] &  & $\circ$ \\[0.3em]
      $N^P_0$ & [$-1$, 3] & $\circ$ $\bullet$ & $\circ$ \\[0.3em]
      $N^P_{2,0}$ [$\mathrm{Mpc}^2$] & [$-5$, 5] $\times\,10^3$ & $\circ$ $\bullet$ & $\circ$ \\[0.3em]
      $N^P_{2,2}$ [$\mathrm{Mpc}^2$] & [$-5$, 5] $\times\,10^3$ & $\circ$ $\bullet$ & $\circ$ \\[0.3em]
      $M^B_0$ & [$-3$, 3] & $\vartriangle$ $\blacktriangle$ & $\vartriangle$ \\[0.3em]
      $N^B_0$ & [$-1$, 3] & $\vartriangle$ $\blacktriangle$ & $\vartriangle$ \\
    \end{tabular}
  \end{ruledtabular}
  \label{tab:priors}
\end{table}

In all cases, we vary the three cosmological parameters $h$, $\omega_c$, and $A_s$, fixing the baryon density and spectral index to their fiducial values (see Table~\ref{tab:minerva_params}).  This is similar to the study presented in \cite{DADonLew2405}, although instead of fixing the value of $\omega_b$ they use a tight BBN-prior, but differs slightly from the analyses in \cite{IvaPhiNis2203,PhiIva2202}, which also fit for $n_s$.  Since the spectral index can only be poorly constrained from galaxy clustering data on its own, varying $n_s$ without a \emph{Planck}-like prior leads to an increase of degeneracies with the remaining cosmological parameters and thus larger uncertainties in general.  However, by comparing results with and without an $n_s$ prior, \cite{PhiIva2202} did not find differences in the relative gains from adding the bispectrum data to the power spectrum.

We further include the linear and second-order galaxy bias parameters, while fixing the third-order parameter $\gamma_{21}$ to the value given by its coevolution relation \cite{EggScoSmi1906},
\begin{equation}
  \label{eq:pipe.g21coevo}
  \gamma_{21}\left(b_1,\gamma_2\right) = \frac{2}{21}\left(b_1-1\right) + \frac{6}{7}\gamma_2\,,
\end{equation}
which was demonstrated to be in excellent agreement with measurements from a collection of different galaxy and halo samples in \cite{EggScoSmi2106}, since it is otherwise strongly degenerate with the second-order tidal bias parameter $\gamma_2$.  In the analyses of \cite{PhiIva2202,DADonLew2405} the parameter equivalent to $\gamma_{21}$ (denoted as $b_{\Gamma_3}$, see Appendix~\ref{sec:app.bias_bases}) is instead varied within a narrow Gaussian prior centred on the \emph{local Lagrangian} prediction (i.e., as a function of $b_1$ only), which is more restrictive than assigning a value based on Eq.~(\ref{eq:pipe.g21coevo}).  For all models we also include the three power spectrum counter terms $c_0$, $c_2$, and $c_4$, as well as the three stochastic parameters relevant for the analysis of the power spectrum alone, and a further two used in combined fits with the bispectrum.  Unlike \cite{IvaPhiNis2203,PhiIva2202}, we do not impose the relation $N^B_0 = (N^P_0 + 1)^2 - 1$ on the constant bispectrum shot noise term.

The model parametrisations differ for the description of the FoG effect: for the EFT model, we consider one variant that matches the description of \cite{IvaPhiNis2203}, which introduces the parameter $c_{\mathrm{nlo}}$ to model beyond next-to-leading order effects in the power spectrum, as well as the two parameters $c^B_1$ and $c^B_2$ for next-to-leading order effects in the bispectrum.  In the second variant of the EFT model, we replace the latter two parameters by a different term with parameter $c^B_{\mathrm{nlo}}$ (see Sec.~\ref{sec:theory.counterterms} for their definitions).  The \VDG model instead captures the impact of small-scale velocities via the parameter $a_{\mathrm{vir}}$ in the VDG function.  In total, for joint fits of the power spectrum and bispectrum, the three models thus have 14, 13, and 12 free parameters respectively (in addition to the cosmological parameters), while 10 free parameters are required in each case when limiting the data vector to the the power spectrum.

For each of the aforementioned parameters, we use uniform priors whose bounds are given in Table~\ref{tab:priors}.  They have been chosen to be sufficiently wide so that they do not inform the final parameter posteriors, although we impose the physically motivated cuts $N^P_0,\,N^B_0 \geq -1$ and $a_{\mathrm{vir}} \geq 0$.  This is in strong contrast to the analyses of \cite{IvaPhiNis2203,PhiIva2202,DADonLew2405}, which employ narrow Gaussian priors on either all or the majority of the galaxy bias, counterterm and stochastic parameters.  The narrow priors that restrict these parameters to values of $\sim {\cal O}(1)$ are motivated by retaining the perturbative nature of the models, which might be violated if certain parameter values become so large that the corresponding next-to-leading order contributions would strongly dominate.  On the flip side, such large values or their running with the cutoff scale when allowing for sufficient freedom in the priors is an important diagnostic for the model performance when fitting observed data where the true cosmological parameters are not known.  Moreover, recent studies \cite{CarMorPou2301,SimZhaPou2306,HolHerSim2309} have shown that --- at least for the statistical precision of the BOSS survey --- the narrow Gaussian priors are informative and affect the cosmological constraints.  For these reasons, we opt against the usage of such priors in our default analyses, although we will present a comparison in Sec.~\ref{sec:results.pk.proj+prior}.

\subsection{Performance metrics}
\label{sec:metrics}

To conduct a quantitative comparison between the EFT and \VDG models, and between various choices of scale cuts, we employ a set of metrics that we introduce in the following. 

\subsubsection{Goodness-of-fit}
\label{sec:pvals}

As described in Sec.~\ref{sec:data}, we fit the power spectrum and bispectrum measurements averaged over the entire ensemble of \textsc{Minerva} simulations, assuming statistical uncertainties that roughly match those of a Stage-IV galaxy survey.  Since this corresponds to a volume that is smaller than from all simulations combined, the resulting chi-square values (Eq.~\ref{eq:pipe.like_ind}) are therefore artificially small and can not be directly interpreted as an (absolute) measure of the goodness-of-fit.  For that reason, we compute the chi-square, $\chi^2_{\mathrm{bf},n}$, for each realisation $n$ using the best-fit model parameters and rescaling the covariance matrix back to the original volume of the simulation snapshots:
\begin{equation}
  \label{eq:pipe.chi2}
  \begin{split}
    \chi^2_{\mathrm{bf},n} = \frac{V_{\mathrm{eff}}^{\mathrm{sample}}}{V_{\mathrm{eff}}^{\mathrm{Euclid}}} \, \sum_{i,j=1}^{N_{\mathrm{bin}}^X} \left(\hat{X}_i^{(n)} - X_{\mathrm{bf},i}^{(n)}\right) \, C_{X,ij}^{-1} \\ \times \, \left(\hat{X}_j^{(n)} - X_{\mathrm{bf},j}^{(n)}\right)\,,
  \end{split}
\end{equation}
where $V_{\mathrm{eff}}^{\mathrm{sample}}$ stands for the effective volume of either the CMASS or LOWZ sample (see Sec.~\ref{sec:covariances}), and $\hat{X}_i^{(n)}$ is the measurement in bin $i$ of the $n$th realisation.  From this we determine the corresponding $p$-value using the degrees of freedom (dof) of the fit and average these values over all realisations, obtaining:
\begin{equation}
  \label{eq:pipe.pvalue}
  \bar{p}\mathrm{-value} \equiv 1 - \frac{1}{N_{\mathrm{sim}}} \sum_{n=1}^{N_{\mathrm{sim}}} \int_0^{\chi^2_{\mathrm{bf},n}} \mathrm{d}\chi^2 \, {\cal P}(\chi^2\,|\,\mathrm{dof})\,,
\end{equation}
where ${\cal P}$ denotes the chi-square distribution.  In other words, we quantify the probability that a given realisation was generated under the hypothesis that the best-fit model is the underlying truth and the assumption that the rescaled covariance matrix describes the statistical uncertainties in the measurement.  For the true model each $p$-value between 0 and 1 is equally likely, which means that the average ($\bar{p}$-value) will tend to 0.5.  Conversely, we consider $p$-values below the threshold of 0.05 as sufficient indication to reject the null-hypothesis and interpret these cases accordingly as a failure of the theory model.

\subsubsection{Figure of bias and figure of merit}
\label{sec:fobfom}

A complementary way to assess the performance of the model is the consistency of the recovered parameters with their true values, provided they are known.  This is the case for the three cosmological parameters that we vary in all of our Markov chains and so we define the figure of bias (FoB, see also \cite{EggScoCro2011}) as the difference between their posterior means, $\bar{\theta}_{\alpha}$, and fiducial values, $\theta_{\mathrm{fid},\alpha}$:
\begin{equation}
  \label{eq:pipe.fob}
  \mathrm{FoB} \equiv \left[\sum_{\alpha,\beta} \left(\bar{\theta}_{\alpha} - \theta_{\mathrm{fid},\alpha}\right) \, S^{-1}_{\alpha\beta} \, \left(\bar{\theta}_{\beta} - \theta_{\mathrm{fid},\beta}\right)\right]^{1/2}\,.
\end{equation}
The role of the parameter covariance matrix $S_{\alpha\beta}$, which we obtain from \texttt{getdist}, is to convert the parameter differences into levels of significance, such that the value $\mathrm{FoB} = 1.88(2.83)$ can be interpreted as the 68(95)\,\% confidence level in this three-dimensional parameter space.  Since statistical fluctuations in our fitted data vectors are suppressed compared to the adopted measurement uncertainties, the dominant contribution to the FoB metric can be expected to arise due to modelling systematics.  We therefore consider FoB values exceeding 1.88 as an indicator for the failure of the model (except in cases of significant prior volume projection effects, see Sec.~\ref{sec:results.pk.proj+prior}).

Finally, as a summary of the model's constraining power we compute the figure of merit (FoM):
\begin{equation}
  \label{eq:pipe.fom}
  \mathrm{FoM} \equiv \frac{1}{\sqrt{\mathrm{det} \, S_{\alpha\beta}}}\,,
\end{equation}
using only the three varied cosmological parameters, as for the FoB.

\subsubsection{Tension measure}
\label{sec:tension}

Inconsistencies (even if only affecting ``nuisance'' parameters) in the analysis of two different datasets, such as the power spectrum and bispectrum, can drive biases in the cosmological parameters when the two datasets are analysed together.  We therefore want to quantify the consistency of our power spectrum only and joint power spectrum and bispectrum Markov chains. This can serve as an additional metric to assess the performance of the models used in the analysis.

In order to estimate the tension between two sets of posteriors we follow the method proposed by \cite{RavZacHu2005}, which computes the probability of parameter differences and is implemented in the \texttt{tensiometer} code\footnote{\url{https://github.com/mraveri/tensiometer}}.  The central quantity in this method is the probability density function of parameter differences,
\begin{equation}
  \label{eq:pipe.Pdiff}
  {\cal P}_{\Delta}(\Delta \boldsymbol{\theta}) = \int_{V_p} \mathrm{d}\boldsymbol{\theta} \, {\cal P}(\boldsymbol{\theta}, \boldsymbol{\theta} - \Delta\boldsymbol{\theta})\,,
\end{equation}
given by an $N$-dimensional integral ($N$ being the dimension of the parameter space common to the two posteriors) of the joint posterior values over the prior volume $V_p$.  Once this function has been obtained, one can compute the probability of a parameter shift by integrating over the parameter space region where ${\cal P}_{\Delta}(\Delta\boldsymbol{\theta}) > {\cal P}_{\Delta}(0)$, giving:
\begin{equation}
  \label{eq:pipe.shiftProb}
  p_\Delta = \int_{{\cal P}_{\Delta}(\Delta\boldsymbol{\theta}) > {\cal P}_{\Delta}(0)} \mathrm{d}(\Delta\boldsymbol{\theta}) \, {\cal P}_{\Delta}(\Delta\boldsymbol{\theta})\,.
\end{equation}
We report this probability in terms of the number of standard deviations that an equivalent event would have had for a Gaussian distribution, i.e.
\begin{equation}
  \label{eq:pipe.shiftSigmas}
  n_{\sigma}^{\mathrm{eff}}(p_\Delta) \equiv \sqrt{2}\,\mathrm{Erf}^{-1}(p_\Delta)\,.
\end{equation}
In practice, \texttt{tensiometer} evaluates Eq.~(\ref{eq:pipe.Pdiff}) by computing the difference between parameter samples from the Markov chains corresponding to the two posteriors, assuming that the two sets of samples are uncorrelated.  This will not be entirely correct in our case, since the two posteriors share the same underlying power spectrum data, which is why the results should rather be interpreted as a lower limit (assuming that the samples are positively correlated).

\section{Approximations and validation}
\label{sec:validation}

An in-depth study of the performance of the theoretical model templates requires repeated execution of the analysis pipeline.  This process can be streamlined significantly by ensuring that the pipeline operates efficiently, which is why in our Python package \texttt{COMET} we approximate Alcock-Paczy\'nski (AP) distortions, IR resummation, and discreteness and binning effects for the bispectrum.  In the following, we provide a detailed description and validation of these approximations.

\subsection{Alcock-Paczy\'nski distortions}
\label{sec:APdist}

The analysis of galaxy clustering data begins by converting the observed redshifts and angular positions of the galaxies into comoving distance scales.  This conversion depends on an assumed cosmological background model and any differences between this so-called fiducial cosmology and the true cosmology lead to an anisotropic distortion of the clustering statistics, also known as Alcock-Paczy\'nski (AP) distortions \cite{AlcPac7910}.  They can be parametrised in terms of the distance scale ratios parallel and perpendicular to the LOS in the two cosmologies,
\begin{equation}
  q_{\parallel} \equiv \frac{H_{\mathrm{fid}}(z)}{H(z)}\,, \quad q_{\perp} \equiv \frac{D_{\mathrm{M}}(z)}{D_{\mathrm{M},\mathrm{fid}}(z)}\,,
\end{equation}
where $H$ and $D_{\mathrm{M}}$ denote the Hubble rate and comoving transverse distance.  When evaluating our model predictions, we need to take into account the AP distortions before projecting the clustering statistics into multipoles.  The observed power spectrum multipoles, i.e., the power spectrum multipoles expressed in the fiducial cosmology (indicated in the following by primed quantities), are given in terms of the true power spectrum evaluated at the distorted wave vector components as follows:
\begin{equation}
  \label{eq:pipe.PwAP}
  P'_{\ell}(k') = \frac{2\ell + 1}{2 q_{\perp}^2\,q_{\parallel}} \int_{-1}^1 \mathrm{d}\mu' \, {\cal L}_{\ell}(\mu') \, P\left[k(k',\mu'), \mu(k',\mu')\right]\,.
\end{equation}
The relations between the true and fiducial wave vector components can be written as \cite{BalPeaHea9610}
\begin{align}
  k(k',\mu') &= \frac{k'}{q_{\perp}} \left[1 + \mu'^2\left(F^{-2}-1\right)\right]^{1/2}\,,   \label{eq:pipe.kAP} \\
  \mu(k',\mu') &= \frac{\mu'}{F} \left[1 + \mu'^2\left(F^{-2}-1\right)\right]^{-1/2}\,,     \label{eq:pipe.muAP}
\end{align}
where $F \equiv q_{\parallel}/q_{\perp}$.  In \texttt{COMET}, the integration in Eq.~(\ref{eq:pipe.PwAP}) is implemented using Gauss-Legendre quadrature (see, however, Sec.~\ref{sec:binning} for the inclusion of discreteness effects), which can be performed very efficiently.

A similar expression also holds for the bispectrum multipoles in the fiducial cosmology and following the definition of the triangle orientation with respect to the LOS in Sec.~\ref{sec:measurements} one obtains
\begin{equation}
  \label{eq:pipe.BwAP}
  \begin{split}
    B'_{\ell}(k'_1,k'_2,k'_3) = \frac{2\ell + 1}{q_{\perp}^4\,q_{\parallel}^2} \int_{0}^{\pi} \frac{\mathrm{d}\theta'\,\sin{\theta'}}{2} \int_{0}^{2\pi} \frac{\mathrm{d}\phi'}{2\pi} \, {\cal L}_{\ell}(\cos{\theta'}) \\
    \times \, B(k_1,k_2,k_3,\mu_1,\mu_2,\mu_3)\,.
  \end{split}
\end{equation}
Each $k_i$ and $\mu_i \equiv \hat{\kv}_i \cdot \hat{\boldsymbol{z}}$ is distorted analogously to Eqs.~(\ref{eq:pipe.kAP}) and (\ref{eq:pipe.muAP}), and the relation between $\mu'_i$ and the angles $\theta'$ and $\phi'$ is given by
\begin{equation}
  \label{eq:pipe.mu123}
  \begin{split}
    &\mu'_1 = \cos{\theta'},\, \\
    &\mu'_2 = \mu'_1\,\mu'_{12} - \sqrt{1 - (\mu'_1)^2} \sqrt{1-(\mu'_{12})^2}\,\cos{\phi'}\,, \\
    &\mu'_3 = -\frac{k'_1\,\mu'_1 + k'_2\,\mu'_2}{k'_3}\,,
  \end{split}
\end{equation}
having defined $\mu'_{12} \equiv \hat{\kv}'_1 \cdot \hat{\kv}'_2$.  We can evaluate Eq.~(\ref{eq:pipe.BwAP}) with Gauss-Legendre quadrature, but compared to the power spectrum it is less computationally efficient, especially if it needs to be done for a large number of triangle configurations.  To retain a speedy computation of our likelihood we therefore Taylor expand the dependence on the AP parameters around $q_{\parallel} = 1 = q_{\perp}$ up to linear order, as previously also considered by the related works \cite{SugSaiBeu2101, KhoSam2311}.  This approximation is warrented since the cosmological parameters are already known to per-cent precision, such that the deviations of the AP parameters from unity are typically of a similar order.  Expanding Eq.~(\ref{eq:pipe.BwAP}) brings the advantage that the average over the solid angle can be precomputed for all individual terms in the bispectrum model with different powers of $\mu_1$, $\mu_2$, and $\mu_3$, so that the full model evaluation during the sampling procedure is extremely efficient.  For instance, let us consider a generic term that appears in the tree-level bispectrum model, i.e.,
\begin{equation}
  \begin{split}
  B_{n_1 n_2 n_3}(\kv_1,\kv_2,\kv_3) \equiv \; &\mu_1^{n_1}\,\mu_2^{n_2}\,\mu_3^{n_3} \, {\cal K}(k_1,k_2,k_3) \\ &\times\,\tilde{P}_{\rm lin,IR}(k_1)\,\tilde{P}_{\rm lin,IR}(k_2) + \, \mathrm{cyc.}\,, \label{eq:pipe.Bn1n2n3}
  \end{split}
\end{equation}
letting ${\cal K}$ stand for an arbitrary kernel function (for a decomposition of the full tree-level bispectrum into all such terms, see \cite{RizMorPar2301}), and $\tilde{P}_{\rm lin,IR}(k)$ is the infrared resummed linear power spectrum with fixed LOS dependence (see Sec.~\ref{sec:IRresum_approx}).  If we now evaluate this term at the fiducial wave vectors and Taylor expand up to linear order in $q_{\parallel}$ and $q_{\perp}$ we obtain
\begin{equation}
  \label{eq:pipe.Bn1n2n3APexp}
  \begin{split}
    B'_{n_1 n_2 n_3}&(\kv'_1,\kv'_2,\kv'_3) \approx \frac{1}{q_{\perp}^4\,q_{\parallel}^2}\Bigg\{ 1 + \left(1-q_{\parallel}\right) n_{123} \\ &\hspace{-1.5em} + \sum_{i=1}^3\Big[1 - q_{\perp} + \left(q_{\perp}-q_{\parallel}\right)\,\left(\mu'_i\right)^2\Big] \, \Big[\partial_{\ln{k'_i}} - n_i\Big] \Bigg\} \\
                    &\hspace{-1.5em}\times\,(\mu'_1)^{n_1}\,(\mu'_2)^{n_2}\,(\mu'_3)^{n_3} \, {\cal K}(k'_1,k'_2,k'_3) \\
    &\hspace{-1.5em}\times\, \tilde{P}_{\rm lin,IR}(k_1')\,\tilde{P}_{\rm lin,IR}(k'_2) + \, \mathrm{cyc.}\,,
  \end{split}
\end{equation}
where $n_{123} = n_1 + n_2 + n_3$.  The integrations over the solid angle can now be done analytically and stored as a function of the required configurations $\{k'_1,k'_2,k'_3\}$ before fitting the model to data.  In case of the \VDG model, the averages over the solid angle involve the damping function, which depends on $a_{\mathrm{vir}}$ and on cosmological parameters through $\sigma_v$, such that these averages can no longer be precomputed.  However, even in this case the Taylor expansion turns out to be beneficial as it suffices to perform a certain number of integrations for different combinations of $n_1$, $n_2$, and $n_3$, while the kernel functions ${\cal K}$ (and their derivatives) still only need to be computed once.

\begin{figure}
  \centering
  \includegraphics{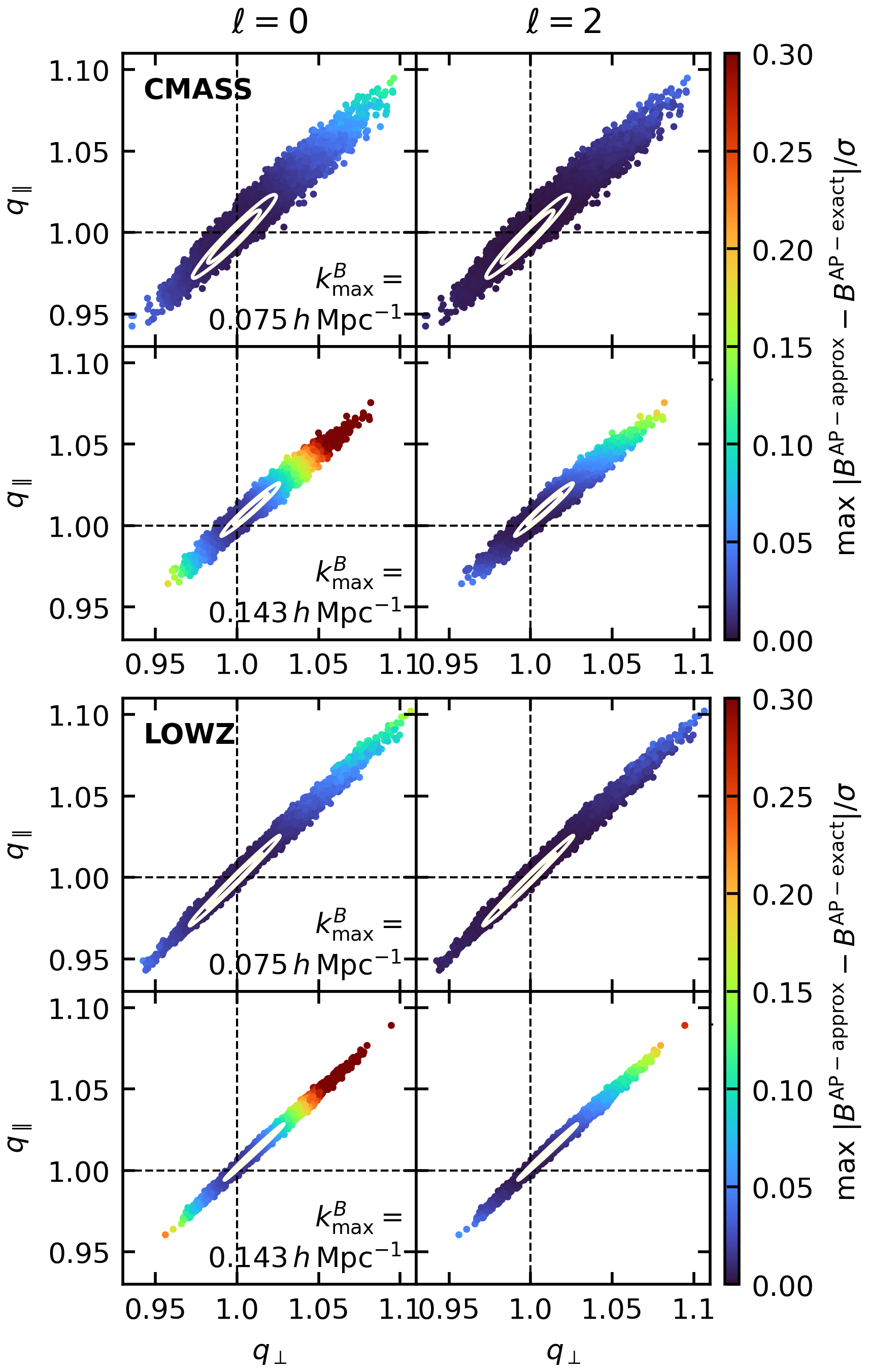}
  \caption{Demonstration of the accuracy of expanding the AP distortions in the bispectrum up to linear order in $q_{\parallel}$ and $q_{\perp}$.  Each plotted point is taken from a Markov chain originating from jointly analysing the power spectrum and bispectrum with different cutoff choices $k_{\mathrm{max}}^B$.  The colour of the points indicates the maximum of the difference (in units of the measurement uncertainties) between the exact and approximate AP treatment over all triangle configurations participating in the fits.  The left column depicts these for the monopole, the right column for the quadrupole.  White contours show the 1- and 2-$\sigma$ constraints on the AP parameters.}
  \label{fig:AP_approximation_test}
\end{figure}

In order to test the accuracy of the AP Taylor expansion, we jointly analyse the power spectrum and bispectrum for the CMASS and LOWZ data sets with two different cutoffs for the bispectrum data vector: 1) $k_{\mathrm{max}}^B = 0.075\,\hinvMpc$, and 2) $k_{\mathrm{max}}^B = 0.143\,\hinvMpc$, whereas the power spectrum cutoff in both cases is $k_{\mathrm{max}}^P = (0.2, 0.15, 0.15)\,\hinvMpc$. These two cases represent the two opposite extremes of those that we will explore in Sec.~\ref{sec:results}: the first case is less constraining on cosmological parameters and thus the AP parameters can deviate more strongly from unity, but it only involves large-scale modes, where the larger measurement uncertainties allow for less precise modelling; in the second case this situation is exactly the opposite.  After running the corresponding Markov chains, we evaluate the bispectrum models for the monopole and quadrupole for each sample in the chain with the exact inclusion of the AP parameters (Eq.~\ref{eq:pipe.BwAP}) and compare to the approximate one.  We compute the difference in units of the respective standard deviations and take the maximum over all involved triangle configurations in the fits, which we plot as a function of $q_{\parallel}$ and $q_{\perp}$ in Fig.~\ref{fig:AP_approximation_test}.  We see that, as expected, larger values of the AP parameters are explored by the chain for the lower $k_{\mathrm{max}}^B$ cutoff, but due to the smaller measurement uncertainties the significance of the differences is larger when smaller scale triangle configurations are included.  Moreover, we note that the significance of the differences in the qadrupole are smaller than for the monopole.  In total, we find that the $2$-$\sigma$ confidence regions (indicated by the white ellipses in the figure) lies in a region of the parameter space, where the differences do not surpass a maximum of $0.1\,\sigma_{\ell}$, from which we conclude that any inaccuracies of our AP treatment are negligible.

\subsection{Bispectrum infrared resummation approximation}
\label{sec:IRresum_approx}

\begin{figure}
  \centering
  \includegraphics{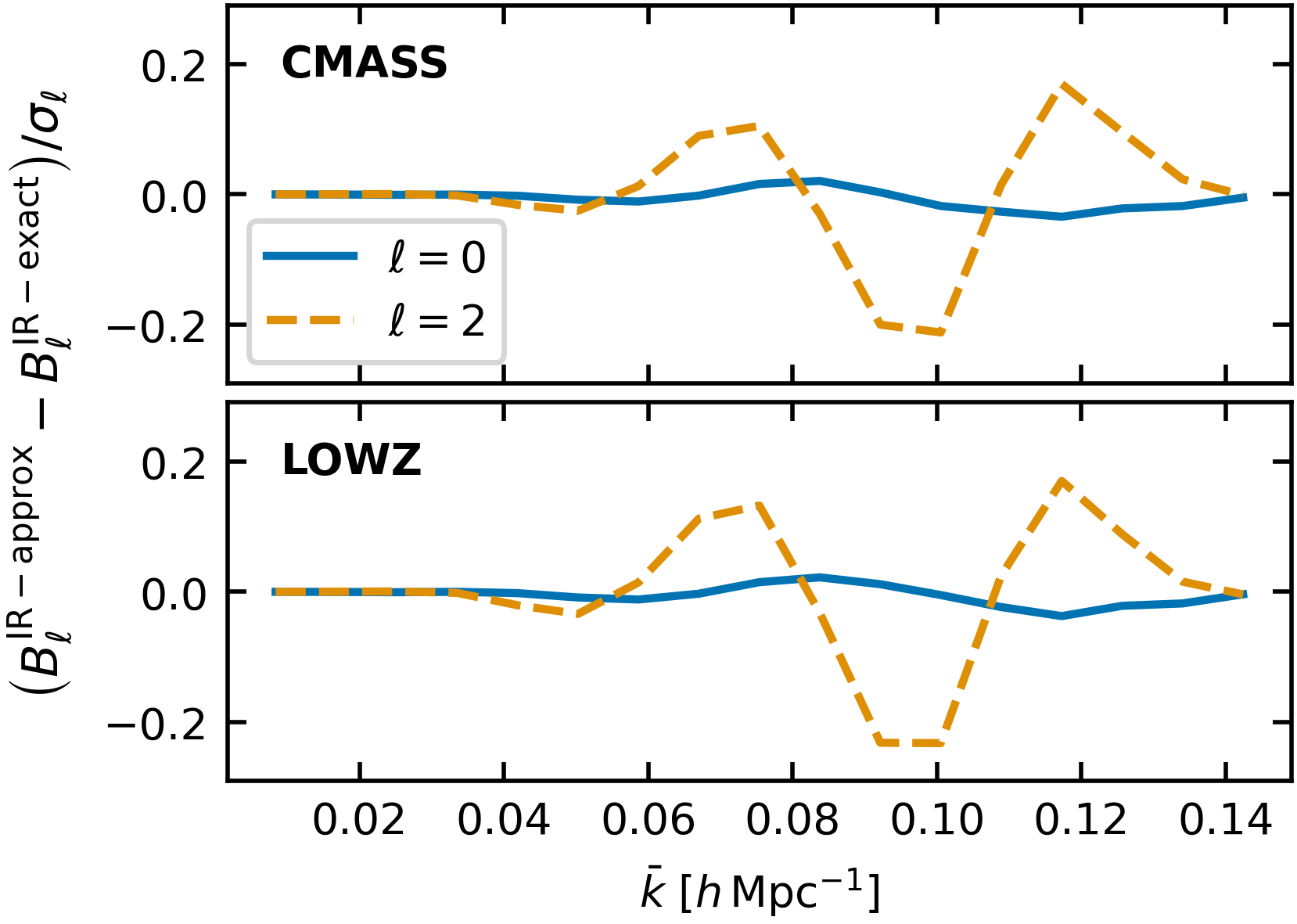}
  \caption{Difference between the approximate and exact treatment of infrared resummation in the bispectrum for the EFT monopole (solid) and quadrupole (dashed), in units of the respective measurement uncertainties.  Results are shown as a function of $\bar{k} = \sqrt{(k_1^2+k_2^2+k_3^2)/3}$ by rebinning the original triangle configurations.  Model parameters used for this comparison correspond to the best-fit values obtained by fitting the approximate EFT model with $k^P_{\mathrm{max}} = (0.2,0.15,0.15)\,\hinvMpc$ and $k^B_{\mathrm{max}} = 0.134\,\hinvMpc$.}
  \label{fig:IRresum_approximation_test}
\end{figure}

In order to compute the bispectrum model efficiently, it is useful to isolate and precompute quantities that do not depend on cosmological parameters.  This can be done for the various kernel functions involved in the bispectrum, and also, as already mentioned in the previous section, for the average over the solid angle when projecting the bispectrum into multipoles.  However, the latter requires an approximation since the infrared resummed linear power spectrum, $P_{\rm lin,IR}(\bk)$, which enters $B_{\rm tree}(\bk_1,\bk_2,\bk_3)$, depends on the LOS, leading to a cosmology dependence of the angular averages.  To circumvent this complication, we fix the LOS and compute the bispectrum with the following power spectrum:
\begin{equation}
  \label{eq:pipe.PIRapprox}
  \tilde{P}_{\rm lin,IR}(k) \equiv P_{\rm lin,IR}(k, \mu=0.6)\,.
\end{equation}
In Fig.~\ref{fig:IRresum_approximation_test} we demonstrate the impact of this approximation by comparing to the exact solution for the monopole and the quadrupole in the EFT model, using the best-fit parameters obtained by fitting the combination of the power spectrum and bispectrum with cutoffs $k^P_{\mathrm{max}} = (0.2,0.15,0.15)\,\hinvMpc$ and $k^B_{\mathrm{max}} = 0.134\,\hinvMpc$.  For a better visualisation of the scale-dependence of the differences we re-organise the respective model predictions from the original triangle configurations into 17 linearly spaced bins of $\bar{k} \equiv \sqrt{(k_1^2+k_2^2+k_3^2)/3}$ with binwidth $\Delta k = 2 k_{\mathrm{f}}$.  The plot shows that the differences are negligible: for the monopole they stay well below $0.1\,\sigma_{0}$, while for the quadrupole they follow a distinctive BAO feature as a consequence of the lacking LOS dependence, but reach at most $0.2\,\sigma_2$.  The value $\mu=0.6$ in Eq.~(\ref{eq:pipe.PIRapprox}) has been found by minimising the chi-square difference of the combined monopole and quadrupole, and is consistent between the two different galaxy samples.

\subsection{Binning and discreteness effects}
\label{sec:binning}

Comparison between the measured power spectrum and bispectrum multipoles and our theoretical models is complicated by the finite binwidth of the measurements, as well as the fact that the Fourier modes are not continuous, but multiples of the fundamental frequency of the overdensity grid.  Since these effects may bias the inference of cosmological parameters, it is important to adequately take them into account, meaning that the theoretical templates must be averaged in the same way as done in the estimators (Eqs.~\ref{eq:data.Pl} and \ref{eq:data.Bl}), which leads to:
\begin{align}
  P^{\mathrm{disc}}_{\ell}(k) &= \frac{2\ell + 1}{N_P(k)} \sum_{\qv \in k} {\cal L}_{\ell}(\hat{\qv} \cdot \hat{\boldsymbol{z}}) \, P(\qv)\,, \label{eq:pipe.Pl_disc}\\
  B^{\mathrm{disc}}_{\ell}(k_1,k_2,k_3) &= \frac{2\ell + 1}{N_B(k_1,k_2,k_3)} \sum_{\qv_1 \in k_1} \sum_{\qv_2 \in k_2} \sum_{\qv_3 \in k_3} \delta^{\mathrm{K}}_{\qv_{123},0} \nonumber \\ &\hspace{1em} \times\, {\cal L}_{\ell}(\hat{\qv}_1 \cdot \hat{\boldsymbol{z}}) \,  B(\qv_1,\qv_2,\qv_3)\,,  \label{eq:pipe.Bl_disc}
\end{align}
where we have left out the AP distortions to simplify the notation.  For an exact computation of Eq.~(\ref{eq:pipe.Pl_disc}) we make use of the technique described in \cite{EggCamPez2212}, which is implemented in \texttt{COMET} and can be performed nearly without any additional cost.

Evaluating Eq.~(\ref{eq:pipe.Bl_disc}) for the bispectrum is, however, much more challenging due to the large number of closed fundamental triangles $\{\qv_1,\qv_2,\qv_3\}$ that fall into a given $\{k_1,k_2,k_3\}$ bin.  For that reason, various approximations have been used in the past, most commonly the effective bin approximation \cite[e.g.][]{OddSefPor2003}, where the discrete average over fundamental triangles is replaced with an evaluation of the theoretical template at the effective triangle sides, $k_{\mathrm{eff},i}$, as defined in Eq.~(\ref{eq:data.keff}).  Other works, such as \cite{EggScoSmi1906,DADonLew2405} have improved on that by applying the continuous approximation and replacing the sums over closed triangles in Eq.~(\ref{eq:pipe.Bl_disc}) with integrations over spherical shells.  It was shown in \cite{OddRizSef2111,IvaPhiNis2203} that this approximation can still lead to substantial errors of up to $\sim 30\,\%$, which is why the two works propose different methods of dealing with the discreteness effect.  The former employs a Taylor expansion of the linear power spectra evaluated at the discrete modes $\qv_i$ around the effective modes, which allows them to factor out the cosmology-dependent terms from the bin average, such that the remaining pieces can be precomputed exactly.  The latter uses instead the continuous approximation, but corrects it with discreteness weights obtained by the ratio of the discrete and continuous averages evaluated at a certain fixed cosmology.

The first of these two methods has the drawback that a large number of terms (especially in redshift space) need to be precomputed, leading not only to an initially large computational overhead, but also when reconstructing the final bispectrum model from those terms.  The second method is equally subject to a computational bottleneck as it requires the evaluation of the continuous bin average at each step of a likelihood analysis.  In order to avoid these difficulties, we have implemented a combination of the two methods.  In particular, for a given generic term in the bispectrum model, $B_{n_1 n_2 n_3}(\kv_1,\kv_2,\kv_3)$ (see Eq.~\ref{eq:pipe.Bn1n2n3}), we approximate the bin average as follows:
\begin{align}
  \label{eq:pipe.Bn1n2n3_disc_approx}
  &B^{\mathrm{disc}}_{\ell, n_1 n_2 n_3}(k_1,k_2,k_3) \approx     \frac{\PLIR(k_{\mathrm{eff},1})\,\PLIR(k_{\mathrm{eff},2})}{\tilde{P}^{\mathrm{fid}}_{\rm lin,IR}(k_{\mathrm{eff},1})\,\tilde{P}^{\mathrm{fid}}_{\rm lin,IR}(k_{\mathrm{eff},2})} \nonumber \\
     &\times \frac{2\ell + 1}{N_B(k_1,k_2,k_3)} \, \sum_{\qv_1 \in k_1} \sum_{\qv_2 \in k_2} \sum_{\qv_3 \in k_3} \delta^{\mathrm{K}}_{\qv_{123},0} \, {\cal L}_{\ell}(\hat{\qv}_1 \cdot \hat{\boldsymbol{z}}) \nonumber \\ &\times \, \mu_1^{n_1} \, \mu_2^{n_2} \, \mu_3^{n_3} \, {\cal K}(q_1,q_2,q_3) \, \tilde{P}^{\mathrm{fid}}_{\rm lin,IR}(q_1)\,\tilde{P}^{\mathrm{fid}}_{\rm lin,IR}(q_2) \, + \, \mathrm{cyc.}\,,
\end{align}
where the exact discrete bin average is precomputed for power spectra in a fixed fiducial cosmology for each term $B_{n_1 n_2 n_3}$.  During the likelihood analysis these terms are then rescaled by the power spectrum ratio of the desired and fiducial cosmologies evaluated at the effective modes, which can be done very efficiently.  We can also easily account for AP distortions in this scheme, provided they are Taylor expanded as discussed in Sec.~\ref{sec:APdist} --- in that case, we include any derivatives of the linear power spectrum in the rescaling.

\begin{figure}
  \centering
  \includegraphics{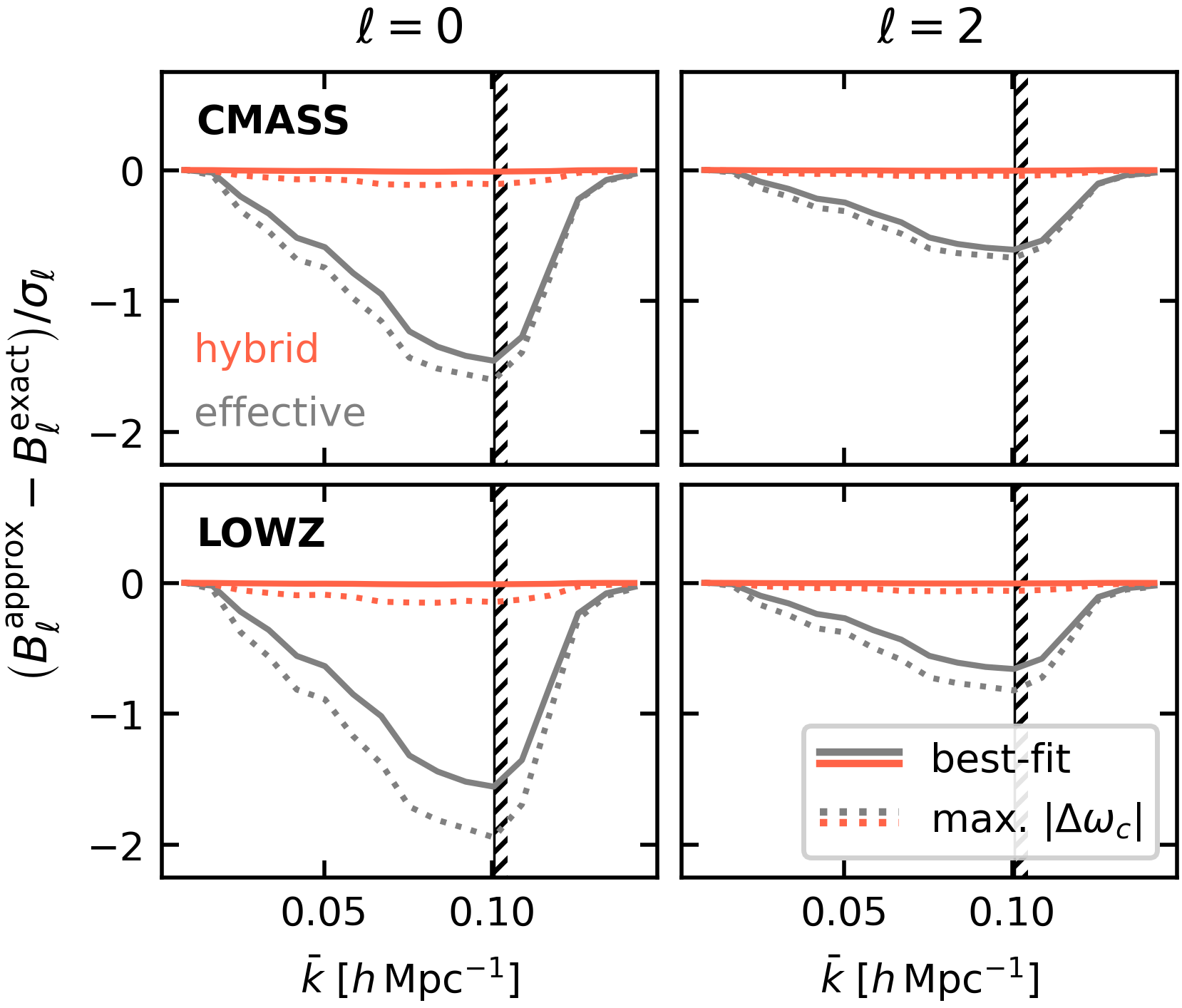}
  \caption{Impact of different approximations for binning and discreteness effects in the bispectrum monopole and quadrupole as a function of $\bar{k}$ (see Fig.~\ref{fig:IRresum_approximation_test}).  Differences between the effective (grey) or hybrid (red) approximations and exact results are evaluated for two different parameter combinations.  Solid lines: best-fit parameters from a Markov chain with $k^P_{\mathrm{max}} = (0.2,0.15,0.15)\,\hinvMpc$ and $k^B_{\mathrm{max}} = 0.143\,\hinvMpc$, dotted lines: parameter combination from the same chain that maximises the deviation of $\omega_c$ from its fiducial value.  Points to the right of the hatched region indicate non-completeness of the binning in $\bar{k}$, i.e., an increasing dominance of equilateral configurations.}
  \label{fig:Discrete_binning_test}
\end{figure}

Our approach is exact (up to negligible differences due to IR resummation) for the variation of parameters that only affect the amplitude of the linear power spectrum, while errors might occur when the shape of the power spectrum changes.  We quantify the size of such errors in Fig.~\ref{fig:Discrete_binning_test}, where we show the difference between either the effective bin approximation (grey colour) or our hybrid approximation from Eq.~(\ref{eq:pipe.Bn1n2n3_disc_approx}) with fiducial cosmology set to the Minerva cosmology (red colour) and the exact bin average.  For this comparison we took parameter combinations from a Markov chain obtained for the EFT model with $k^P_{\mathrm{max}} = (0.2,0.15,0.15)\,\hinvMpc$ and $k^B_{\mathrm{max}} = 0.143\,\hinvMpc$, using either the best-fit parameters (solid lines), or the chain sample that maximises the difference with respect to the fiducial value of $\omega_c$ (dotted lines).  To highlight the overall scale dependence we again rebin the differences in terms of $\bar{k}$, as explained in Sec.~\ref{sec:IRresum_approx}.

\begin{figure}
  \centering
  \includegraphics{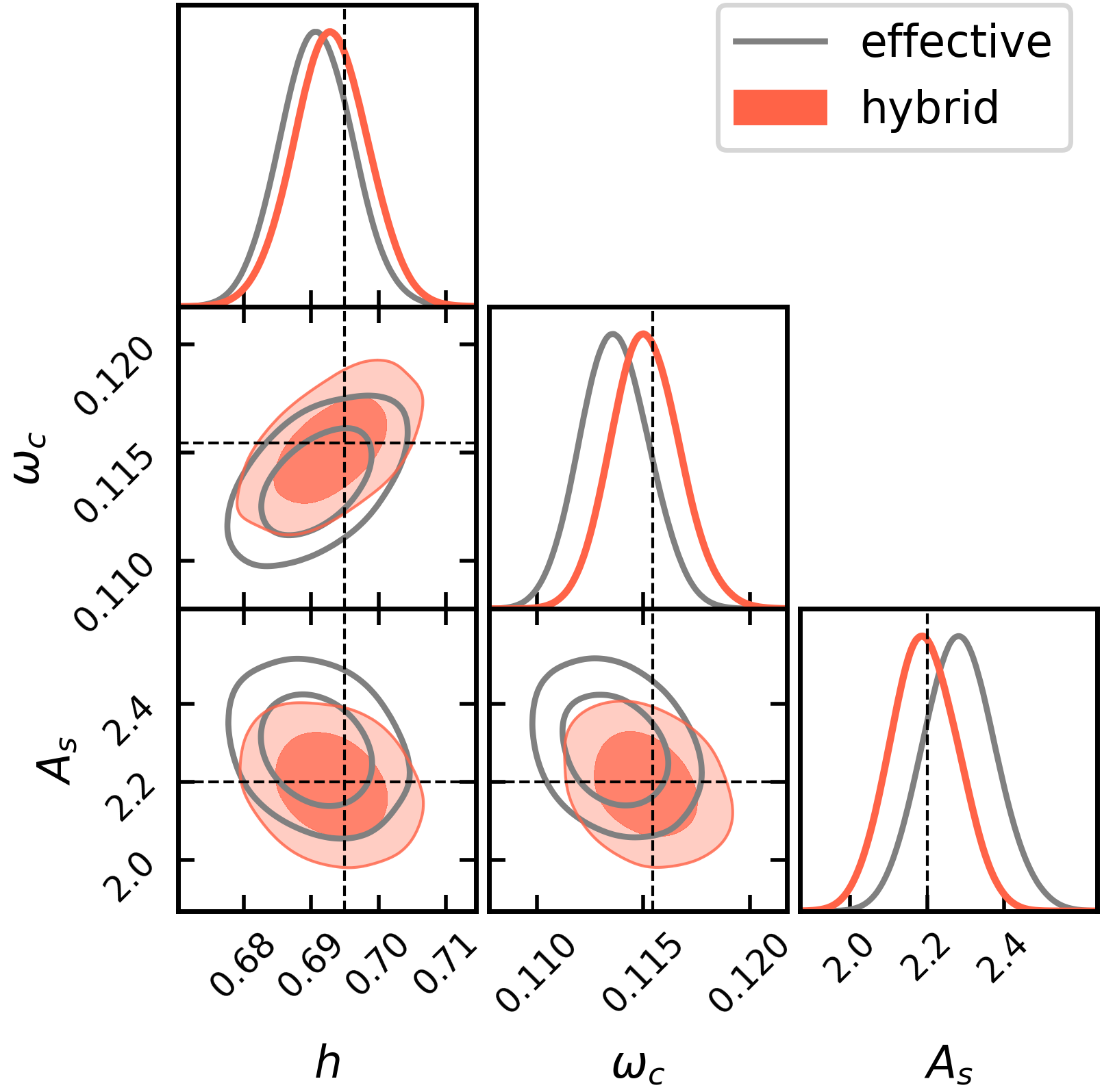}
  \caption{Shifts in cosmological constraints resulting from the effective and hybrid schemes for dealing with binning and discreteness effects in the bispectrum (see also Fig.~\ref{fig:Discrete_binning_test}). The results shown derive from an analysis of the CMASS power spectrum and bispectrum measurements with the \VDG model using $k_{\mathrm{max}}^P = (0.3,0.25,0.25)\,\hinvMpc$ and $k_{\mathrm{max}}^B = 0.143\,\hinvMpc$. Black dotted lines indicate the fiducial parameter values of the simulations.}
  \label{fig:Discrete_binning_posteriors_cmass}
\end{figure}

We see that simply evaluating the bispectrum model at the effective modes leads to a significant and systematic underestimation of the order $1.5$ - $2\,\sigma_0$ for the monopole, and $0.6$ - $0.8\,\sigma_2$ for the quadrupole, which we have checked can lead to shifts in the posterior means by up to $1\,\sigma$.  On the other hand, our hybrid approximation is very accurate: for the best-fit parameter combination the difference vanishes almost entirely as expected because the parameter values are close to the fiducial ones, but also in case of a large deviation from the fiducial $\omega_c$ (which alters the shape of the linear power spectrum) the differences are at most $0.1\,\sigma_0$ for the monopole and even less for the quadrupole.  Another notable feature of Fig.~\ref{fig:Discrete_binning_test} is that in all cases the differences quickly decrease beyond $k_{\mathrm{sph}} > 0.1\,\hinvMpc$.  However, this is a consequence of the binning scheme, since for the fixed mode cutoff, $k_1 \leq k_{\mathrm{max}}$, that we impose in our measurements, bins of $\bar{k}$ larger than a certain threshold (indicated by the hatched vertical line) do not include all possible triangle configurations and are increasingly dominated by equilateral configurations, for which the effective approximation is accurate.  Finally, it is important to note that our findings here strongly depend on the fundamental frequency of the measurement grid and the binwidth, so they need to be verified for any given case at hand.

\begin{figure*}
  \centering
  \includegraphics{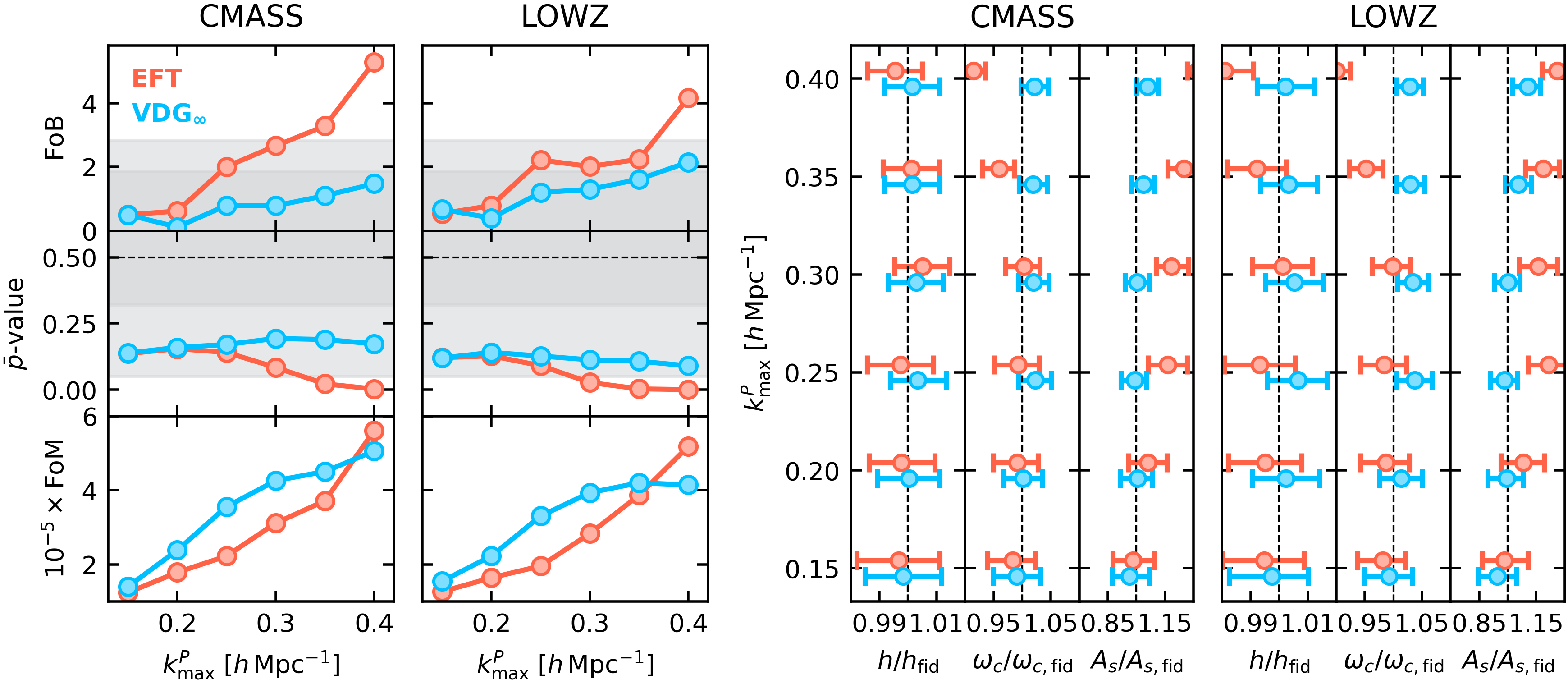}
  \caption{Left: performance metrics derived from MCMC analyses of the power spectrum multipoles using the EFT and \VDG model for different scale cuts; the same scale cuts were applied to all three multipoles.  Grey shaded areas indicate the critical values corresponding to the 68 and 95\,\% confidence levels.  Right: posterior means and 68\,\% credible intervals for the three fitted cosmological parameters (data points are slightly displaced vertically for better readability).  Dashed vertical lines indicate the fiducial parameters of the \textsc{Minerva} suite.}
  \label{fig:pk_metrics_diff0p0}
\end{figure*}

One further complication arises for the \VDG model, where the various terms $B_{n_1 n_2 n_3}$ are multiplied by the damping function $W^B_{\infty}$, which strongly depends on the LOS orientation.  That means it cannot be pulled out of the bin averages in Eq.~(\ref{eq:pipe.Bl_disc}) without introducing a significant error.  The averages would thus have to be recomputed at each step when sampling the likelihood, since $W^B_{\infty}$ depends on cosmological parameters (through $\sigma_v$) and $a_{\mathrm{vir}}$.  In order to avoid the resulting computational bottleneck we expand the damping function as
\begin{equation}
  \label{eq:pipe.WB_expansion}
  \begin{split}
    W^B_{\infty}(k_1,k_2,k_3,&\mu_1,\mu_2,\mu_3\,|\,\sigma_v,a_{\mathrm{vir}}) \\ & \approx 1 - \beta(\sigma_v,a_{\mathrm{vir}}) \, \sum_{i=1}^3 k_i^p \, \mu_i^2\,,
  \end{split}
\end{equation}
such that its effect can be absorbed in a few additional terms of the form $B_{n_1 n_2 n_3}$, to which we can apply the discrete binning operation in Eq.~(\ref{eq:pipe.Bn1n2n3_disc_approx}).  We find that setting $p = 7/4$ provides an accurate approximation for the VDG damping function on scales of interest to our analysis (more so than the formal Taylor expansion for small $k$, which would yield $p = 2$) and we determine the coefficient $\beta$ by fitting the bispectrum monopole and quadrupole of the exact \VDG model to the approximate one using Eq.~(\ref{eq:pipe.WB_expansion}) across a dense, regular grid of $\sigma_v$ and $a_{\mathrm{vir}}$, spanning values from 2\,Mpc to 10\,Mpc, and 0.01\,$\mathrm{Mpc}^2$ to 10\,$\mathrm{Mpc}^2$, respectively\footnote{More precisely, we parametrise $\beta(\sigma_v,a_{\mathrm{vir}})$ as $\tilde{\beta}(\sigma_v,a_{\mathrm{vir}})\,(a_{\mathrm{vir}}^p + \sigma_v^p/2)$ and fit for $\tilde{\beta}$, which varies more slowly as a function of $\sigma_v$ and $a_{\mathrm{vir}}$.}.  From these fits we construct a linear interpolator that provides $\beta$ as a function of $\sigma_v$ and $a_{\mathrm{vir}}$.  Comparing the exact and approximate models for all samples in a chain run with $k_{\mathrm{max}}^P = (0.2,0.15,0.15)\,\hinvMpc$ and $k_{\mathrm{max}}^B = 0.143\,\hinvMpc$ we find the maximum difference (across all triangle bins) for the monopole and quadrupole to be $0.1\,\sigma$ and $0.24\,\sigma$, respectively, and therefore do not expect this approximation to have any significant impact on our analysis.

Finally, in Fig.~\ref{fig:Discrete_binning_posteriors_cmass} we demonstrate the difference in cosmological parameter posteriors obtained from using either the effective or hybrid schemes.  As an example, we show the results for the CMASS power spectrum and bispectrum measurements, fitted with the \VDG model using the cutoff scales $k_{\mathrm{max}}^P = (0.3,0.25,0.25)\,\hinvMpc$ and $k_{\mathrm{max}}^B = 0.143\,\hinvMpc$.  We see that the underestimation of the bispectrum observed in Fig.~\ref{fig:Discrete_binning_test} for the effective binning approximation is compensated by shifts of $\sim 1\,\sigma$ for both, $\omega_c$ and $A_s$, while $h$ is mostly unaffected.  These shifts result in disagreements with the fiducial parameter values that disappear when employing the hybrid scheme.  By further comparing the significance of the differences between the effective and hybrid approximations with the exact computation (see Fig.~\ref{fig:Discrete_binning_test}) it is evident that any additional impact on the posteriors from the hybrid scheme will be negligible.


\vspace*{-0.5em}
\section{Results}
\label{sec:results}

Based on the setup described in Sec.~\ref{sec:pipeline}, we analyse the mock CMASS and LOWZ data and compare the performance of the EFT and \VDG models.  We first report the results for the power spectrum alone, before considering joint fits together with the bispectrum.

\vspace*{-0.5em}
\subsection{Power spectrum analysis}
\label{sec:results.pk}

\subsubsection{Performance metrics as a function of $\kmax$}
\label{sec:results.pk.nominal}

To begin with, we apply the same scale cut to the monopole, quadrupole, and hexadecapole of the power spectrum and fit the resulting datasets for six different values of $\kmax^P$, ranging from $0.15$ to $0.4\,\hinvMpc$.  The two left-hand panels of Fig.~\ref{fig:pk_metrics_diff0p0} display the FoB, $\bar{p}$-value, and FoM for the two different galaxy samples and both models.  As $\kmax^P$ increases, nonlinear effects become more significant and measurements uncertainties decrease, making it more challenging to accurately model the data.  This is reflected by the rising FoB values for both models, although their behaviour is very different: the FoB of the \VDG model remains below the 68\,\% critical value (indicated by the grey band) across all scales, except for $\kmax^P = 0.4\,\hinvMpc$ in the case of the LOWZ sample.  In contrast, the EFT model exceeds this threshold already starting from $\kmax^P = 0.25\,\hinvMpc$ for both samples, reaching FoB values more than twice as large as for the \VDG model at $\kmax^P = 0.4\,\hinvMpc$.  The goodness-of-fit criterion also signals an earlier breakdown of the EFT model: initially consistent with the VDG$_{\infty}$, the $\bar{p}$-values begin to diverge at $\kmax^P = 0.25\,\hinvMpc$, from which point the $\bar{p}$-value of the \VDG model remains approximately constant, whereas the EFT falls below the 0.05 threshold.

\begin{figure*}
  \centering
  \includegraphics{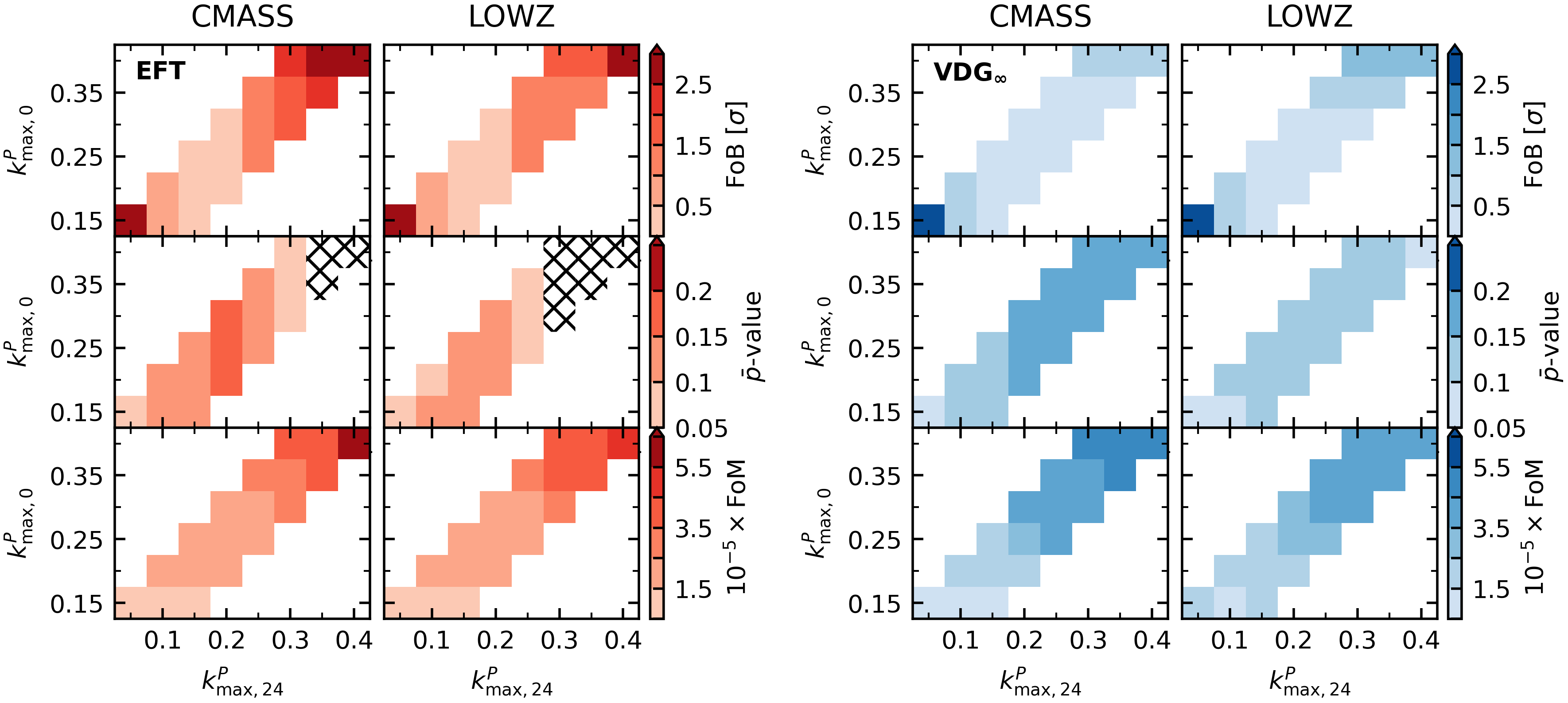}
  \caption{Performance metrics of the EFT and \VDG models, plotted for different values of the monopole scale cutoff, $k_{\mathrm{max},0}^P$, and the quadrupole and hexadecapole scale cutoff, $k_{\mathrm{max},24}^P$ (all cutoff values are given in units of $h\,\mathrm{Mpc}^{-1}$).  The FoB values are shown here in units of critical values corresponding to different confidence levels.  The hatched cells indicate $\bar{p}$-values lower than 0.05.}
  \label{fig:pk_metrics_diffvar}
\end{figure*}

The large FoB values indicate increasing deviations between the posterior means and the fiducial cosmological parameters.  These deviations are more clearly illustrated in the two right-hand panels of Fig.~\ref{fig:pk_metrics_diff0p0}, where the fully marginalised constraints on the individual parameters are plotted.  In the EFT model, the parameter most affected is $A_s$, which is overestimated for scale cuts beyond $\kmax^P = 0.2\,\hinvMpc$.  At even larger scale cuts ($\kmax^P > 0.3\,\hinvMpc$) the values for $\omega_c$ also become biased, while $h$ remains mostly unaffected.  In the \VDG model, on the other hand, all parameters are recovered roughly within the 68\,\% credible intervals, only $\omega_c$ consistently differs by about $1\sigma$ for $\kmax^P > 0.2\,\hinvMpc$ in case of the LOWZ sample.

The constraining power (FoM) of the two models diverges with increasing $\kmax^P$, despite both sampling the same number of parameters (the \VDG model varies $a_{\rm vir}$ instead of $c_{\rm nlo}$ in the EFT model, see Table~\ref{tab:priors}).  The \VDG model consistently delivers tigther constraints for both galaxy samples, except at $\kmax^P = 0.4\,\hinvMpc$, where the EFT has, however, already failed.  Notably, when comparing the FoM obtained at the maximum scales cuts where the FoB is still below the 68\,\% critical value, the \VDG model demonstrates an improvement by a factor 2 or more. 

A maximum scale cut of $\kmax^P \sim 0.2\,\hinvMpc$ for the EFT model is consistent with several previous studies in the literature \cite[e.g.,][]{dAGleKok2005,NisDAIva2012,ChuDolIva2101,MauLaiNor2406}.  However, our results appear to be in conflict with the recent work \cite{ChuIva2302}, which concludes that the EFT outperforms a model based on the same damping function that we have applied here.  We note that there are nevertheless a number of important differences that likely affect their conclusions: 1) the damping function in \cite{ChuIva2302} is applied to the full SPT power spectrum, missing the correction term in Eq.~(\ref{eq:DeltaP}), which leads to an incorrect double counting of velocity dispersion terms; 2) no counterterms are included, such that potentially relevant stress-tensor corrections are ignored; 3) the power spectrum on BAO scales is not accurately modelled, as no IR resummation is applied.

\subsubsection{Different multipole scale cut choices}
\label{sec:results.pk.varcuts}

The quadrupole and hexadecapole moments of the power spectrum are more sensitive to RSD than the monopole, making them potentially more vulnerable to inaccuracies in the RSD mapping on larger scales.  We therefore want to explore whether the analysis of the previous section can be enhanced by applying different scale cuts to the various multipoles.

Specifically, we retain identical cutoffs for the quadrupole and hexadecapole\footnote{We have checked, however, that the hexadecapole has little overall impact (due to its small signal-to-noise), so changing its cutoff independently from the quadrupole does not change the conclusions of this section.} (denoted by $k_{\mathrm{max},24}^P$), but consider constant differences, $\Delta k_{\mathrm{max}}$, with respect to the monopole cutoff $k_{\mathrm{max},0}^P$, such that $k_{\mathrm{max},24}^P = k_{\mathrm{max},0}^P - \Delta k_{\mathrm{max}}$.  We allow for three different values, $\Delta k_{\mathrm{max}} = 0.0$, $0.05$, $0.1\,\hinvMpc$, and use the same range of cutoffs for the monopole as before.  The resulting performance metrics for all combinations of $k_{\mathrm{max},0}^P$ and $k_{\mathrm{max},24}^P$ are shown as heatmaps in Fig.~\ref{fig:pk_metrics_diffvar}, where the hatched cells indicate $\bar{p}$-values lower than the threshold of 0.05.

Most strikingly, we observe that both the FoB and the $\bar{p}$-value of the EFT model are primarily functions of $k_{\mathrm{max},24}^P$, with only a minor dependence on $k_{\mathrm{max},0}^P$.  In particular, for both galaxy samples, $k_{\mathrm{max},24}^P = 0.2\,\hinvMpc$ represents a critical threshold, beyond which the FoB indicates biases in the cosmological parameters exceeding $1\sigma$, and the $\bar{p}$-value begins to decline.  This suggests that the breakdown of the EFT model is more likely due to inadequate modelling of the RSD mapping rather than inaccuracies in the treatment of the non-linear evolution of the matter and velocity fields, or galaxy bias.  If the latter were the case, one would also expect a stronger impact on the monopole.  In contrast, the \VDG model shows no dependence of the FoB and $\bar{p}$-value on $k_{\mathrm{max},24}^P$ (for the behaviour at small $k_{\mathrm{max},0}^P$, see the discussion in Sec.~\ref{sec:results.pk.proj+prior}), demonstrating the robustness of the non-perturbative VDG function for the galaxy samples considered here. 

The constraining power generally increases both as a function of $k_{\mathrm{max},0}^P$ and $k_{\mathrm{max},24}^P$.  The EFT model therefore benefits from choosing a larger $k_{\mathrm{max},0}^P$, while keeping $k_{\mathrm{max},24}^P = 0.2\,\hinvMpc$ fixed, such that the FoB remains below the $1\sigma$ threshold.  However, the resulting improvement in FoM is at most $16\,\%$ and thus still modest compared to the overall FoM achieved by the \VDG model.

\subsubsection{Projection and prior effects}
\label{sec:results.pk.proj+prior}

Recent studies \cite{CarMorPou2301,SimZhaPou2306,HolHerSim2309} have demonstrated that cosmological constraints derived from current galaxy clustering measurements can be subject to so-called \emph{prior volume projection effects}.  These effects occur when marginalising over high-dimensional, non-Gaussian posteriors---i.e., when projecting into a lower-dimensional subspace of parameters---and can lead to substantial shifts in the marginalised posterior means.  Moreover, these shifts have been found to depend sensitively on the size of the priors.  Strongly non-Gaussian posteriors often result from data that is not sufficiently constraining, making projection effects potentially irrelevant at the statistical precision of upcoming data sets, such as those assumed in this work.  Nevertheless, we aim to verify that our results from the previous sections are robust against such effects.

\begin{figure}
  \centering
  \includegraphics{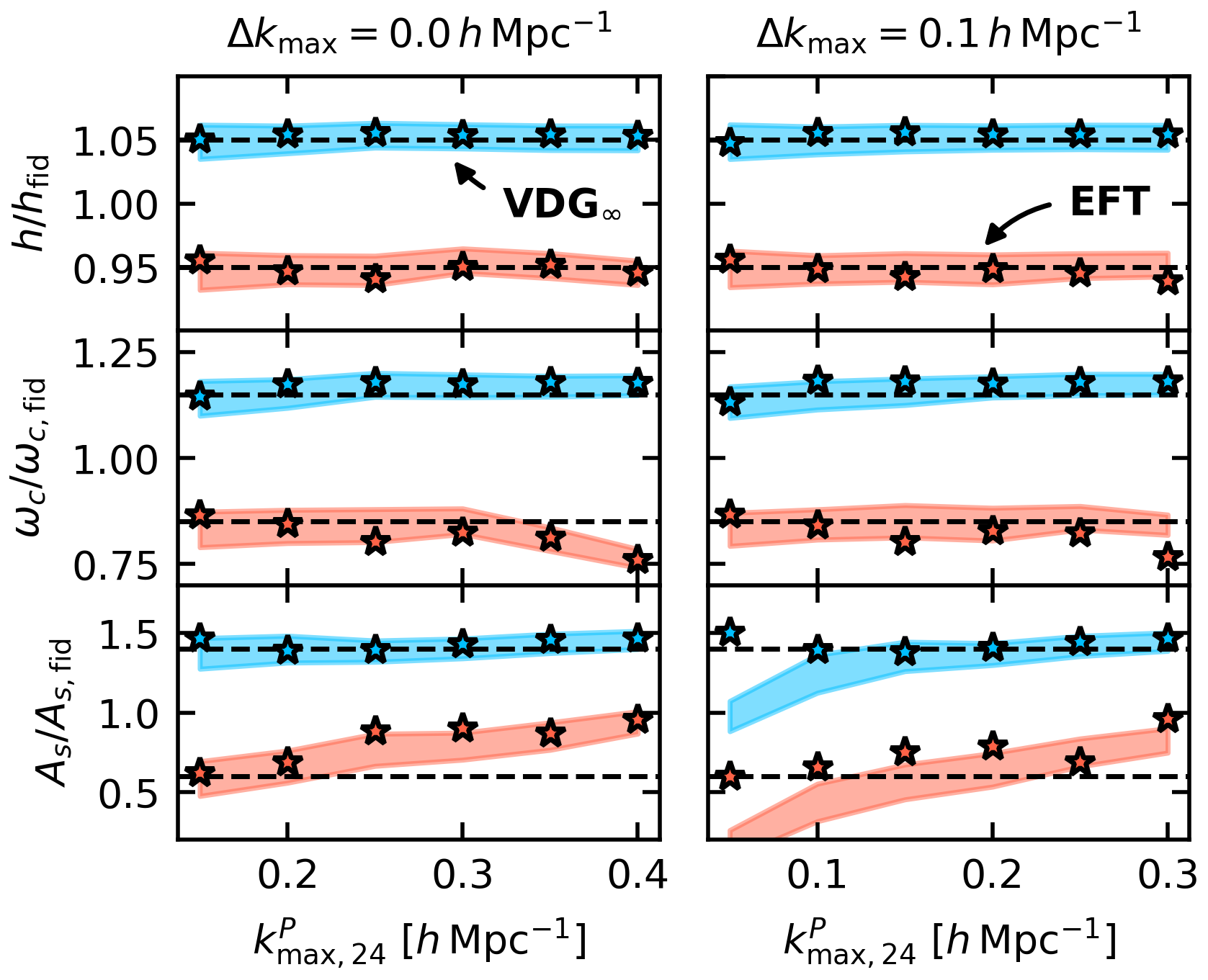}
  \caption{MAP cosmological parameters (stars) vs. 68\,\% credible intervals (error bands) for the CMASS sample.  Constraints are shown as a function of the quadrupole and hexadecapole scale cut for two constant differences with respect to the monopole, $\Delta k_{\mathrm{max}} = k_{\mathrm{max},0}^P - k_{\mathrm{max},24}^P$.  For better readability the constraints have been displaced by a small amount along the $y$-axis; fiducial values are given in each case by the dashed horizontal lines.}
  \label{fig:mean_vs_bestfit}
\end{figure}

A strong indicator for projection effects are significant differences between the marginalised posterior means and the maximum a posteriori (MAP) parameter values.  We therefore determine the MAP parameters in an independent procedure using the minimisation function given in the \texttt{Python} \texttt{Scipy} package (scipy.optimize.minimize), which is called with the Sequential Least Squares Programming (SLSQP) method, for each of our Markov chains.  In Fig.~\ref{fig:mean_vs_bestfit}, we compare these values (stars) with the marginalised 68\,\% credible intervals (shaded error bands) for the three cosmological parameters extracted from the CMASS sample, considering both models (EFT in red, \VDG in blue) and two different values for $\Delta k_{\mathrm{max}}$, the difference between the monopole and quadrupole/hexadecapole cutoffs.

\begin{figure}
  \centering
  \includegraphics{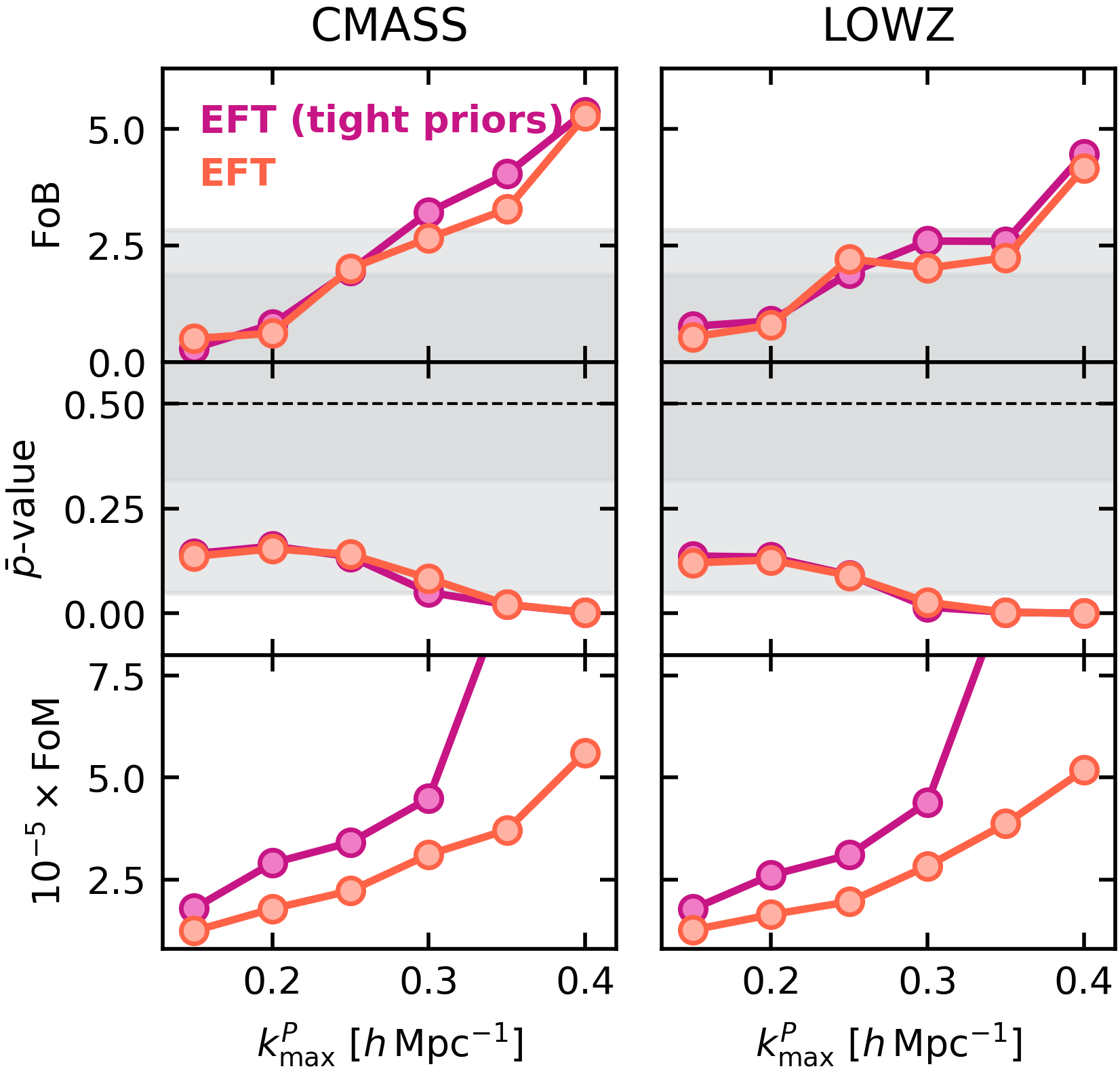}
  \caption{Comparison of performance metrics for the EFT model using two different sets of priors: wide uniform priors (see Table~\ref{tab:priors}; red), priors used in \cite{PhiIva2202} (purple).}
  \label{fig:performance_metrics_Pk_diff0p0_varpriors}
\end{figure}

For most scale cuts, the MAP parameters are in excellent agreement with the 68\,\% credible intervals.  This is especially true for the high $\kmax^P$ values, where the EFT model becomes biased, demonstrating that this bias is not due to projection effects.  Only for $\Delta k_{\mathrm{max}} = 0.1\,\hinvMpc$ and the lowest $k_{\mathrm{max},24}^P$ values, we observe a discrepancy between the MAP estimates and the marginalised posteriors, i.e., the same scale cut combinations that give rise to a larger FoB in Fig.~\ref{fig:pk_metrics_diffvar}.  In these cases, the MAP values are close to the fiducial parameter values, suggesting that the marginalised posteriors appear biased because of projection.  This is to be expected, as the constraining power of the data is significantly reduced by the scale cuts.  Although not shown in Fig.~\ref{fig:mean_vs_bestfit}, we obtain qualitatively identical results also for the LOWZ sample.  We caution that the absence of projection effects for the majority of scale cuts studied here does not easily generalise: extended cosmological parameter spaces that involve, for instance, the scalar spectral index, or dark energy equation of state parameters, may induce strong degeneracies that can in turn lead to projection effects \cite{CarMorPou2301}.  Hence, their impact must be studied for any given case at hand.

Finally, we check whether any prior choices influence our results.  To do so, we apply the same narrow Gaussian prior distributions for bias, stochasticity and counterterm parameters as in \cite{PhiIva2202}, after transforming our basis into the one adopted by \texttt{Class-PT} \cite{ChuIvaPhi2009} using the relations given in Appendix~\ref{sec:app.bias_bases}.  We then repeat the analysis for the EFT model and compare the performance metrics with our baseline analysis that instead applied wide uniform priors (see Tab.~\ref{tab:priors}) in Fig.~\ref{fig:performance_metrics_Pk_diff0p0_varpriors}.  Notably, the performance of the model is barely affected by the choice of priors, leading to the same breakdown at cutoff scales $\kmax^P \approx 0.25\,\hinvMpc$ and a fully equivalent goodness-of-fit.  This implies that for our final Stage-IV-like measurement uncertainties, the constraining power of the data clearly supersedes the prior information\footnote{Indeed, we find that the 68\,\% credible intervals are up to a factor of a few smaller than the standard deviations of the priors.}, which is consistent with our previous conclusion that projection effects are irrelevant in this case.  However, the tighter priors on ``nuisance'' parameters give rise to a moderate increase of the FoM by up to $\sim 50\,\%$ on scales where the model is still valid.

\subsubsection{Comparison with phenomenological damping functions}
\label{sec:results.pk.VDGdamping}

The VDG is the key ingredient in the \VDG model, as it describes how non-linearities in the RSD mapping affect the clustering statistics in the large scale limit.  While the parametrisation used in the baseline \VDG model is motivated by resumming quadratic non-linearities in the infinite separation limit (see Sec.~\ref{sec:theory.VDG}), there are various prescriptions in the literature that phenomenologically reproduce the damping effect, most commonly based on either a Gaussian or Lorentzian ansatz (Eq.~\ref{eq:theory.VDG.Winfty_G} and \ref{eq:theory.VDG.Winfty_Lor}).

\begin{figure}
  \centering
  \includegraphics{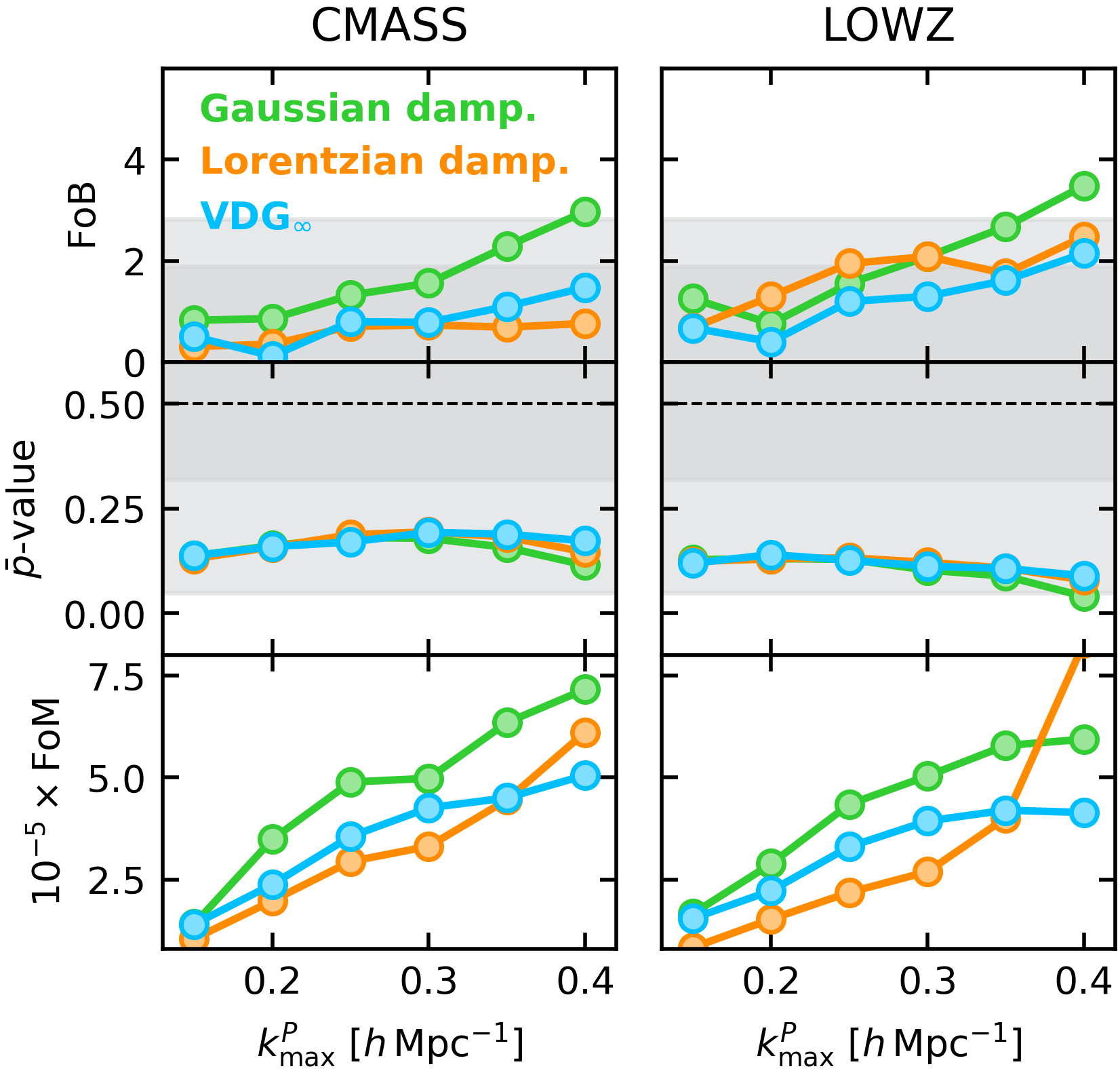}
  \caption{Comparison of performance metrics using different prescriptions for the VDG: baseline (Eq.~\ref{Winfty}, blue), a Gaussian damping function (Eq.~\ref{eq:theory.VDG.Winfty_G}, green), and Lorentzian damping function (Eq.~\ref{eq:theory.VDG.Winfty_Lor}, orange).}
  \label{fig:performance_metrics_Pk_diff0p0_vardamping}
\end{figure}

We analyse the data with such Gaussian and Lorentzian damping functions, treating the velocity dispersion, $\sigma_v$, as a free parameter, which is fitted using a uniform prior with bounds $[0,\,100]\,\Mpc$.  Since these models do not depend on $a_{\rm vir}$, the total number of fitting parameters remains the same as in our baseline.  We compare their respective performance metrics in Fig.~\ref{fig:performance_metrics_Pk_diff0p0_vardamping}, which demonstrates that the Gaussian damping function generally results in the highest FoB values, and we find that these are caused by a consistent overestimation of $\omega_c$, starting from $\kmax^P = 0.25\,\hinvMpc$.  On the other hand, this model is the most constraining, such that the maximum attainable FoM before model breakdown (based on the FoB crossing the $1\sigma$ threshold) is comparable to what we obtain with our baseline prescription.  Finally, application of the Lorentzian damping function shows some improvement over the baseline for the CMASS sample but is less robust for the LOWZ sample.

Compared to the analysis presented in \cite{ChuIvaSim2102}, the performance of the model with Gaussian damping function is significantly better.  As already noted at the end of Sec.~\ref{sec:results.pk.nominal}, this can be attributed to a number of differences in their modelling.

\subsubsection{Constraints on counterterms for different models}
\label{sec:results.pk.counterterms}

\begin{figure}
  \centering
  \includegraphics{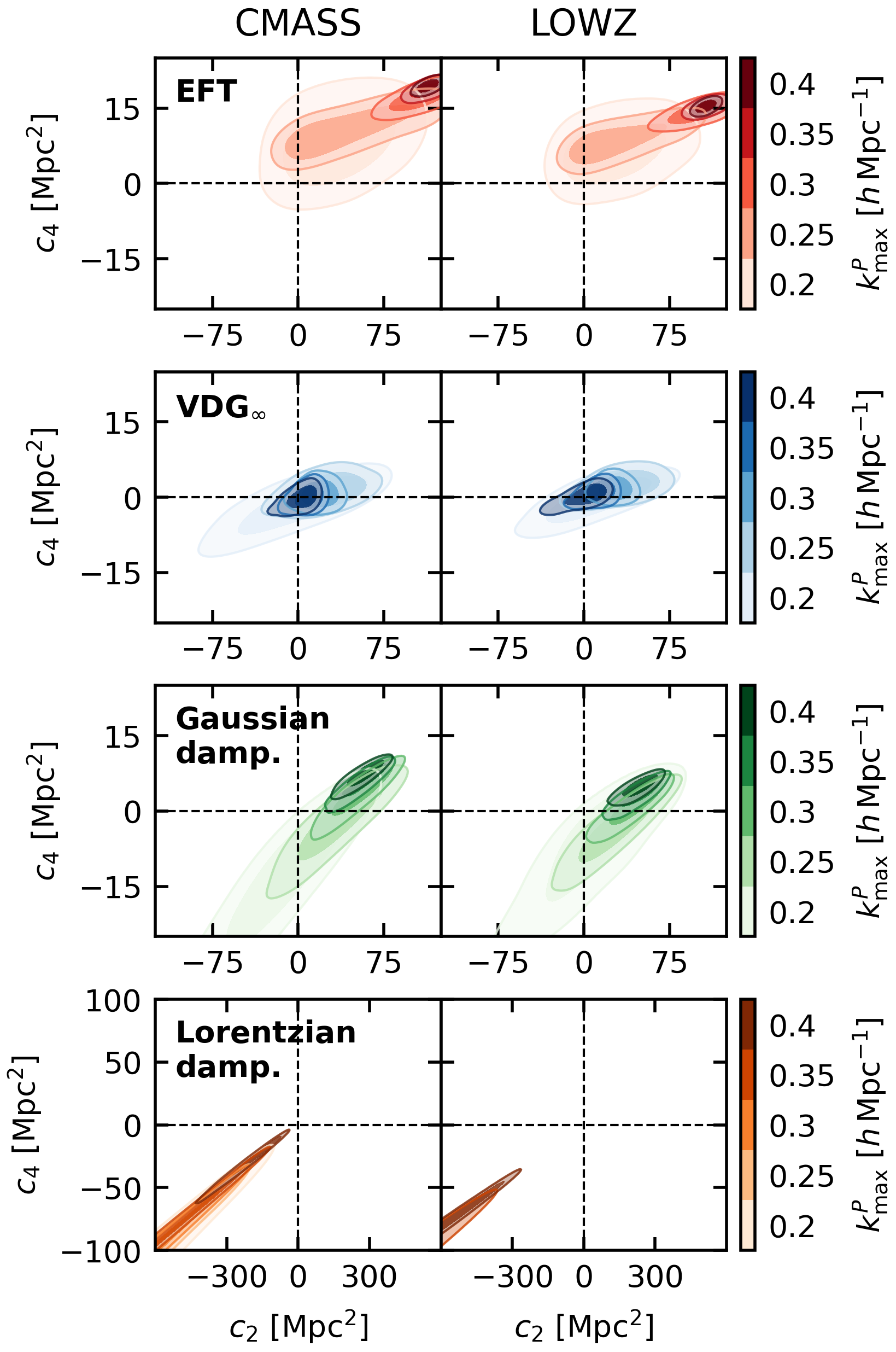}
  \caption{Constraints on the redshift-space counterterms for four different modelling options as a function of scale cut (same for all three power spectrum multipoles).  Note the different plotting ranges for the model with Lorentzian damping function.}
  \label{fig:counterterm_constraints_varkmax}
\end{figure}

\begin{figure*}[t]
  \centering
  \includegraphics{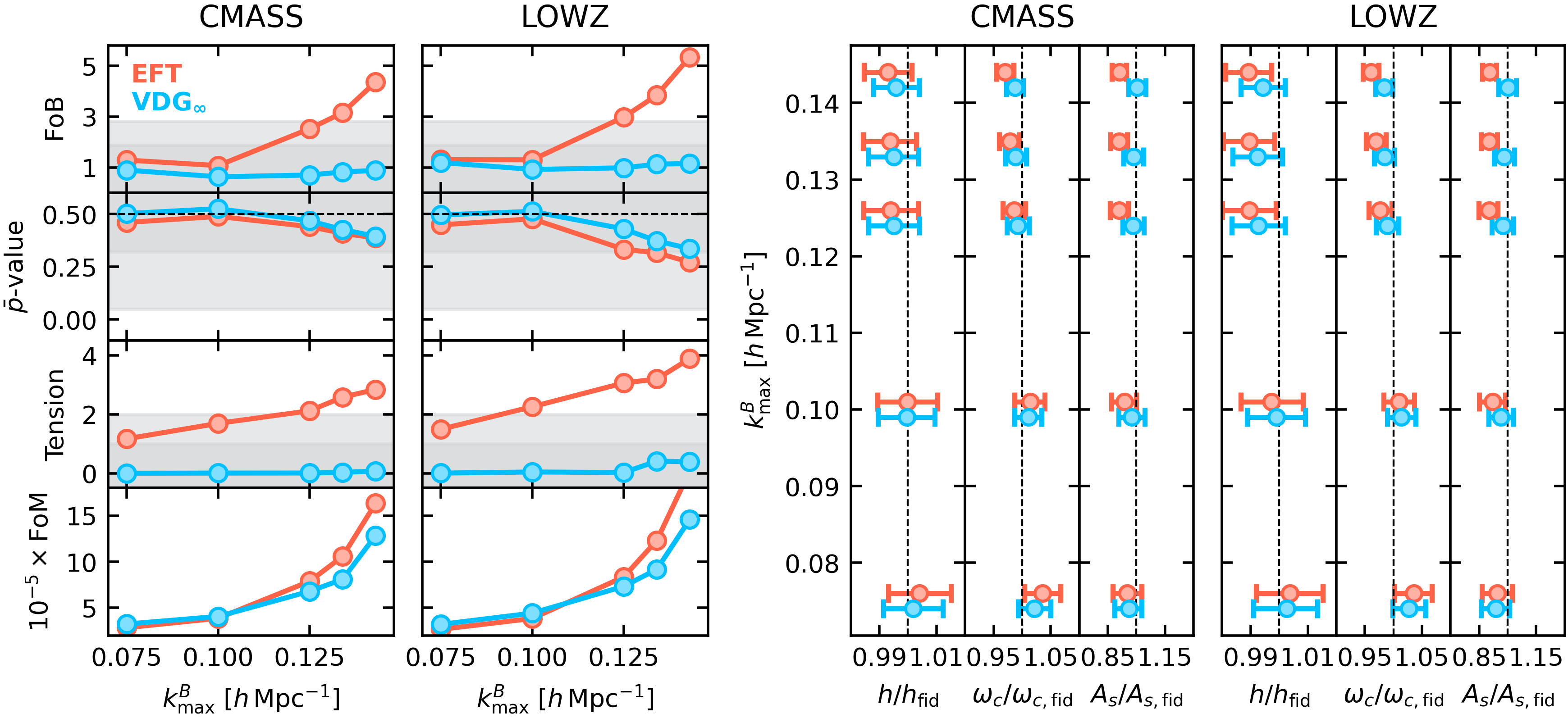}
  \caption{Same as Fig.~\ref{fig:pk_metrics_diff0p0}, but for the joint analysis of the power spectrum and bispectrum.  All quantities are shown as a function of the bispectrum scale cut (identical for monopole and quadrupole), while the power spectrum scale cut was fixed to $\kmax^P = (0.25,0.2,0.2)\,\hinvMpc$.  The performance metrics include an additional panel showing the tension between the posteriors for the power spectrum alone and for the joint analysis.  The fourth panel shows the increase of the joint FoM over the power spectrum alone.}
  \label{fig:performance_metrics_hwcAs_PkBk_kmaxP_0p25_0p2_0p2}
\end{figure*}

The performance metrics do not indicate a clear preference for any of the three models of the VDG considered in the previous section.  Therefore, we now approach the issue from a different perspective, focusing on the redshift-space counterterms $c_2$ and $c_4$.  As discussed in Sec.~\ref{sec:theory.counterterms}, these counterterms are designed to absorb any non-linear effects of the RSD mapping that are not captured by the perturbative expansion or by the non-perturbative damping function in case of the EFT and \VDG models, respectively.

Fig.~\ref{fig:counterterm_constraints_varkmax} displays the two-dimensional posteriors for $c_2$ and $c_4$ for each model, with different colours representing various $\kmax^P$ values.  In the EFT model, the constraints favour positive values, which, according to our sign convention, implies a suppression of the quadrupole and hexadecapole, as expected from the finger-of-god effect.  Additionally, we observe a dependence of the posterior means on $\kmax^P$ starting from $0.3\,\hinvMpc$, which means that the non-linear corrections that need to be absorbed by the counterterms become increasingly relevant.  This is yet another indicator for the failure of the model and aligns well with the increasing FoB values that we identified in Sec.~\ref{sec:results.pk.nominal}.

In the \VDG model, the values of the counterterms are highly sensitive to the modelling of the VDG function.  Fig.~\ref{fig:counterterm_constraints_varkmax} shows that only our baseline prescription results in counterterms that are consistent with zero for both galaxy samples and without any $\kmax^P$ trends, suggesting that it captures the non-linear RSD effects most accurately.  In contrast, as $\kmax^P$ increases, the Gaussian damping function requires positive $c_2$ and $c_4$ values (albeit smaller than those in the EFT), whereas the Lorentzian damping function consistently favours large, negative values.  This behaviour arises from their different damping effects for various LOS orientations compared to the baseline model (see discussion in Sec.~\ref{sec:theory.VDG}).  Specifically, when $\sigma_v$ in the Gaussian (Lorentzian) model is adjusted to match the baseline monopole for a given $a_{\rm vir}$ value, its quadrupole and hexadecapole are enhanced (suppressed) with respect to the baseline.  These differences are compensated accordingly by the counterterms.

\subsection{Joint power spectrum and bispectrum analysis}
\label{sec:results.joint}

\begin{figure*}
  \centering
  \includegraphics{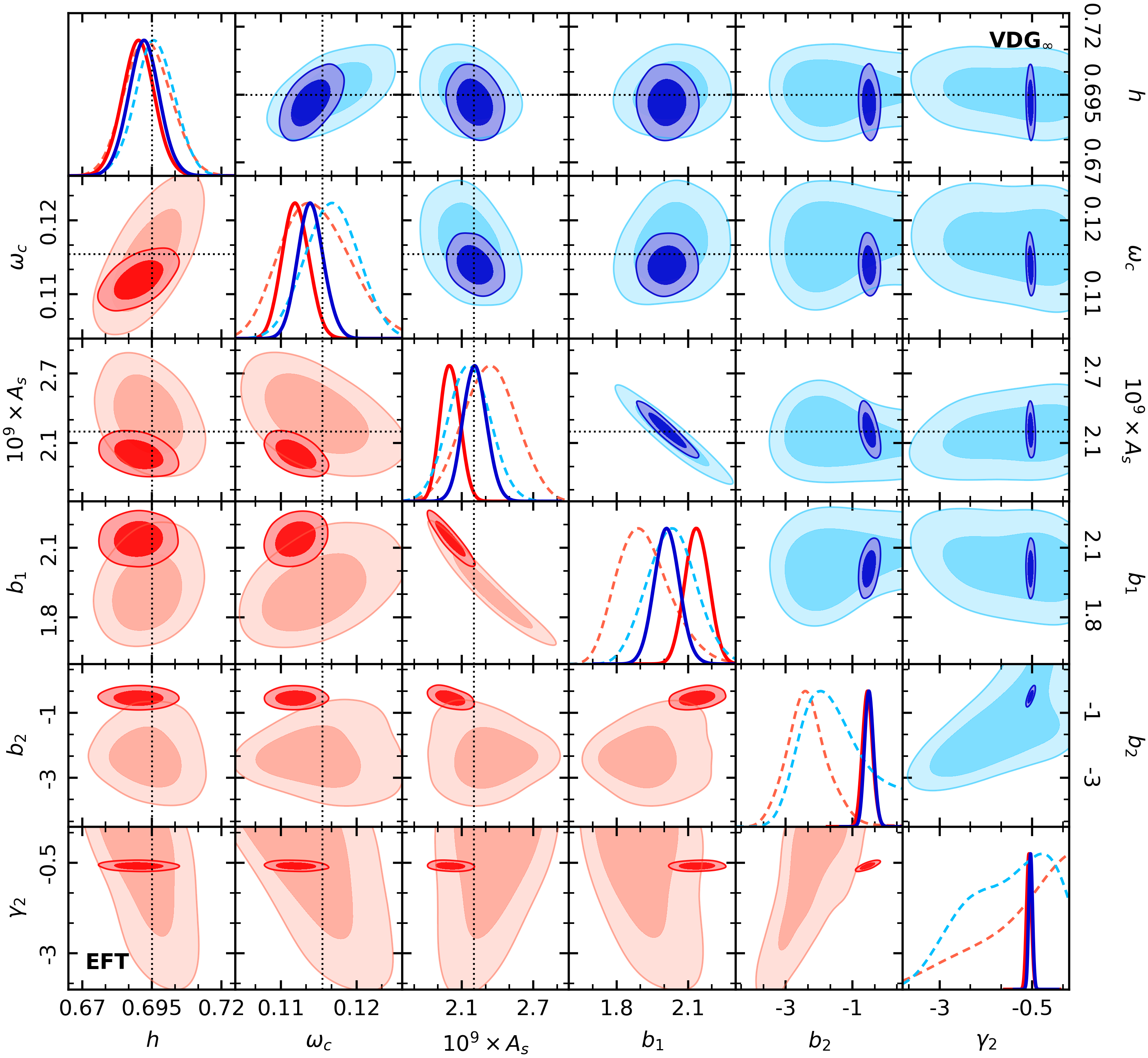}
  \caption{Posterior constraints for the CMASS sample, shown for a subset of the fitted model parameters (fiducial values for the three cosmological parameters are indicated by dotted lines).  The lower half triangle (red colours) corresponds to the EFT model, the upper half triangle (blue colours) to the \VDG model.  Lighter shades and dashed lines represent constraints from the power spectrum alone using a scale cut $k_{\mathrm{max}}^P = (0.25,0.2,0.2)\,\hinvMpc$, while darker shades and solid lines represent constraints from the joint power spectrum and bispectrum analysis using the same power spectrum scale cut and $\kmax^B = 0.143\,\hinvMpc$.}
  \label{fig:PB_posteriors_cmass_big_kmaxP_0p25_0p2_0p2_kmaxB_0p143}
\end{figure*}

\subsubsection{Performance at fixed power spectrum scale cut}
\label{sec:results.bk.nominal}

In this section, we analyse the monopole and quadrupole of the bispectrum in conjunction with the previously studied power spectrum multipoles.  Our initial goal is to evaluate the performance of the EFT and \VDG bispectrum models while keeping the power spectrum scale cut fixed.  Specifically, we select the combination $\kmax^P = (0.25,0.2,0.2)\,\hinvMpc$, where the EFT model still provides unbiased cosmological constraints, and vary the bispectrum multipole cutoffs, $\kmax^B$, across five different values ranging from $0.075$ to $0.143\,\hinvMpc$.  The resulting FoB, $\bar{p}$-value, the tension between the joint $P_{\ell}$ + $B_{\ell}$ posteriors and those of $P_{\ell}$ alone (refer to Sec.~\ref{sec:tension} for the method to compute this tension), as well as the FoM, are presented in Fig.~\ref{fig:performance_metrics_hwcAs_PkBk_kmaxP_0p25_0p2_0p2}.

As with the power spectrum alone, we see that increasing $\kmax^B$ enlarges the difference in FoB between the two models.  Independent of the galaxy sample, the FoB of the \VDG model remains relatively stable across the tested range of scale cuts, whereas the EFT model exceeds the critical $68\,\%$ threshold already at values beyond $\kmax^B = 0.1\,\hinvMpc$.  Despite this, both models fit the data equally well, as indicated by the $\bar{p}$-values\footnote{Note that the $\bar{p}$-values of the joint fits are dominated by the bispectrum due to its significantly larger data vector compared to the power spectrum.}.  The large FoB values in case of the EFT are mainly driven by the parameters $\omega_c$ and $A_s$, as shown in the two right-hand panels of Fig.~\ref{fig:performance_metrics_hwcAs_PkBk_kmaxP_0p25_0p2_0p2}.  Both parameters are biased low, differing from the trend we observed for $A_s$ when considering the power spectrum alone.  These findings are in good agreement with the recent studies \cite{IvaPhiNis2203,PhiIva2202,IvaPhiCab2304}, which suggest a scale cut of $\kmax^B = 0.08\,\hinvMpc$ to prevent significant biases in the cosmological analyses of BOSS data.

\begin{figure*}
  \centering
  \includegraphics[width=0.94\textwidth]{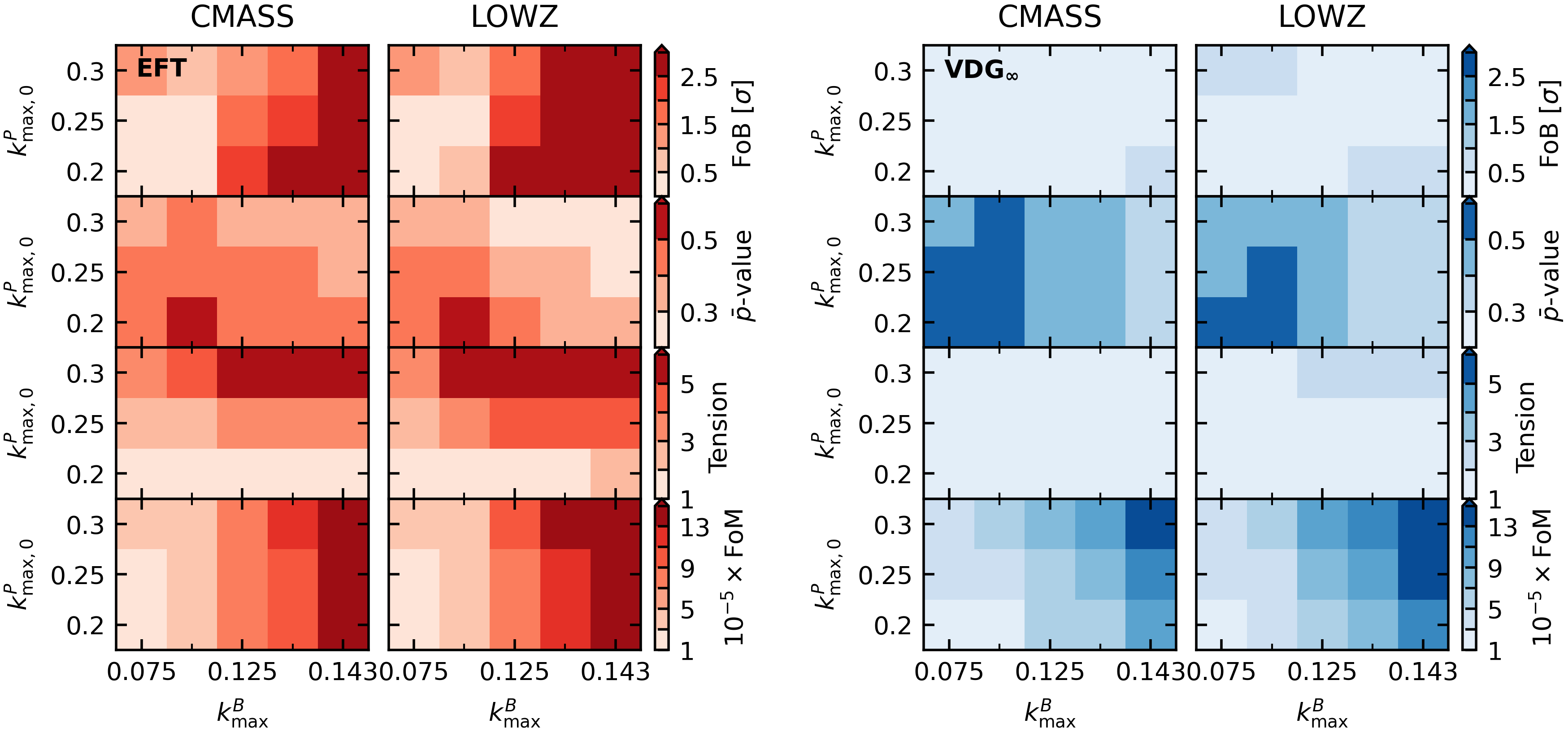}
  \caption{Same performance metrics as in Fig.~\ref{fig:PB_posteriors_cmass_big_kmaxP_0p25_0p2_0p2_kmaxB_0p143}, for different values of $\kmax^B$ and $k_{\mathrm{max},0}^P$.  The bispectrum scale cut for monopole and quadrupole is identical, for the power spectrum the constant difference $\Delta \kmax = k_{\mathrm{max},0} - k_{\mathrm{max},24} = 0.05\,\hinvMpc$ is used.  All scale cuts are given in units of $\hinvMpc$.}
  \label{fig:performance_metrics_PkBk_varkmaxP_heatmap}
\end{figure*}

\begin{figure}
  \centering
  \includegraphics{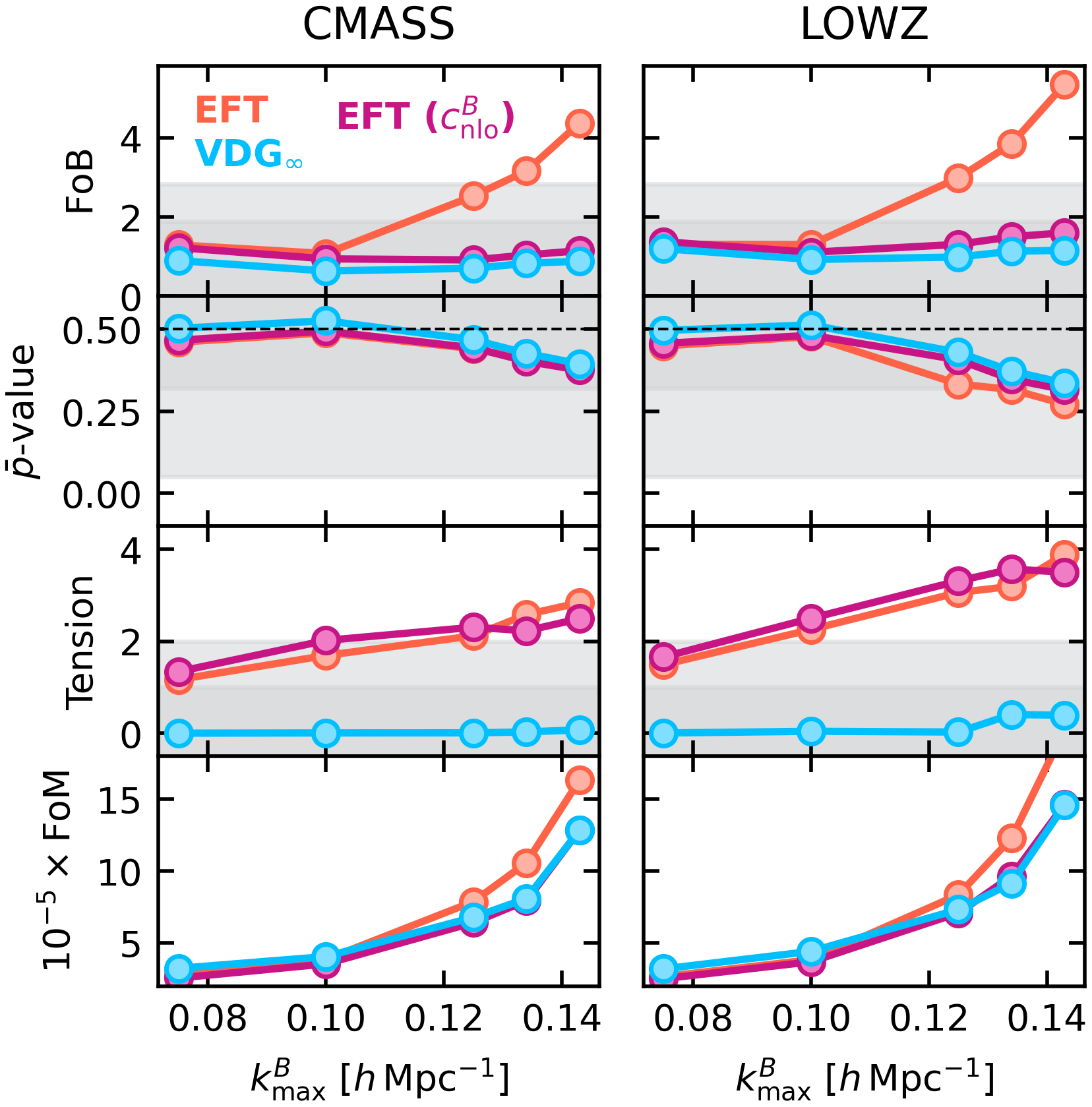}
  \caption{Performance metrics for joint power spectrum and bispectrum analyses, using two different prescriptions for the bispectrum counterterms: default choice of \cite{PhiIva2202} (Eq.~\ref{eq:theory.ctr.Z1ctr}; red) and the $c_{\rm nlo}^B$ counterterm (Eq.~\ref{eq:theory.ctr.Bctr_EFT}; violet). For reference, the \VDG is shown as well.  In all cases the power spectrum scale cut $\kmax^P = (0.25,0.2,0.2)\,\hinvMpc$ is adopted.}
  \label{fig:performance_metrics_PkBk_cnloB}
\end{figure}

Fig.~\ref{fig:performance_metrics_hwcAs_PkBk_kmaxP_0p25_0p2_0p2} moreover demonstrates that the posteriors derived from the joint fit and from the power spectrum alone are increasingly in tension for the EFT model, but remain perfectly consistent for the \VDG model.  This tension is computed from the full parameter space shared between the power spectrum and bispectrum models and is primarily due to the galaxy bias parameters rather than the cosmological parameters.  An illustrative example is given in Fig.~\ref{fig:PB_posteriors_cmass_big_kmaxP_0p25_0p2_0p2_kmaxB_0p143}, which compares the power spectrum posteriors at $\kmax^P = (0.25,0.2,0.2)\,\hinvMpc$ (depicted by the lighter shades and dashed lines), with the joint ones at $\kmax^B = 0.143\,\hinvMpc$ (darker shades and solid lines) for the CMASS sample.  The parameters most in tension for the EFT model are $b_1$ and $b_2$, whereas all parameters are fully consistent for the \VDG model.  Interestingly, the joint constraints on $b_2$, which are predominantly coming from the bispectrum, align very well between the two models.  This indicates that even at a scale cut where the EFT power spectrum still provides unbiased constraints on cosmological parameters, inaccuracies in the modelling may already lead to shifts in the extended parameter space.

The enhanced performance of the \VDG model significantly boosts the overall constraining power of the joint power spectrum and bispectrum analysis.  Although the FoM of the EFT model is larger for higher $\kmax^B$ values, a comparison  of the \VDG model at $\kmax^B = 0.143\,\hinvMpc$ with the EFT model at $\kmax^B = 0.1\,\hinvMpc$---the largest value before surpassing the 68\,\% threshold in both galaxy samples---reveals FoM differences between factors of 2 to 3.  This is because the FoM initially increases only gradually, but beyond $\kmax^B = 0.1\,\hinvMpc$, it rises much more steeply.  We will return to a more precise quantification of the gains from the bispectrum in Sec.~\ref{sec:results.bk.gain}.

\vspace*{-0.2em}
\subsubsection{Varying power spectrum and bispectrum scale cuts}
\label{sec:results.bk.varykmaxP}

We now examine two additional power spectrum scale cuts: $\kmax^P = (0.3,0.25,0.25)\,\hinvMpc$ and $\kmax^P = (0.2,0.15,0.15)\,\hinvMpc$, while varying $\kmax^B$ across the same values as before.  This analysis serves two purposes: firstly, to assess whether joint fits of the \VDG model remain valid when extended to larger $\kmax^P$; and secondly, to investigate whether the inconsistencies in the galaxy bias parameters between the EFT power spectrum and bispectrum, as revealed by the previous section, contribute to the biases in the cosmological parameters.

The four performance metrics for both samples and models are shown in Fig.~\ref{fig:performance_metrics_PkBk_varkmaxP_heatmap}, where the $y$-axis represents the power spectrum monopole cutoff.  The FoB and $\bar{p}$-value demonstrate that the \VDG model can be applied across all tested $\kmax^B$ values, even when incorporating additional small-scale information from the power spectrum.  This leads to a further  $\sim\,25\,\%$ enhancement of the overall FoM.  In contrast, the FoB of the EFT model does not improve with a reduced power spectrum scale cut, despite a significant decrease in tension.  Since the tension is primarily due to discrepancies in the bias parameters, this implies that the bias in the cosmological parameters is unaffected by these discrepancies, suggesting instead a breakdown of the bispectrum model. 

\subsubsection{Testing bispectrum counterterm prescriptions}
\label{sec:results.bk.counterterms}

Unlike for the power spectrum, in the regime where we analyse the bispectrum measurements, the effect of the VDG can still be captured perturbatively.  We have already demonstrated this in Sec.~\ref{sec:binning}, where we exploit this fact to efficiently compute the discreteness corrections for the \VDG bispectrum.  Consequently, the breakdown of the EFT bispectrum model must have a different cause than the power spectrum, potentially linked to the specific choice of counterterms, which do not originate from a low-$k$ expansion of the \VDG damping function (see Sec.~\ref{sec:theory.counterterms}).

This is fully supported by the data presented in Fig.~\ref{fig:performance_metrics_PkBk_cnloB}, where we compare the results of the nominal EFT model (with fixed $\kmax^P = (0.25,0.2,0.2)\,\hinvMpc$) to an analysis in which the two counterterms associated with $c_1^B$ and $c_2^B$ are replaced by a single counterterm, $c^B_{\rm nlo}$ that is consistent with the \VDG bispectrum damping function in the low-$k$ limit (cf. Eq.~\ref{eq:theory.ctr.Bctr_EFT}).  This replacement (depicted by the violet lines) significantly reduces the FoB to values that are only slightly larger than for the \VDG model, and we also find consistency in the $\bar{p}$-values and FoM.  However, the tension between the power spectrum and the joint posteriors remains unchanged, which is in agreement with our previous conclusion that this tension is driven by the power spectrum posteriors and primarily due to discrepancies in the galaxy bias rather than cosmological parameters.

\subsubsection{Information gain from the bispectrum}
\label{sec:results.bk.gain}

\begin{figure*}
  \centering
  \includegraphics{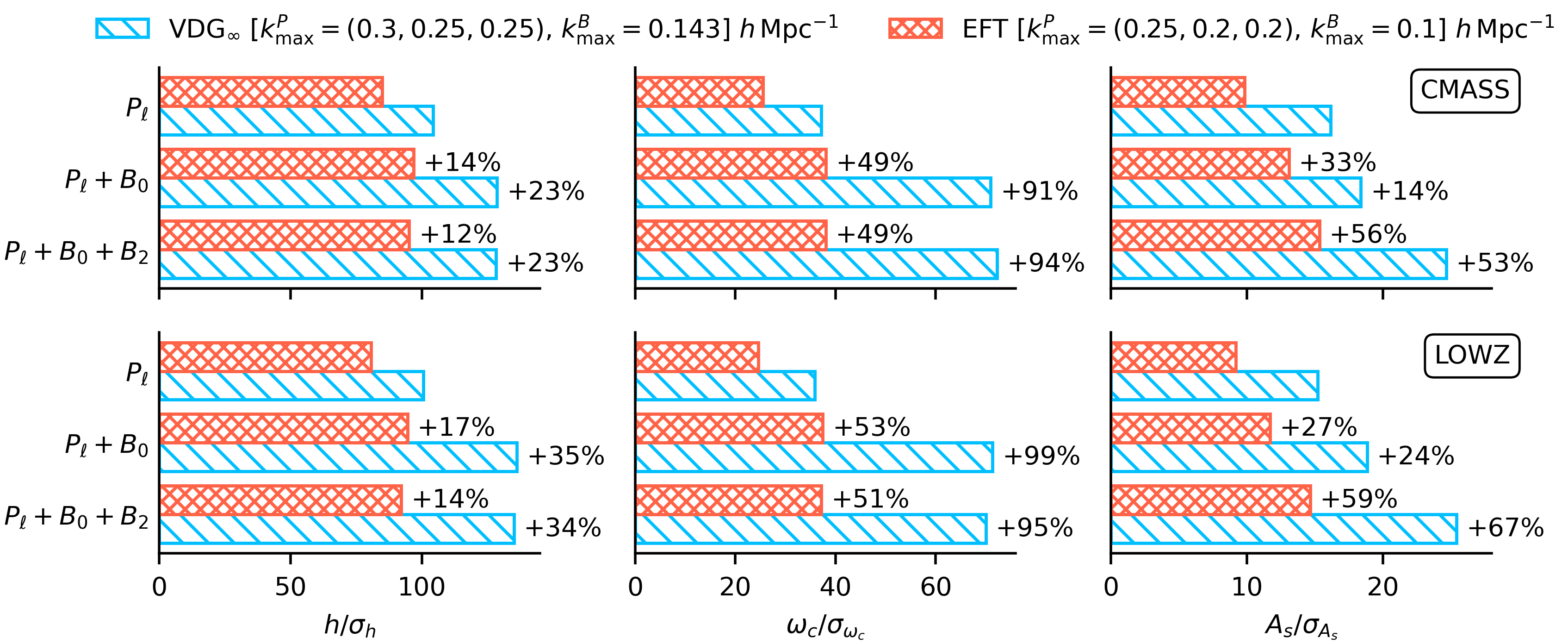}
  \caption{Inverse relative uncertainties on the three cosmological parameters $h$, $\omega_c$, and $A_s$ obtained for the CMASS and LOWZ samples (top and bottom rows, respectively).  Each panel depicts three cases corresponding to the power spectrum multipoles alone and in combination with either the bispectrum monopole, or bispectrum monopole and quadrupole.  The percentages indicate the relative improvement over the power spectrum alone.  Scale cuts for each model were chosen to maximise constraining power under the condition that $\mathrm{FoB} < 1\sigma$.}
  \label{fig:gains_PB0B2}
\end{figure*}
Finally, we explore the extent to which cosmological constraints can be enhanced by combining information from the bispectrum multipoles with the power spectrum.  For this comparison, we select the scale cut combinations that maximise the FoM without the FoB surpassing the 68\,\% critical threshold, resulting in\footnote{We note that for the \VDG model $\kmax^P = (0.3,0.25,0.25)\,\hinvMpc$ is the largest cutoff that we tested in combination with the bispectrum, although the power spectrum itself remains unbiased at even larger values.  Since the power spectrum FoM flattens off beyond $\kmax^P \sim 0.3\,\hinvMpc$, we expect that this restriction does not considerably impact the reported gain factors from the bispectrum.}:
\begin{itemize}
\item EFT: $\left\{\begin{array}{l} \kmax^P = (0.25,0.2,0.2)\,\hinvMpc \\[0.25em]
                     \kmax^B = 0.1\,\hinvMpc \end{array}\right.$
\item \VDG: $\left\{\begin{array}{l} \kmax^P = (0.3,0.25,0.25)\,\hinvMpc \\[0.25em]
                     \kmax^B = 0.143\,\hinvMpc \end{array}\right.$
\end{itemize}
In Fig.~\ref{fig:gains_PB0B2} we plot the inverse of the relative uncertainties for the three cosmological parameters individually and distinguish between three cases: constraints from the power spectrum multipoles alone and in combination with either the bispectrum monopole, or the bispectrum monopole and quadrupole (in this case, the $\kmax^B$ value given above is applied to both).  Independent of the employed model, we see that the bispectrum monopole tightens constraints on all three cosmological parameters, while the bispectrum quadrupole only provides further information on $A_s$.  This is because the quadrupole helps breaking the degeneracy between $A_s$ and $b_1$, but we find that---at least for the considered $\kmax^B$ values---the monopole has exhausted the information on the growth rate (hence no improvement in $h$ and $\omega_c$).

On the other hand, the overall gain from the bispectrum depends strongly on $\kmax^B$ and thus on the modelling approach: in case of the EFT model and for both samples, the bispectrum enhances constraints on $h$ and $\omega_c$ by roughly $15$ and $50\,\%$, respectively, whereas the \VDG model doubles these improvements to $\sim 30$ and $100\,\%$.  Only for $A_s$ the gain from the bispectrum is more comparable between the two models, totalling about $60\,\%$.  If we also take into account that the \VDG model already provides tighter constraints at the level of the power spectrum alone and directly compare the constraints from $P_{\ell} + B_0 + B_2$, we find that the \VDG model allows us to shrink the uncertainties obtained from the EFT model (based on $\sigma_{\mathrm{VDG}_{\infty}}/\sigma_{\rm EFT}-1$) by about $25$, $50$, and $40\,\%$ for $h$, $\omega_c$, and $As$, respectively.


\section{Discussion and conclusions}
\label{sec:conclusions}

In this paper, we compared two approaches for modelling RSD:
\begin{enumerate}
\item the EFT, which expands the mapping from real to redshift space perturbatively and aims to account for the FoG effect, caused by small-scale, virialised velocities, through a series of counterterms with adjustable amplitudes;
\item an approach centred on the VDG, which encapsulates all non-perturbative effects of small-scale velocities and which is modelled here in terms of an effective damping function (specifically referred to as the \VDG model).
\end{enumerate}
We applied these models to the first three Legendre multipoles of the power spectrum, as well as the monopole and quadrupole of the bispectrum, measured from two synthetic samples of LRGs.  These samples mimic the BOSS CMASS and LOWZ selections at redshifts $z=0.57$ and $z=0.34$, but with measurement uncertainties scaled to those expected for Stage-IV galaxy surveys at the end of their mission.

Our analysis shows that RSD due to virialised velocities become a dominant factor from quasi-linear scales onward, such that ignoring their non-perturbative nature can lead to severe modelling inaccuracies.  This is the case for the EFT approach, where fits of the power spectrum multipoles result in increasingly biased cosmological parameter constraints (in this work, we focused on the dimensionless Hubble rate $h$, the dark matter density $\omega_c$, and the amplitude of scalar fluctuations $A_s$) when applying scale cuts with $\kmax > 0.2\,\hinvMpc$ (Fig.~\ref{fig:pk_metrics_diff0p0}).  This is consistent with several previous analyses that studied the performance of the EFT model \cite{dAGleKok2005,NisDAIva2012,ChuDolIva2101,MauLaiNor2406}, as well as cutoff choices applied to real data, including the recent DESI results \cite{AdaAguAhl2411}.  We demonstrated that these parameter biases are largely driven by the power spectrum quadrupole (Fig.~\ref{fig:pk_metrics_diffvar}), which is a strong indicator for deficiencies in the RSD modelling, as it is most sensitive to the anisotropic RSD signal.  In contrast, these deficiencies are overcome in the VDG approach, which provides unbiased constraints, even up to scales of $\kmax = 0.35\,\hinvMpc$ (Fig.~\ref{fig:pk_metrics_diff0p0}), aligning the validity range of the redshift-space models more closely with that typically found in real space \cite[see, e.g.,][]{EggScoCro2011,PezCroEgg2108,EucPezMor2407}.

Joint fits of the power spectrum and bispectrum revealed a similar discrepancy between the two models: in agreement with previous studies \cite{IvaPhiNis2203,IvaPhiCab2304}, our findings indicate that the EFT is limited to scale cuts below $\kmax = 0.1\,\hinvMpc$ (at fixed power spectrum cutoff).  Conversely, the \VDG model maintains unbiased cosmological constraints up to $\kmax = 0.14\,\hinvMpc$ (Fig.~\ref{fig:PB_posteriors_cmass_big_kmaxP_0p25_0p2_0p2_kmaxB_0p143}), consistent with the performance of the tree-level bispectrum model for tracers in real space \cite{EggScoSmi2106}.  However, on the scales where we analyse the bispectrum measurements, the FoG effect is still within the perturbative regime, prompting questions about the origin of the difference between the two models.  We found that this is caused by the counterterm prescription used in \cite{IvaPhiNis2203,PhiIva2202,IvaPhiCab2304}, which cannot be linked to a perturbative expansion of the VDG and fails to adequately capture the FoG effect in the bispectrum.  Instead, using an alternative counterterm that does originate from an expansion of the VDG, we demonstrated that results comparable to the full \VDG model can be achieved (Fig.~\ref{fig:performance_metrics_PkBk_cnloB}).

The extended validity ranges of the \VDG model compared to the EFT yield immediate improvements in constraining power and thus our ability to distinguish between cosmological models.  For the power spectrum alone, the reductions in the fully marginalised uncertainties on the three studied cosmological parameters range from 20 to 40\,\%.  The gain from including the information in the bispectrum is notably enhanced for the \VDG model (Fig.~\ref{fig:gains_PB0B2}), since even the modest cutoff increase from $0.1$ to $0.14\,\hinvMpc$ unlocks a significant extra part of the steeply rising bispectrum signal-to-noise.  In total, the joint power spectrum and bispectrum analysis with the \VDG model therefore results in uncertainty reductions between 25 and 50\,\% relative to the EFT.

Another validity test of the FoG modelling in the VDG approach is provided by the constraints on the counterterm parameters.  These parameters are designed to absorb any residual FoG contributions, yet they remain consistent with zero in the \VDG model for both galaxy samples, showing no significant dependence on the scale cuts (Fig.~\ref{fig:counterterm_constraints_varkmax}).  Given that the counterterms have been identified as a potential driver of strong prior volume projection effects if the constraining power of the data is insufficient \cite{AdaAguAhl2411}, the \VDG model might offer an intriguing simplification by omitting the redshift-space counterterms.  This simplification is not feasible with the EFT, where the counterterm parameters attain non-zero and sample-dependent values.

We note that the \VDG model bears strong similarity with what is known as the ``TNS'' model in the literature \cite{TarNisSai1009}.  However, we pointed out several crucial differences that impact performance: 1) the TNS model makes use of a phenomenological damping function (typically either a Gaussian or Lorentzian), while the damping function of the \VDG model is derived by resumming quadratic non-linearities; 2) the stochastic contributions in the TNS model are typically multiplied by the damping function \cite[e.g.,][]{BeuSaiSeo1409,BeuSeoSai1704}, whereas we have shown here that in the context of the VDG approach the FoG effect leads to a LOS dependent stochastic contribution, but that stochastic terms are not subject to any additional damping; 3) counterterms are traditionally ignored in the TNS model.  We demonstrated that different prescriptions of the damping function can indeed yield comparable results (Fig.~\ref{fig:performance_metrics_Pk_diff0p0_vardamping}), but only if any mismodelling of the FoG effect can be absorbed by counterterms (Fig.~\ref{fig:counterterm_constraints_varkmax}).  Some other works in the literature \cite{ChuIvaSim2102,ChuIva2302} have also studied models similar to the VDG approach, but come to the conclusion that they are outperformed by the EFT.  We clarified that this is because the models used in these works lack crucial ingredients and apply the damping function incorrectly.

This paper has also introduced several technical advances that enable an efficient, yet robust evaluation of the bispectrum likelihood.  Specifically, we validated a scheme that Taylor expands the effect of the Alcock-Paczy\'nski distortions, showing that inclusion of the linear-order terms are sufficient for the monopole and quadrupole.  Moreover, we introduced a novel approach to approximately account for binning and discreteness effects in the bispectrum.  After  a one-off computational overhead for a given binning and Fourier grid setup, the analysis can be performed at no additional cost, allowing us to evaluate the joint power spectrum and bispectrum likelihood in ${\cal O}(10)\,\mathrm{ms}$.  At the same time, we showed that any resulting inaccuracies of the approximations made are insignificant for the adopted range of scales and measurement uncertainties in this work.  All these advances have been incorporated into our \texttt{Python} package \texttt{COMET}, which is publicly available.

As a final remark, we stress that all results in this paper were derived from just two different LRG samples based on the HOD methodology.  Although these samples remain relevant for Stage-IV galaxy surveys (e.g., LRGs represent a significant fraction of the targets observed by DESI), it is highly desirable to extend the tests presented here to other galaxy populations, such as actively star-forming galaxies identified through specific emission lines, and to higher redshifts.  Furthermore, it is intriguing to explore whether the performance gains offered by the \VDG model persist for more general HOD catalogues that incorporate assembly and velocity bias, or through entirely different approaches (e.g., subhalo abundance matching or semi-analytic models).  We plan to investigate these scenarios in future works.

\begin{acknowledgments}
  AE is supported at the Argelander Institut für Astronomie by an Argelander Fellowship. NL was supported by the James Arthur Graduate Associate Fellowship from the Center for Cosmology and Particle Physics at New York University and the Horizon Fellowship from Johns Hopkins University. MC acknowledges support form the Spanish Ministerio de Ciencia, Innovación y Universidades, project PID2021-128989NB, while AGS is supported by the Deutsche Forschungsgemeinschaft (DFG, German Research Foundation) under Germany´s Excellence Strategy – EXC 2094 – 390783311. This work was supported in part through the NYU IT High Performance Computing resources, services, and staff expertise. Our research made use of \texttt{matplotlib}, a Python library for publication quality graphics \cite{Hun07}.
\end{acknowledgments}

\appendix

\section{Perturbation theory kernels}
\label{sec:app.PTkernels}

For reference, we provide the expressions for the perturbation theory kernels, $Z_n$, used in our computation of the one-loop power spectrum (see Eq.) and the tree-level bispectrum (see Eq.).  This requires kernels up to third order and in our galaxy bias convention they are given by:
\begin{equation}
    Z_1(\kv) = b_1+f\mu^2\,,
\end{equation}
\begin{widetext}
  \begin{equation}
    Z_2(\kv_1,\kv_2) =  {\cal K}_2(\kv_1,\kv_2) + f\mu^2\,G_2(\kv_1,\kv_2) + \frac{1}{2}fk\mu \left[\frac{\mu_1}{k_1}\left(b_1+f\mu_2^{\,2}\right)+\frac{\mu_2}{k_2}\left(b_1+f\mu_1^{\,2}\right)\right]\,,
  \end{equation}
  and
  \begin{equation}
    \begin{split}
      Z_3(\kv_1,\kv_2,\kv_3) = \; {\cal K}_3(\kv_1,\kv_2,\kv_3) + f\mu^2\,G_3(\kv_1,\kv_2,\kv_3)
      + fk\mu\frac{\mu_3}{k_3}\left[b_1F_2(\kv_1,\kv_2)+f\mu_{12}^{\,2}\,G_2(\kv_1,\kv_2)\right]  \\
        + \frac{1}{2}f^2k^2\mu^2\frac{\mu_2\,\mu_3}{k_2\,k_3}\left(b_1+f\mu_1^{\,2}\right) 
         + fk\mu\frac{\mu_{23}}{k_{23}}\left(b_1+f\mu_1^{\,2}\right)G_2(\kv_2,\kv_3) 
         +fk\mu\frac{\mu_1}{k_1}\left[\frac{b_2}{2}+\gamma_2\,K(\kv_2,\kv_3)\right]\,,
    \end{split}
  \end{equation}
\end{widetext}
where we have defined $\mu_i \equiv k_{i,z}/k_i$, $k \equiv \left|\sum_{i=1}^n \kv_i\right|$, and $\mu \equiv \sum_{i=1}^n k_{i,z}/k$.  The ${\cal K}_n$ kernels represent the real-space contributions including galaxy bias:
\begin{equation}
  \label{eq:K2}
    {\cal K}_2(\kv_1,\kv_2)=b_1\,F_2(\kv_1,\kv_2)+\frac{b_2}{2}+\gamma_2\, K(\kv_1,\kv_2)\,,
\end{equation}
\begin{equation}
  \label{eq:K3}
    \begin{split}
        {\cal K}_3(\kv_1,\kv_2,\kv_3)= \;& b_1\,F_3(\kv_1,\kv_2,\kv_3) + b_2\,F_2(\kv_1,\kv_2) \\
        &\hspace{-5em}+2\gamma_2\,K(\kv_1,\kv_2+\kv_3)\,G_2(\kv_2,\kv_3) \\
        &\hspace{-5em}+2\gamma_{21}\,K(\kv_1,\kv_2+\kv_3)\,K(\kv_2,\kv_3)\,,
    \end{split}
\end{equation}
with 
\begin{equation}
    K(\kv_1,\kv_2)=\frac{\left(\kv_1\cdot\kv_2\right)^2}{k_1^{\,2}k_2^{\,2}}-1\,,
\end{equation}
where the functions $F_n$ and $G_n$ are the usual kernels for the non-linear evolution of the matter and velocity divergence fields \cite[see e.g.,][]{BerColGaz0209}.  Note that the above expressions for both, the $Z_3$ and ${\cal K}_3$ kernels, still need to be symmetrised over their three arguments.

\section{Gaussian bispectrum covariance in redshift space}
\label{sec:app.Bcov}

In this work, we employ a Gaussian model for the covariance matrix of the bispectrum multipoles between configurations $\left\{k_1^{(i)},k_2^{(i)},k_3^{(i)}\right\}$ and $\left\{k_1^{(j)},k_2^{(j)},k_3^{(j)}\right\}$.  In this approximation, we obtain from the definition of the bispectrum estimator (Eq.~\ref{eq:data.Bl}):
\begin{widetext}
  \begin{equation}
    \label{eq:app.Bcov.Gcov}
    \begin{split}
      \mathrm{Cov}\left[\hat{B}_{\ell_1}^{(i)}\,,\hat{B}_{\ell_2}^{(j)}\right] = \frac{(2\ell_1+1)\,(2\ell_2+1)}{N_B^2\left(k_1^{(i)},k_2^{(i)},k_3^{(i)}\right)} \, \delta_{ij}^{\rm K} \sum_{\bq_1 \in k_1^{(i)}} \sum_{\bq_2 \in k_2^{(i)}} \sum_{\bq_3 \in k_3^{(i)}} \delta^{\rm K}_{\bq_{123},0} \, P^{\rm tot}(\bq_1) \, P^{\rm tot}(\bq_2) \, P^{\rm tot}(\bq_3) \\
      \times \, \Big[ \left(1 + \delta^{\rm K}_{k_2^{(i)},k_3^{(i)}}\right) {\cal L}_{\ell_1}(\mu_{\bq_1})\,{\cal L}_{\ell_2}(-\mu_{\bq_1}) + \left(\delta^{\rm K}_{k_1^{(i)},k_2^{(i)}} + \delta^{\rm K}_{k_2^{(i)},k_3^{(i)}}\right) {\cal L}_{\ell_1}(\mu_{\bq_1})\,{\cal L}_{\ell_2}(-\mu_{\bq_2}) + 2 \delta^{\rm K}_{k_1^{(i)},k_3^{(i)}} \, {\cal L}_{\ell_1}(\mu_{\bq_1})\,{\cal L}_{\ell_2}(-\mu_{\bq_3}) \Big]\,,
    \end{split}
  \end{equation}
  where $\mu_{\bq} \equiv q_z/q$ and we have assumed that $k_1^{(i)} \geq k_2^{(i)} \geq k_3^{(i)}$ (similarly for configuration $j$).  The functions $P^{\rm tot}$ denote the anisotropic power spectra in redshift space, including the Poisson shot noise contribution, and following the original derivation in \cite{RizMorPar2301}, we can expand them in Legendre moments, giving:
  \begin{equation}
    \label{eq:app.Bcov.Gcov_Legendre}
    \begin{split}
      \mathrm{Cov}\left[\hat{B}_{\ell_1}^{(i)}\,,\hat{B}_{\ell_2}^{(j)}\right] = \frac{(2\ell_1+1)\,(2\ell_2+1)}{N_B^2\left(k_1^{(i)},k_2^{(i)},k_3^{(i)}\right)} \, \delta_{ij}^{\rm K} \sum_{\ell_3,\ell_4,\ell_5} \sum_{\bq_1 \in k_1^{(i)}} \sum_{\bq_2 \in k_2^{(i)}} \sum_{\bq_3 \in k_3^{(i)}} \delta^{\rm K}_{\bq_{123},0} \, P_{\ell_3}^{\rm tot}(q_1) \, P_{\ell_4}^{\rm tot}(q_2) \, P_{\ell_5}^{\rm tot}(q_3) \\ \times \, {\cal L}_{\ell_3}(\mu_{\bq_1}) \, {\cal L}_{\ell_4}(\mu_{\bq_2}) \, {\cal L}_{\ell_5}(\mu_{\bq_3}) \Big[ \left(1 + \delta^{\rm K}_{k_2^{(i)},k_3^{(i)}}\right) {\cal L}_{\ell_1}(\mu_{\bq_1})\,{\cal L}_{\ell_2}(-\mu_{\bq_1}) \Big. \\ \Big. + \left(\delta^{\rm K}_{k_1^{(i)},k_2^{(i)}} + \delta^{\rm K}_{k_2^{(i)},k_3^{(i)}}\right) {\cal L}_{\ell_1}(\mu_{\bq_1})\,{\cal L}_{\ell_2}(-\mu_{\bq_2}) + 2 \delta^{\rm K}_{k_1^{(i)},k_3^{(i)}} \, {\cal L}_{\ell_1}(\mu_{\bq_1})\,{\cal L}_{\ell_2}(-\mu_{\bq_3}) \Big]\,.
    \end{split}
  \end{equation}
  In order to simplify this expression one can make two further approximations: the continuum, and the thin-shell limit.  The former neglects effects due to the discreteness of the wave modes, allowing us to replace the sums over the shells with corresponding integrals.  The latter moreover assumes that the width of the shells is small compared to the scale over which the integrand varies.  As a result, we can evaluate the right-hand side of  Eq.~(\ref{eq:app.Bcov.Gcov_Legendre}) at fixed wave modes (here we choose the effective wave modes $k_{\mathrm{eff},i}$ defined in Eq.~\ref{eq:data.keff}) and average the angular dependence over the orientation of the triangle configuration with respect to the LOS.  We then have \cite{RizMorPar2301}:
  \begin{equation}
    \label{eq:app.Bcov.Bcov_approx}
    \mathrm{Cov}\left[\hat{B}_{\ell_1}^{(i)}\,,\hat{B}_{\ell_2}^{(j)}\right] \approx \frac{(2\ell_1+1)\,(2\ell_2+1)}{N_B\left(k_1^{(i)},k_2^{(i)},k_3^{(i)}\right)} \, \delta_{ij}^{\rm K} \sum_{\ell_3,\ell_4,\ell_5} P_{\ell_3}^{\rm tot}\left(k_{\mathrm{eff},1}^{(i)}\right) \, P_{\ell_4}^{\rm tot}\left(k_{\mathrm{eff},2}^{(i)}\right) \, P_{\ell_5}^{\rm tot}\left(k_{\mathrm{eff},3}^{(i)}\right) \, R_{\ell_1 \ell_2 \ell_3 \ell_4 \ell_5}^B\left(k_{\mathrm{eff},1}^{(i)},k_{\mathrm{eff},2}^{(i)},k_{\mathrm{eff},3}^{(i)}\right)
  \end{equation}
  using the multipole mixing coefficient
  \begin{align}
    \label{eq:app.Bcov.RB}
    R_{\ell_1 \ell_2 \ell_3 \ell_4 \ell_5}^B(k_{\mathrm{eff},1}^{(i)},k_{\mathrm{eff},2}^{(i)},k_{\mathrm{eff},3}^{(i)}) &\approx \prod_{i=1}^5 \sum_{n_i=1}^{\ell_i/2} C_{\ell_i,n_i} \, \Big[ \left(1 + \delta^{\rm K}_{k_2^{(i)},k_3^{(i)}}\right) \, {\cal I}_{\ell_1 + \ell_2 + \ell_3 -2(n_1+n_2+n_3),\ell_4-2n_4,\ell_5-2n_5}(k_{\mathrm{eff},1}^{(i)},k_{\mathrm{eff},2}^{(i)},k_{\mathrm{eff},3}^{(i)}) \Big. \nonumber \\ &\Big. + \left(\delta^{\rm K}_{k_1^{(i)},k_2^{(i)}} + \delta^{\rm K}_{k_2^{(i)},k_3^{(i)}}\right) \, {\cal I}_{\ell_1+\ell_3-2(n_1+n_3),\ell_2+\ell_4-2(n_2+n_4),\ell_5-2n_5}(k_{\mathrm{eff},1}^{(i)},k_{\mathrm{eff},2}^{(i)},k_{\mathrm{eff},3}^{(i)})  \nonumber \\ &+ 2 \delta^{\rm K}_{k_1^{(i)},k_3^{(i)}} \, {\cal I}_{\ell_1+\ell_3-2(n_1+n_3),\ell_4-2n_4,\ell_2+\ell_5-2(n_2+n_5)}(k_{\mathrm{eff},1}^{(i)},k_{\mathrm{eff},2}^{(i)},k_{\mathrm{eff},3}^{(i)}) \Big]\,.
  \end{align}
\end{widetext}
In the computation of the mixing coefficient we have exploited that Legendre polynomials of even order can be expressed as
\begin{align}
  \label{eq:app.Bcov.Leg_exp}
  {\cal L}_{\ell}(\mu) &= \frac{1}{2^{\ell}} \sum_{n=1}^{\ell/2} \frac{(-1)^n\,(2\ell-2n)!}{n!\,(\ell-n)!\,(\ell-2n)!} \, \mu^{\ell-2n} \\ &\equiv \sum_{n=1}^{\ell/2} C_{\ell,n}\,\mu^{\ell-2n}\,,
\end{align}
and we have defined the average of generic (integer) powers of $\mu_i = k_i/k$ over the triangle orientation as
\begin{equation}
  \label{eq:app.Bcov.In1n2n3}
  {\cal I}_{n_1,n_2,n_3}(k_1,k_2,k_3) = \frac{1}{4\pi} \int_{-1}^1 \mathrm{d}\mu_1 \int_0^{2\pi} \mathrm{d}\phi \, \mu_1^{n_1} \, \mu_2^{n_2} \, \mu_3^{n_3}\,.
\end{equation}
Given our choice of coordinate system (see Sec.~\ref{sec:measurements}), we can write $\mu_2$ and $\mu_3$ in terms of $\mu_1$ and $\phi$ as follows:
\begin{align}
  \mu_2 &= \mu_1 \, \mu_{12} - \sqrt{1-\mu_1^2} \, \sqrt{1-\mu_{12}^2} \, \cos{\phi}\,, \\
  \mu_3 &= -\frac{k_1}{k_3}\,\mu_1 -\frac{k_2}{k_3}\,\mu_2\,,
\end{align}
with $\mu_{12} \equiv \bk_1 \cdot \bk_2/(k_1\,k_2)$.

\section{Relation to galaxy bias and counterterm parameters in \texttt{Class-PT}}
\label{sec:app.bias_bases}

As part of our power spectrum analysis, we compared our set of priors on galaxy bias and counterterm parameters with those used in \cite{PhiIva2202} and related works.  Since they employ \texttt{Class-PT} \cite{ChuIvaPhi2009}, which uses different definitions for the galaxy bias operators,  counterterms, and stochastic terms, we first needed to transform our parameters to ensure a correct application of their priors.

The two parameters that model the effect of large-scale tides are called $b_{{\cal G}_2}$ and $b_{\Gamma_3}$ in \texttt{Class-PT}.  They can be expressed in terms of our $\gamma_2$ and $\gamma_{21}$ as follows:
\begin{align}
  b_{{\cal G}_2} &= \gamma_2\,, \\
  b_{\Gamma_3} &= -\frac{7}{4}\gamma_{21} - \gamma_2\,.
\end{align}
Furthermore, if we denote all counterterm parameters in \texttt{Class-PT} with a tilde, then their relation to the corresponding parameters in \texttt{COMET} are given by:
\begin{align}
  &\tilde{c}_0 = c_0\,,  &&\tilde{c}_2 = \frac{3}{2f} \, c_2\,, \\
  &\tilde{c}_4 = \frac{35}{8 f^2} \, c_4\,, &&\tilde{c}_{\rm nlo} = -c_{\rm nlo}\,.
\end{align}
Finally, the relation to the scale-dependent stochastic parameters $a_0$ and $a_2$ in \texttt{Class-PT} are:
\begin{equation}
  a_0 = k_{\rm NL}^2 \, \left(N^P_{20} - \frac{N^P_{22}}{2}\right) \,, \quad
  a_2 = \frac{3}{2} \, k_{\rm NL}^2 \, N^P_{22}\,,
\end{equation}
where they defined $k_{\rm NL} = 0.45\,\hinvMpc$.

\bibliography{references}

\end{document}